\RequirePackage{lineno} 
\documentclass[prc,twocolumn,superscriptaddress,showpacs,amssymb,amsmath,amsfonts,aps,nofootinbib]{revtex4-1}

\usepackage{array,mathtools,amssymb,booktabs}
\newcolumntype{C}{>{$}c<{$}}
\AtBeginDocument{
\heavyrulewidth=.08em
\lightrulewidth=.05em
\cmidrulewidth=.03em
\belowrulesep=.65ex
\belowbottomsep=0pt
\aboverulesep=.4ex
\abovetopsep=0pt
\cmidrulesep=\doublerulesep
\cmidrulekern=.5em
\defaultaddspace=.5em
}
\usepackage{graphicx}
\usepackage{dcolumn}
\newcolumntype{d}[1]{D{.}{.}{-1}} 

\usepackage{latexsym}
\usepackage{amsmath} 
\usepackage{url}
\usepackage{natbib}
\usepackage{verbatim}
\usepackage{mathtools}
\usepackage{framed}
\usepackage{diagbox}
\usepackage{multirow}
\usepackage{hyperref}
\usepackage{float}

\usepackage[margin=20mm]{geometry}
\usepackage[autostyle=true]{csquotes}

\newcolumntype{P}[1]{>{\centering\arraybackslash}p{#1}}

\newcolumntype{C}[1]{>{\centering\arraybackslash}m{#1}}

\usepackage{lineno}

\begin{document}


\date{\today}
\title{\large Double-pion electroproduction off protons in deuterium: \\Quasifree cross sections and final-state interactions \\}




   \newcommand*{\ANL}{Argonne National Laboratory, Argonne, Illinois 60439, USA} 
   \newcommand*{\ANLindex}{64}
   \affiliation{\ANL}
   \newcommand*{\ASU}{Arizona State University, Tempe, Arizona 85287-1504, USA}
   \newcommand*{\ASUindex}{9}
   \affiliation{\ASU}
   \newcommand*{\CANISIUS}{Canisius College, Buffalo, New York 14208, USA} 
   \newcommand*{\CANISIUSindex}{94}
   \affiliation{\CANISIUS}
   \newcommand*{\CMU}{Carnegie Mellon University, Pittsburgh, Pennsylvania 15213, USA}
   \newcommand*{\CMUindex}{28}
   \affiliation{\CMU}
   \newcommand*{\CUA}{Catholic University of America, Washington, DC 20064, USA }
   \newcommand*{\CUAindex}{21}
   \affiliation{\CUA}
   \newcommand*{\CNU}{Christopher Newport University, Newport News, Virginia 23606, USA }
   \newcommand*{\CNUindex}{27}
   \affiliation{\CNU}
   \newcommand*{\WM}{College of William and Mary, Williamsburg, Virginia 23187-8795, USA} 
   \newcommand*{\WMindex}{6}
   \affiliation{\WM}
   \newcommand*{\DUKE}{Duke University, Durham, North Carolina 27708-0305, USA }
   \newcommand*{\DUKEindex}{30}
   \affiliation{\DUKE}
   \newcommand*{\DUQUESNE}{Duquesne University, Pittsburgh, Pennsylvania 15282, USA  }
   \newcommand*{\DUQUESNEindex}{128}
   \affiliation{\DUQUESNE}
   \newcommand*{\FU}{Fairfield University, Fairfield, Connecticut 06824, USA }
   \newcommand*{\FUindex}{82}
   \affiliation{\FU}
   \newcommand*{\FIU}{Florida International University, Miami, Florida 33199, USA }
   \newcommand*{\FIUindex}{17}
   \affiliation{\FIU}
   \newcommand*{\FSU}{Florida State University, Tallahassee, Florida 32306, USA }
   \newcommand*{\FSUindex}{11}
   \affiliation{\FSU}
   \newcommand*{\GWUI}{The George Washington University, Washington, DC 20052, USA }
   \newcommand*{\GWUIindex}{13}
   \affiliation{\GWUI}
   \newcommand*{\GSI}{GSI Helmholtzzentrum f{\"u}r Schwerionenforschung GmbH, 64291 Darmstadt, Germany }
   \newcommand*{\GSIindex}{143}
   \affiliation{\GSI}
   \newcommand*{\INFNFE}{INFN, Sezione di Ferrara, 44100 Ferrara, Italy }
   \newcommand*{\INFNFEindex}{97}
   \affiliation{\INFNFE}
   \newcommand*{\INFNFR}{INFN, Laboratori Nazionali di Frascati, 00044 Frascati, Italy }
   \newcommand*{\INFNFRindex}{8}
   \affiliation{\INFNFR}
   \newcommand*{\INFNGE}{INFN, Sezione di Genova, 16146 Genova, Italy }
   \newcommand*{\INFNGEindex}{5}
   \affiliation{\INFNGE}
   \newcommand*{\INFNPAV}{INFN, Sezione di Pavia, 27100 Pavia, Italy }
   \newcommand*{\INFNPAVindex}{134} 
   \affiliation{\INFNPAV}
   \newcommand*{\INFNRO}{INFN, Sezione di Roma Tor Vergata, 00133 Rome, Italy }
   \newcommand*{\INFNROindex}{90}
   \affiliation{\INFNRO}
   \newcommand*{\INFNTUR}{INFN, Sezione di Torino, 10125 Torino, Italy }
   \newcommand*{\INFNTURindex}{108}
   \affiliation{\INFNTUR}
   \newcommand*{\SACLAY}{IRFU, CEA, Universit\'{e} Paris-Saclay, F-91191 Gif-sur-Yvette, France }
   \newcommand*{\SACLAYindex}{4}
   \affiliation{\SACLAY}
   \newcommand*{\KNU}{Kyungpook National University, Daegu 41566, Republic of Korea }
   \newcommand*{\KNUindex}{37}
   \affiliation{\KNU}
   \newcommand*{\LAMAR}{Lamar University, Beaumont, Texas 77710, USA }
   \newcommand*{\LAMARindex}{129}
   \affiliation{\LAMAR}
   \newcommand*{\MISS}{Mississippi State University, Mississippi State, Mississippi 39762-5167, USA }
   \newcommand*{\MISSindex}{96}
   \affiliation{\MISS}
\newcommand*{\GATCHINA}{National Research Centre ``Kurchatov Institute" B. P. Konstantinov Petersburg Nuclear Physics Institute, Gatchina, St. Petersburg, 188300, Russia}
\affiliation{\GATCHINA}
   \newcommand*{\NSU}{Norfolk State University, Norfolk, Virginia 23504, USA }
   \newcommand*{\NSUindex}{36}
   \affiliation{\NSU}
   \newcommand*{\OHIOU}{Ohio University, Athens, Ohio  45701, USA }
   \newcommand*{\OHIOUindex}{20}
   \affiliation{\OHIOU}
   \newcommand*{\ODU}{Old Dominion University, Norfolk, Virginia 23529, USA }
   \newcommand*{\ODUindex}{12}
   \affiliation{\ODU}
   \newcommand*{\JLU}{II Physikalisches Institut der Universitaet Giessen, 35392 Giessen, Germany }
   \newcommand*{\JLUindex}{116}
   \affiliation{\JLU}
   \newcommand*{\RPI}{Rensselaer Polytechnic Institute, Troy, New York 12180-3590, USA }
   \newcommand*{\RPIindex}{2}
   \affiliation{\RPI}
   \newcommand*{\RUHR}{Ruhr University Bochum, 44801 Bochum, Germany}
   \newcommand*{\RUHRindex}{227}
   \affiliation{\RUHR}

   \newcommand*{\MSU}{Skobeltsyn Institute of Nuclear Physics, Lomonosov Moscow State University, 119234 Moscow, Russia }
   \newcommand*{\MSUindex}{47}
   \affiliation{\MSU}
   \newcommand*{\TEMPLE}{Temple University,  Philadelphia, Pennsylvania 19122, USA  }
   \newcommand*{\TEMPLEindex}{109}
   \affiliation{\TEMPLE}
   \newcommand*{\JLAB}{Thomas Jefferson National Accelerator Facility, Newport News, Virginia 23606, USA }
   \newcommand*{\JLABindex}{10}
   \affiliation{\JLAB}
   \newcommand*{\BRESCIA}{Universit\`{a} degli Studi di Brescia, 25123 Brescia, Italy }
   \newcommand*{\BRESCIAindex}{132}
   \affiliation{\BRESCIA}
   \newcommand*{\UCR}{University of California Riverside, Riverside, California 92521, USA }
   \newcommand*{\UCRindex}{138}
   \affiliation{\UCR}
   \newcommand*{\UCONN}{University of Connecticut, Storrs, Connecticut 06269, USA }
   \newcommand*{\UCONNindex}{31}
   \affiliation{\UCONN}
   \newcommand*{\FERRARAU}{Universit\'{a} di Ferrara, 44121 Ferrara, Italy }
   \newcommand*{\FERRARAUindex}{117}
   \affiliation{\FERRARAU}
   \newcommand*{\GLASGOW}{University of Glasgow, Glasgow G12 8QQ, United Kingdom }
   \newcommand*{\GLASGOWindex}{35}
   \affiliation{\GLASGOW}
   \newcommand*{\UNH}{University of New Hampshire, Durham, New Hampshire 03824-3568, USA }
   \newcommand*{\UNHindex}{19}
   \affiliation{\UNH}
   \newcommand*{\ORSAY}{Universit\'{e} Paris-Saclay, CNRS/IN2P3, IJCLab, 91405 Orsay, France }
   \newcommand*{\ORSAYindex}{33}
   \affiliation{\ORSAY}
   \newcommand*{\URICH}{University of Richmond, Richmond, Virginia 23173, USA }
   \newcommand*{\URICHindex}{32}
   \affiliation{\URICH}
   \newcommand*{\ROMAII}{Universit\'{a} di Roma Tor Vergata, 00133 Rome, Italy }
   \newcommand*{\ROMAIIindex}{91}
   \affiliation{\ROMAII}
   \newcommand*{\SCAROLINA}{University of South Carolina, Columbia, South Carolina 29208, USA }
   \newcommand*{\SCAROLINAindex}{15}
   \affiliation{\SCAROLINA}
   \newcommand*{\UTFSM}{Universidad T\'{e}cnica Federico Santa Mar\'{i}a, Casilla 110-V Valpara\'{i}so, Chile }
   \newcommand*{\UTFSMindex}{87}
   \affiliation{\UTFSM}
   \newcommand*{\VIRGINIA}{University of Virginia, Charlottesville, Virginia 22901, USA} 
   \newcommand*{\VIRGINIAindex}{25}
   \affiliation{\VIRGINIA}
   \newcommand*{\YORK}{University of York, York YO10 5DD, United Kingdom} 
   \newcommand*{\YORKindex}{130}
   \affiliation{\YORK}
  
   \newcommand*{\YEREVAN}{Yerevan Physics Institute, 375036 Yerevan, Armenia}
   \newcommand*{\YEREVANindex}{3}
   \affiliation{\YEREVAN}

\author {Iu.~A.~Skorodumina} 
\affiliation{\JLAB}
\affiliation{\SCAROLINA}
\author {G.~V.~Fedotov} 
\affiliation{\GATCHINA}
\author {R.~W.~Gothe} 
\affiliation{\SCAROLINA}

\author {P.~Achenbach}
\affiliation{\JLAB}

\author {Z.~Akbar}
\affiliation{\VIRGINIA} 

\author {J.~S.~Alvarado} 
\affiliation{\ORSAY}

\author {M.~J.~Amaryan}
\affiliation{\ODU} 

\author {Whitney~R.~Armstrong}
\affiliation{\ANL} 

\author {H.~Atac} 
\affiliation{\TEMPLE}

\author {H.~Avakian}
\affiliation{\JLAB} 

% ...

\collaboration{The CLAS Collaboration}
\noaffiliation

\begin{abstract}
The single-differential and fully integrated cross sections for quasifree $\pi^{+}\pi^{-}$ electroproduction off protons bound in deuterium have been extracted for the first time. The experimental data were collected at Jefferson Laboratory with the CLAS detector. The measurements were performed in the kinematic region of the invariant mass $W$ from 1.3 to 1.825~GeV and the photon virtuality $Q^{2}$ from 0.4 to 1.0~GeV$^2$. Sufficient experimental statistics allowed for narrow binning in all kinematic variables, while maintaining a small statistical uncertainty. The extracted cross sections were compared with the corresponding cross sections off free protons, which allowed us to obtain an estimate of the contribution from events in which interactions between the final-state hadrons and the spectator neutron took place.

\end{abstract}


\maketitle


\section{Introduction  }
\label{intro}


Exclusive reactions of meson photo- and electroproduction off protons are intensively studied in laboratories around the world as a very powerful tool to investigate nucleon structure and the principles of the strong interaction. These studies include extraction of various observables from analyses of experimental data, as well as subsequent theoretical and phenomenological interpretations of the extracted observables~\cite{Krusche:2003ik,Aznauryan:2011qj,Skorodumina:2016pnb,Ireland:2019uwn,Thiel:2022xtb}.

Exclusive reactions off free protons have already been studied in considerable detail, and a lot of information on differential cross sections and different single- and double-polarization asymmetries with almost complete coverage of the reaction phase space has become available. A large part of this information comes from analyses of data collected in Hall B at Jefferson Lab (JLab) with the CLAS detector~\cite{Mecking:2003zu,CLAS_DB}.

Meanwhile, reactions occurring in photon and electron scattering off nuclei have been less extensively investigated. Experimental information on these processes is sparse and mostly limited to inclusive measurements of the total nuclear photoproduction cross section~\cite{Mokeev:1995fy,Bianchi:1994ax,Ahrens:1986hn} and the nucleon structure function $F_{\text{2}}$~\cite{Osipenko_2005_note,Osipenko:2005gt,Osipenko:2010sb}, while exclusive measurements off bound nucleons are lacking.

However, information on exclusive reactions off bound nucleons is crucially important to the investigation of nuclear structure and for a deeper understanding of the processes occurring in the nuclear medium because various exclusive channels will have different energy dependences and different sensitivities to reaction mechanisms. This situation creates a strong demand for exclusive measurements off bound nucleons, and the deuteron, being the lightest and most weakly bound nucleus, is the best target with which to begin these efforts.

This paper presents the results of the data analysis of charged double-pion electroproduction off protons bound in a deuteron. The study became possible owing to the experiment of electron scattering off a deuterium target conducted in Hall B at JLab with the CLAS detector~\cite{Mecking:2003zu}. The description of the detector and target setup is given in Sec.\!~\ref{Chapt:experiment} together with information on the overall data analysis strategy.

The analysis covers the second resonance region, where double-pion production plays an important role. The channel opens at the double-pion production threshold $W \approx 1.22$~GeV, contributes significantly to the total inclusive cross section for $W \lesssim 1.6$~GeV, and dominates all other exclusive channels for $W \gtrsim 1.6$~GeV.

Exclusive reactions off bound protons manifest some specific features that are not present in reactions off free protons and originate from (a) Fermi motion of the initial proton and (b) final-state interactions (FSIs) of the reaction final hadrons with the spectator neutron. In this paper, special attention is paid to the detailed description of these issues.

The paper introduces new information on the fully integrated and single-differential cross sections of the reaction $\gamma_{v}p(n) \rightarrow p' (n')\pi^{+}\pi^{-}$, where $\gamma_{v}$ denotes the virtual photon. The cross section measurements were performed in the kinematic region of the invariant mass $W$ from 1.3 to 1.825~GeV and the photon virtuality $Q^{2}$ from 0.4 to 1.0~GeV$^2$. Sufficient experimental statistics allowed for narrow binning, {\it i.e.}~25~MeV in $W$ and 0.05~GeV$^2$ in $Q^2$, while maintaining adequate statistical uncertainties. The extracted cross sections are quasifree, meaning that the admixture of events in which the final hadrons interacted with the spectator neutron is kinematically reduced to the achievable minimum.

The details of the cross section extraction analysis are presented in Secs.~\ref{Sect:select} through~\ref{Sect:uncert}, which encompass the selection of quasifree events, the cross section calculation framework, the description of the corrections applied to the cross sections, and the study of the cross section uncertainties, respectively.

Effects of the initial proton motion (also called Fermi motion) turned out to be tightly interwoven with many analysis aspects, and for this reason their description is scattered throughout the paper. Meanwhile, FSI effects are addressed in separately in Sec.~\ref{Sect:discuss_fsi}, which outlines specificities of FSIs in reactions off bound protons and their differences from FSIs in conventional free proton reactions. Some details on FSI effects in this particular analysis are also presented there.

Section~\ref{Sect:cr_sect_qf} presents the measured cross sections and their comparison with the cross section estimation obtained based on the JLab-MSU (JM) model\footnote[1]{JLab---Moscow State University (Russia) model.}, which is a phenomenological reaction model for the process of double-pion production off free protons~\cite{Mokeev:2008iw,Mokeev:2012vsa,Mokeev:2015lda}.

This study benefits from the fact that the free proton cross sections of the same exclusive reaction have been recently extracted from CLAS data~\cite{Fed_an_note:2017,Fed_paper_2018}. These free proton measurements were performed under the same experimental conditions as in this study, including the beam energy value and the target setup. For this reason, the free proton study~\cite{Fed_an_note:2017,Fed_paper_2018} was naturally used as a reference point for many analysis components. This unique advantage allowed us not only to verify the reliability of those analysis aspects that are similar for reactions off free and bound protons but also to obtain a deeper understanding of those that differ. The latter include the effects of the initial proton motion and FSIs.

Section~\ref{Sect:with_fed_comp} introduces a comparison of the cross sections extracted in this study with their free proton counterparts from Refs.\!~\cite{Fed_an_note:2017,Fed_paper_2018}, which allowed us to estimate the proportion of events in which FSIs between the final hadrons and the spectator neutron took place. Assuming that events affected by FSIs with the spectator are the main cause of the difference between the cross section sets, their contribution to the total number of reaction events was found to be on a level of $\approx$25\% in the most part of the reaction phase space. With further theoretical interpretation of the performed comparison, various FSI mechanisms can be investigated and other potential reasons that may contribute to the difference between the cross section sets can be explored, which includes possible in-medium modifications of properties of nucleons and their excited states~\cite{Mokeev:1995fy,Bianchi:1994ax,Ahrens:1986hn,Krusche:2004xz,Noble:1980my}.

\section{Experiment}
\label{Chapt:experiment}

The electron scattering experiment that provided data for this study was conducted in Hall B at JLab as a part of the ``e1e" run period. A~longitudinally polarized electron beam was produced by the Continuous Electron Beam Accelerator Facility (CEBAF) and then was subsequently scattered off the target, which was located in the center of the CEBAF Large Acceptance Spectrometer (CLAS)~\cite{Mecking:2003zu}. This state-of-the-art detector covered a good fraction of the full solid angle and provided efficient registration of final-state particles originating from the scattering process.

The ``e1e" run period lasted from November 2002 until January 2003 and included several experiments with different beam energies (1 and 2.039 GeV) and target cell contents (liquid hydrogen and liquid deuterium). This study is devoted to the experiment conducted with the liquid-deuterium target and utilizing a 2.039-GeV electron beam. 

\subsection{Detector setup}

The design of the CLAS detector was based on a toroidal magnetic field that was generated by six superconducting coils arranged around the beam line. The magnetic field bent charged particles towards or away from the beam axis (depending on the particle charge and~the direction of the torus current) but left the azimuthal angle essentially unchanged. For this experiment, the torus field setting was to bend negatively charged particles towards the beam line (inbending configuration).

The magnet coils naturally separated the detector into six sectors, each functioning as an independent magnetic spectrometer. Each sector included four subsystems: drift chambers (DC), \v Cerenkov counters (CC), a time-of-flight (TOF) system, and an electromagnetic calorimeter (EC)~\cite{Mecking:2003zu}.

The azimuthal coverage for CLAS was limited only by the magnet coils and was approximately 90\% at backward polar angles and 50\% at forward angles~\cite{Amarian:2001zs}. The polar angle coverage spanned from $8^{\circ}\mathrm{}$ to $45^{\circ}\mathrm{}$ for the \v Cerenkov counters and the electromagnetic calorimeter and from $8^{\circ}\mathrm{}$ to $140^{\circ}\mathrm{}$ for the drift chambers and the time-of-flight system.

The drift chambers were located within the region of the magnetic field and performed charged particle tracking, allowing for determination of the particle momentum from the curvature of their trajectories~\cite{Mestayer:2000we}. All other subdetectors were located outside the magnetic field region, meaning that charged particles traveled through them along a straight line.

The \v Cerenkov counters were located right behind the DC and served the dual function of triggering on electrons and separating electrons from pions~\cite{Adams:2001kk}. 

The TOF scintillators were located radially outside the drift chambers and the \v Cerenkov counters but in front of the calorimeter. The time-of-flight system measured the time when a particle hit a TOF scintillator, thus allowing for determination of the particle velocity. Then, using the particle momentum known from the DC, its mass can be determined, meaning that the particle can be identified~\cite{Smith:1999ii, clas_tof_paddles}.

The main functions of the electromagnetic calorimeter were triggering on and detection of electrons, as well as detection of photons (allowing for the $\pi^{0}$ and $\eta$ reconstructions from their $2\gamma$ decays) and neutrons~\cite{Amarian:2001zs}. 

The six CLAS sectors were equipped with a common data-acquisition (DAQ) system that collected the digitized data and stored the information for later offline analysis.

\subsection{Target setup}
\label{Sect:target_setup}

The ``e1e" target had a conical shape with the diameter varying from 0.35 to 0.6 cm, which served the purpose of effective extraction of gas bubbles formed in the liquid target contents due to the heat that either originated from the beam and/or came from outside through the target walls. Due to the conical shape, the bubbles drained upwards and into a wider area of the target, thus clearing the beam interaction region and allowing the boiled deuterium to be effectively delivered back to the cooling system to be reliquified. 

The target was located at $-$0.4 cm along the $z$ axis (near the center of CLAS) and its interaction region was 2 cm long. The target cell had 15-$\mu$m-thick aluminum entrance and exit windows. In addition, an aluminum foil was located 2.0 cm downstream of the target. This foil was made exactly to the same specifications as the entry and exit windows of the target cell and served the purpose of providing reference event distributions for electron scattering off the target windows (see Sec.\!~\ref{Sect:vertex}).

In this experiment, the target setup was identical to that used for the free proton part of the ``e1e" run period~\cite{Fed_an_note:2017,Fed_paper_2018}. More details on the ``e1e" target assembly can be found in Ref.\!~\cite{target}.

\subsection{Data analysis strategy}
\label{Sect:clas_software}

Events corresponding to the investigated reaction $ep(n) \rightarrow e'p'(n')\pi^{+}\pi^{-}$ were distinguished among all other detected events through the event selection procedure described in detail in Sec.\!~\ref{Sect:select}. The selected exclusive events,  however, represent only a part of the total number of events produced in the reaction, while the remainder were not registered due to (i) geometrical holes in the detector acceptance and (ii) less than 100\% efficient particle detection within the detector acceptance. Therefore, to extract the reaction cross sections, the experimental event yield was adjusted for the geometric acceptance and detection efficiency, thereby accounting for the lost events.

To determine the overall detector efficiency, a Monte Carlo simulation was performed. In this analysis, double-pion events were generated with TWOPEG-D, which is an event generator for double-pion electroproduction off a moving proton~\cite{twopeg-d}. These events are hereinafter called ``generated" events. 

The generated events were passed through a standard multi-stage procedure of simulating the detector response~\cite{Mecking:2003zu}. The procedure included the simulation of the particle propagation through the CLAS detector from the vertex produced by the event generator and the subsequent event reconstruction. Events that survived this process are hereinafter called ``reconstructed" Monte Carlo events. They were analyzed in the same way as real experimental events.

\section{Quasifree event selection}
\label{Sect:select}

For each analyzed event, an electron candidate was defined as the first-in-time particle that had signals in all four subsystems of the CLAS detector (DC, CC, TOF, and EC). To select hadron candidates, signals only in two subdetectors (DC and TOF) were required.

\subsection{Electron identification}

To select good electrons among all electron candidates and to separate them from electronic noise, accidentals, and $\pi^{-}$ contamination, various cuts in the electromagnetic calorimeter (EC) and the \v Cerenkov counters (CC) were applied.

According to Ref.\!~\cite{Egian:007}, the overall EC resolution, as well as uncertainties from the EC output summing electronics, led to fluctuations of the EC response near the hardware threshold. Therefore, to select only reliable EC signals, a minimal cut on the scattered electron momentum $P_{e'}$ was applied in the software. The value of this cut was chosen to be 0.461~GeV, according to the relation suggested in Ref.\!~\cite{Egian:007}.

\begin{figure}[htp]
\begin{center}
 \includegraphics[width=0.45\textwidth,keepaspectratio]{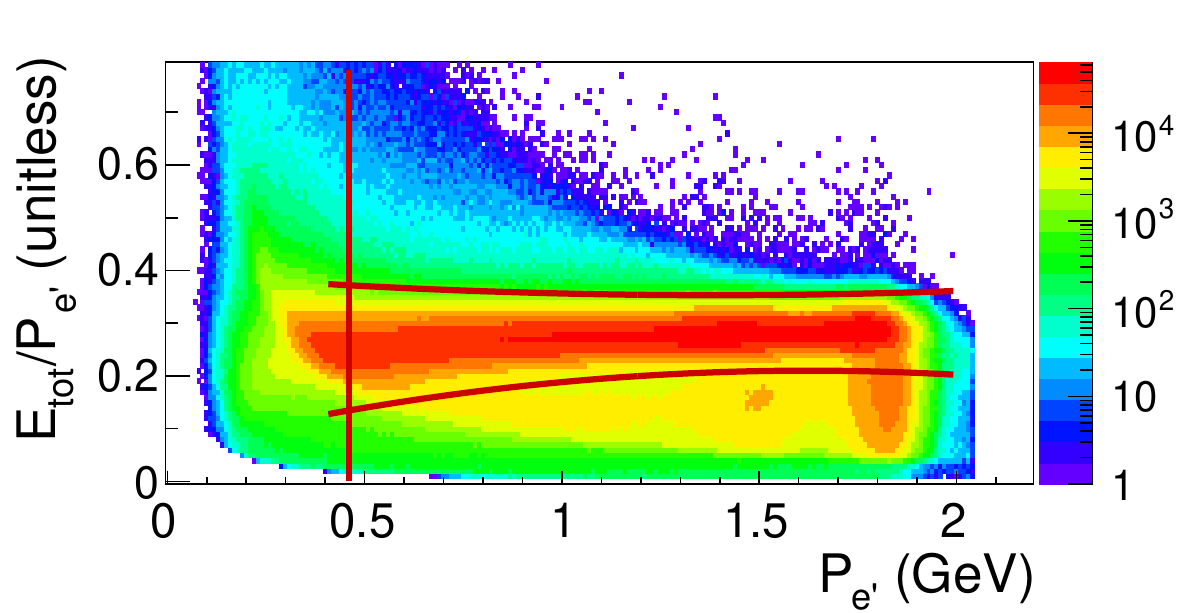}
\caption{EC sampling fraction distribution for the experimental data for CLAS sector 1. The vertical red line shows the position of the minimum momentum cut, while the other two curves correspond to the sampling fraction~cut.}
\label{fig:ec_cut}
\end{center}
\end{figure}

In the next step, to eliminate part of the pion contamination, a sampling fraction cut was applied, which was based on different energy deposition patterns of electrons and pions in the EC. Specifically, when traveling through the EC, an electron produces an electromagnetic shower, where the deposited energy $E_{\text{tot}}$ is proportional to the electron momentum $P_{e'}$, while a $\pi^{-}$ loses a constant amount of energy per scintillation layer independently of its momentum. Hence, for electrons, the quantity $E_{\text{tot}}/P_{e'}$ plotted as a function of $P_{e'}$ is expected to follow a straight line parallel to the $x$ axis and located around the value 1/3 on the $y$ axis (since electrons lose about 2/3 of their energy in the EC lead sheets).

Figure~\ref{fig:ec_cut} shows the EC sampling fraction ($E_{\text{tot}}/P_{e'}$) plotted as a function of the particle momentum for the experimental data. In this figure, the minimum momentum cut is shown by the vertical line, while the other two curves correspond to the sampling fraction cut that isolates the band of good electron candidates. The cut was determined via Gaussian fits to individual momentum slices of the distribution~\cite{my_an_note:2020, my_thesis:2021}.

\begin{figure}[htp]
\begin{center}
 \includegraphics[width=0.45\textwidth,keepaspectratio]{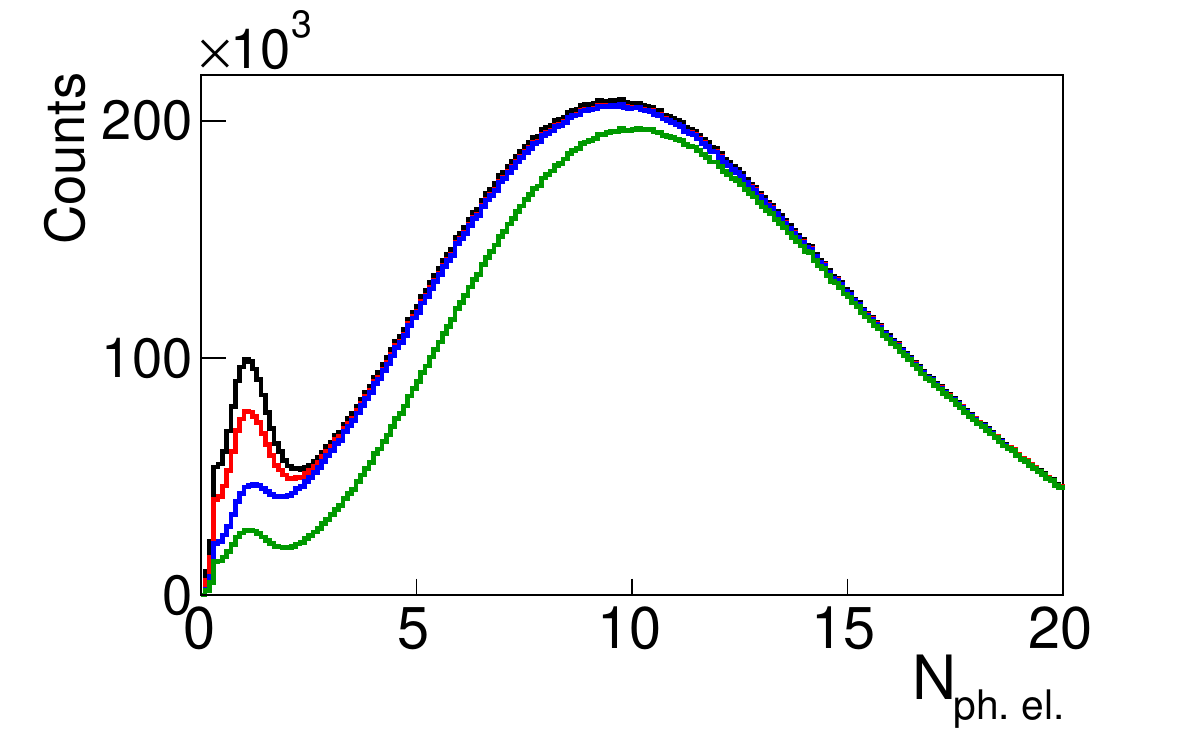}
\caption{Influence of different CC cuts on the photoelectron spectrum for CLAS sector 3. Curves from top to bottom: the black curve -- only the fiducial cut in the CC plane is applied, the red curve -- the $\varphi_{\text{cc}}$ matching cut is added, the blue curve -- the $\theta_{\text{cc}}$ matching cut is added, and the green curve -- the geometrical cut that removes inefficient CC zones is added. The peaking structure on the left is the contamination peak.}
\label{fig:photoel}
\end{center}
\end{figure}

To further improve the quality of electron identification and $\pi^{-}/e^{-}$ separation, event selection in the CC was performed~\cite{Adams:2001kk}. Figure~\ref{fig:photoel} illustrates photoelectron distributions measured in the CC for CLAS sector 3.  As seen in Fig.\!~\ref{fig:photoel}, contamination is present in the CC spectrum in the form of a peak located at values of a few photoelectrons. The contamination is thought to originate from accidental coincidences of measured pion tracks with noise signals from photomultiplier tubes (PMTs)~\cite{Osipenko:2004}. The goal of event selection in the CC was to separate the spectrum of good electron candidates from the contamination peak, while minimizing the loss of good events. To achieve this goal, the following set of CC cuts was applied: 

\begin{itemize}
\item fiducial cut in the CC,
\item $\varphi_{\text{cc}}$ matching cut,
\item $\theta_{\text{cc}}$ matching cut,
\item geometrical cut that removes inefficient zones, and
\item cut on the number of photoelectrons.
\end{itemize}

Each of these cuts, except the last one, was defined in the ``CC projective plane"~\cite{Osipenko:2004}, wherein the polar and azimuthal angles $(\theta_{\text{cc}},\varphi_{\text{cc}})$ were defined. The details on the plane definition and angle calculations can be found in Refs.\!~\cite{my_an_note:2020,Osipenko:2004}.

The analytical shape of the fiducial cut in the CC plane was taken from Ref.\!~\cite{Khetarpal:2010}. The $\varphi_{\text{cc}}$ and $\theta_{\text{cc}}$ matching procedures were based on the studies~\cite{Osipenko:2004} and~\cite{Ungaro:2010} and relied on the anticipation that for real events, there must be a one-to-one correspondence between PMT signals in the CC and the angles in the CC plane (which are calculated from the DC information), while background noise and accidentals should not show such a correlation.

The idea of the $\varphi_{\text{cc}}$ matching cut is that the particle track on the right side of the CC segment should match with the signal from the right-side PMT, and vice versa. Events that did not satisfy these conditions were removed. Events with signals from both PMTs were kept.

\begin{figure}[htp]
\begin{center}
 \includegraphics[width=0.45\textwidth,keepaspectratio]{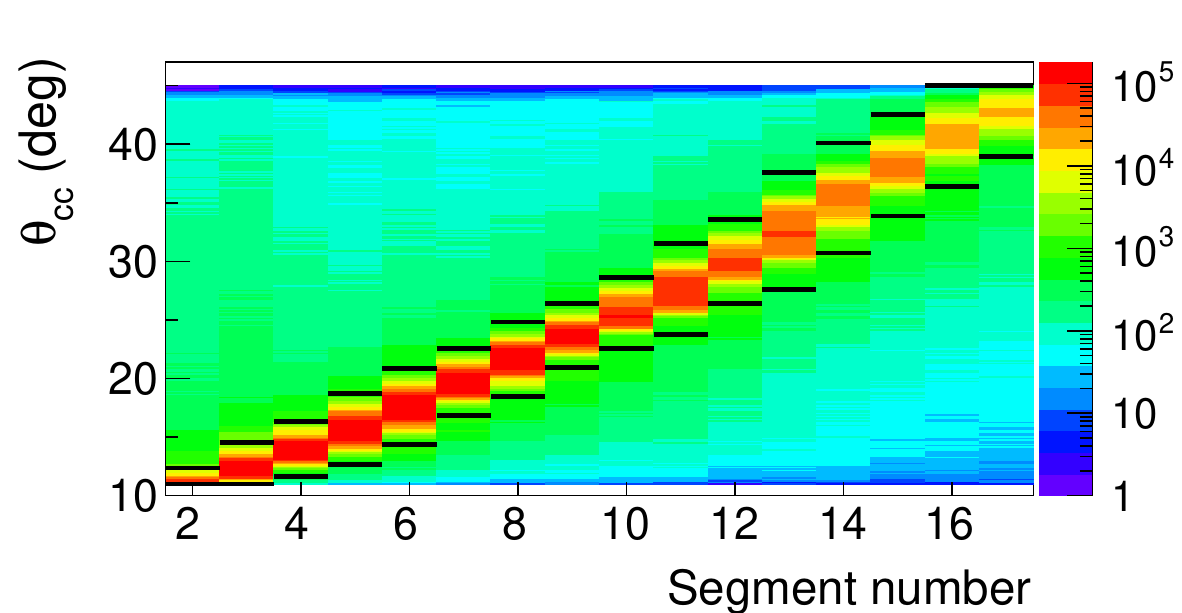}
\caption{$\theta_{\text{cc}}$ versus CC segment number distribution for CLAS sector 2. Events between the horizontal black lines were treated as good electron candidates.}
\label{fig:th_vs_seg}
\end{center}
\end{figure}

To perform $\theta_{\text{cc}}$ matching, a $\theta_{\text{cc}}$ versus segment number cut was applied. Figure~\ref{fig:th_vs_seg} shows the $\theta_{\text{cc}}$ versus segment number distribution for CLAS sector 2. To develop the cut, event distributions in each segment were plotted as a function of $\theta_{\text{cc}}$ and fit with Gaussians. The horizontal black lines correspond to the positions of the fit maxima $\pm4\sigma$. Events between these black lines were treated as good electron candidates~\cite{my_an_note:2020, my_thesis:2021}.

Another important issue was that some specific geometrical zones in the CC showed low detection efficiency. When an electron hit such a zone, the number of detected photoelectrons was significantly less than expected. This led to a systematic overpopulation of the low-lying part of the photoelectron spectrum and an enhancement of the contamination peak. Low-efficiency zones were distributed inhomogeneously in the CC plane and (being too dependent on specific features of the CC design) could not be simulated by the Monte Carlo technique. For this reason, a geometrical cut that removes inefficient zones was applied to both experimental events and reconstructed Monte Carlo events. The details on this cut can be found in Refs.\!~\cite{my_an_note:2020, my_thesis:2021,Fed_an_note:2017,Fed_paper_2018}.

The influence of the above cuts on the photoelectron spectrum is demonstrated in Fig.\!~\ref{fig:photoel}, where the distribution before the matching cuts is plotted in black, after the $\varphi_{\text{cc}}$ matching -- in red, and after the subsequent $\theta_{\text{cc}}$ matching cut -- in blue. As seen in Fig.\!~\ref{fig:photoel}, the matching cuts reduce the contamination peak but do not affect the main part of the photoelectron spectrum. Finally, the green distribution is plotted after adding the cut that removes inefficient CC zones. As expected, this cut leads to an event reduction in the low-lying part of the photoelectron spectrum (including the region of the contamination peak), while leaving the high-lying part of the distribution essentially unchanged.

The applied cuts result in a significant reduction of the contamination peak and its better separation from the main part of the photoelectron spectrum. As a final step, an explicit cut on the number of photoelectrons was applied, which altogether eliminated the remains of the contamination peak. The cut position was individually optimized for each PMT for each CC segment for each CLAS sector. As the Monte Carlo simulation did not reproduce photoelectron distributions well enough, the cut was performed only on the experimental data. To recover good electrons lost in this way, a standard correction procedure was applied, which was based on the fit of the photoelectron distributions by a modified Poisson function. More details on the procedure can be found in Refs.\!~\cite{my_an_note:2020, my_thesis:2021,Fed_an_note:2017,Fed_paper_2018}.

\subsection{Hadron identification}

Hadrons were identified through the timing information provided by the TOF system~\cite{Smith:1999ii, clas_tof_paddles}, which allowed the velocity $(\beta_{h} = v_{h}/c)$ of the hadron candidates to be determined.

A charged hadron can be identified by comparing $\beta_{h}$ determined from TOF information with $\beta_{\text{n}}$ given by 
\begin{equation}
\beta_{\text{n}}=\frac{p_{h}}{\sqrt{p_{h}^{2}+m_{h}^{2}}}.
\label{eq:hadron_hadronmass}
\end{equation}

In Eq.\!~\eqref{eq:hadron_hadronmass}, the quantity $\beta_{\text{n}}$ (which is termed the nominal value) is calculated using the particle momentum ($p_{h}$) known from the DC and the exact particle mass assumption ($m_{h}$).

To develop hadron identification cuts, experimental distributions of $\beta_{h}$ were examined as a function of the particle momentum for each TOF scintillator for each CLAS sector. This examination revealed that for some scintillation counters, the distributions were either shifted from their nominal positions or showed a double-band structure. To correct the timing information for such counters, a special procedure, described in Refs.\!~\cite{my_an_note:2020, my_thesis:2021}, was applied. In addition, a few counters were found to give unreliable signals; they were removed from consideration for both the experimental data and the simulation.

\begin{figure}[htp]
\begin{center}
 \includegraphics[width=0.45\textwidth,keepaspectratio]{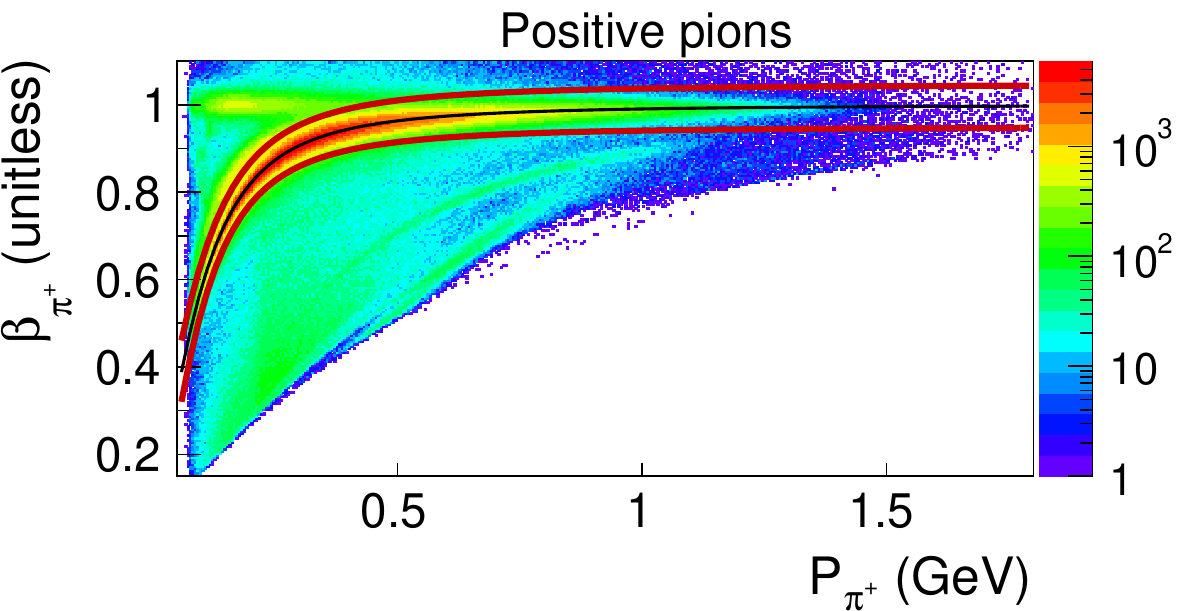}
\caption{$\beta_{h}$ versus momentum distribution for positive pion candidates. The thin black curve in the middle of the event band corresponds to the nominal $\beta_{\text{n}}$ given by Eq.\!~\eqref{eq:hadron_hadronmass}. The red curves show the applied identification cuts. Events between the red curves were treated as good pion candidates. }
\label{fig:hadron_id}
\end{center}
\end{figure}

Figure~\ref{fig:hadron_id} shows the $\beta_{h}$ versus momentum distribution for positive pion candidates plotted for all sectors and all reliable scintillators. The event band of the pion candidates is clearly seen. The thin black curve in the middle of the event band corresponds to the nominal $\beta_{\text{n}}$ given by Eq.\!~\eqref{eq:hadron_hadronmass}. The red curves show the applied identification cuts. Events between the red curves were selected for further analysis. Analogous identification cuts were performed for the proton and negative pion candidates.

\subsection{Momentum corrections}

While traveling through the detector and the target, the final-state particles lose a part of their energy due to interactions with the medium. As a result, the measured particle momentum appears to be lower than the actual value. In the investigated kinematic region, this effect is pronounced only for low-energy protons, while for all other detected particles it is insignificant.

The simulation of the CLAS detector correctly propagates particles through the media, and therefore the effect of the energy loss is already included in the estimated detector efficiency and does not impact the extracted cross sections. Nevertheless, in this study, the proton momentum magnitude was corrected for the energy loss. The simulation of the CLAS detector was used to establish the correction function, which was then applied to both experimental events and reconstructed Monte Carlo events~\cite{my_an_note:2020, my_thesis:2021}.

Additionally, Ref.\!~\cite{KPark:momcorr} provides evidence that particle momenta and angles may have some small systematic deviations from their real values due to slight misalignments in the DC position, small inaccuracies in the description of the torus magnetic field, and other possible reasons. The magnitude of this effect depends on the particle momentum, increasing as the momentum grows. In the investigated kinematic region, the effect was discernible only for scattered electrons. 

Due to undefined origin of the above effect, it cannot be simulated, and therefore, it has become conventional for CLAS data analyses to apply a special momentum correction to the experimental data. This particular study uses the electron momentum corrections that have previously been developed and tested in the analysis of the free proton part of the ``e1e" dataset at the same beam energy~\cite{Fed_an_note:2017,Fed_paper_2018}. To establish them, the approach from Ref.~\cite{KPark:momcorr}, which was based on elastic kinematics, was used. The applied corrections include an electron momentum magnitude correction as well as an electron polar angle correction, which were developed for each CLAS sector individually.

\subsection{Other cuts}

\subsubsection{Fiducial cuts}

The active detection solid angle of the CLAS detector was smaller than $4\pi$, in part due to the space occupied by the torus field coils. This is to say that the angles covered by the coils were not equipped with any detection system and therefore formed a ``dead" area for particle detection~\cite{Mecking:2003zu}. Additionally, the detection area was further limited in the polar angle from 8$^{\circ}\mathrm{}$ up to 45$^{\circ}\mathrm{}$ for electrons and up to 140$^{\circ}\mathrm{}$ for other charged particles~\cite{Mecking:2003zu}. 

As was shown in different data analyses, the edges of the active detection area also do not provide a safe region for particle reconstruction, being affected by rescattering from the coils, field distortions, and similar effects. Therefore, it has become common practice to exclude these regions from consideration by applying specific fiducial cuts. This method guarantees that events accepted in the analysis include only particles detected in ``safe" areas of the detector, where the acceptance is well understood.

In this study, fiducial cuts were applied for all four final-state particles ($e'$, $p'$, $\pi^{+}$, and $\pi^{-}$) for both experimental events and reconstructed Monte Carlo events. The analytical shapes of the cuts are similar to those used in the analysis of the free proton part of the ``e1e" dataset at the same beam energy~\cite{Fed_an_note:2017,Fed_paper_2018}, and more details can be found in Refs.\!~\cite{my_an_note:2020, my_thesis:2021}.

\subsubsection{Data quality checks}

During a long experimental run, variations of the experimental conditions ({\it e.g.}~fluctuations in the target density, deviations in the beam current and position, and/or changes in the detector response) can lead to fluctuations in event yields. To select for the analysis only the parts of the run with relatively stable event rates, cuts on the data-acquisition (DAQ) live time and the number of events per Faraday cup (FC) charge were applied.

The FC charge updated with a given frequency, so that the whole run time can be divided into ``blocks", where each block corresponds to the portion of time between two FC charge readouts. The DAQ live time is the portion of time within the block during which the DAQ system was able to accumulate events. A significant deviation of the live time from the average value indicates event rate alterations.

To establish data quality check cuts, the DAQ live time, as well as the yields of inclusive and elastic events normalized to the FC charge, were examined as a function of the block number. Those blocks for which these quantities demonstrated large fluctuations were excluded from the analysis. In this study, the quality check cuts are similar to those used in Refs.\!~\cite{Fed_an_note:2017,Fed_paper_2018}. For more details, see also Refs.\!~\cite{my_an_note:2020, my_thesis:2021}.

\subsubsection{Vertex cut}
\label{Sect:vertex}

The analyzed dataset included runs with the target cell filled with liquid deuterium, as well as runs with the empty target cell. The latter are needed to account for background events produced by the electron scattering off the target windows.

\begin{figure}[htp]
\begin{center}
 \includegraphics[width=0.45\textwidth,keepaspectratio]{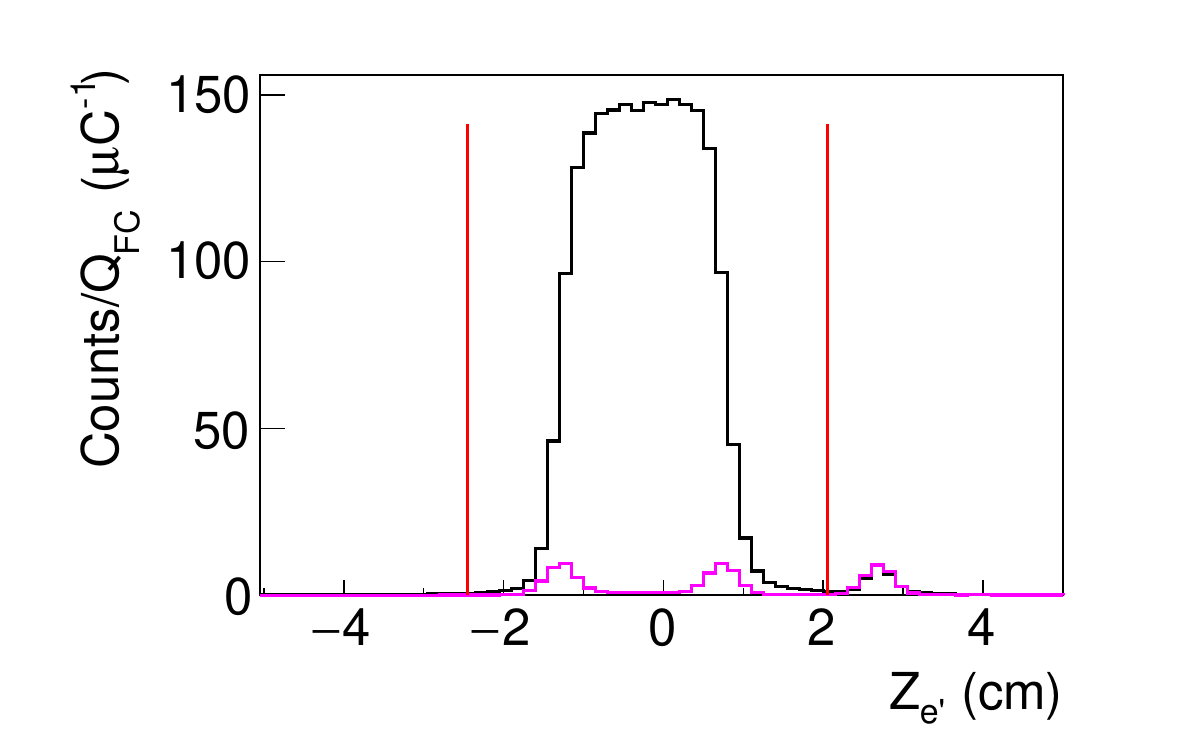}
\caption{Distributions of the electron $z$ coordinate at the vertex for full (black histogram) and empty (magenta histogram) target runs for CLAS sector 4. Vertical red lines show the applied cuts. Both the full and empty target distributions are normalized to the corresponding~FC~charge.  }
\label{fig:z_vertex}
\end{center}
\end{figure}

Figure~\ref{fig:z_vertex} presents the distributions of the electron $z$ coordinate at the interaction vertex for events from full and empty target runs (black and magenta histograms, respectively). Both distributions are normalized to the corresponding charge accumulated in the Faraday cup. The value of the vertex coordinate $z$ is corrected for the effects of the beam offset\footnote[2]{The beam offset is the deviation of the beam position from the CLAS central line $(x,y)=(0,0)$ that can lead to inaccurate determination of the vertex position.}~\cite{my_an_note:2020, my_thesis:2021}. Both distributions in Fig.\!~\ref{fig:z_vertex} demonstrate a well-separated peak around $z_{e'} = 2.6$~cm originating from the downstream aluminum foil. The distribution of events from the empty target runs shows two other similar peaks that correspond to the entrance and exit windows of the target cell (see also Sec.\!~\ref{Sect:target_setup}). 

Empty target events were passed through the same selection procedure that was established for the liquid-deuterium data and eventually were subtracted from the latter as described in Sec.\!~\ref{Sect:cr_sect_formula}. In addition to the empty target event subtraction, a cut on the electron $z$ coordinate was applied. This cut is shown by the two vertical lines in Fig.\!~\ref{fig:z_vertex}: events outside these lines were excluded from the analysis.

\subsection{Exclusivity cut in the presence of Fermi smearing and FSIs}
\label{Sect:excl_cut}

\subsubsection{Reaction topologies}
\label{Sect:topologies}

To identify a certain exclusive reaction, one needs to register the scattered electron and either all final hadrons or all except one. In the latter case, the four-momentum of the unregistered hadron can be deduced using energy-momentum conservation. Thus for the reaction $e p(n) \rightarrow e' p'(n') \pi^{+} \pi^{-} $ one can, in general, distinguish between four ``topologies" depending on the specific combination of registered final hadrons. In this particular analysis, the following two topologies were analyzed ($X$ denotes the undetected part),
\begin{itemize}
\item the fully exclusive topology (all final particles are registered) $e p (n) \rightarrow e' p'(n') \pi^{+} \pi^{-} X$ and
\item the $\pi^{-}$ missing topology $e p(n) \rightarrow e' p'(n') \pi^{+} X$.
\end{itemize}

The statistics of the fully exclusive topology are very limited, mainly because CLAS did not cover the polar angle range $0\,^{\circ}\mathrm{} < \theta_{\textrm{lab}} < 8\,^{\circ}\mathrm{}$~\cite{Mecking:2003zu}. In this experiment, the presence of this forward acceptance hole mostly impacted registration of the negative particles ($e$ and $\pi^{-}$) as their trajectories were bent by the torus magnetic field towards the beam axis. This led to a constraint on the minimum achievable $Q^2$ for electrons and prevented registration of the majority of negative pions. As a consequence, the $\pi^{-}$ missing topology contains the dominant part of the statistics. The contribution of the fully exclusive topology to the total analyzed statistics varies from $\approx$5\% near the reaction threshold to $\approx$25\% at $W$ between 1.7 and 1.8~GeV.

Besides the limited statistics, the fully exclusive topology also suffers from limited acceptance and hence from a very large number of empty cells (see Sec.\!~\ref{Sect:empt_cells} for more details on the empty cells). These circumstances do not allow for any sensible cross section information to be obtained from this topology alone. The $\pi^{-}$ missing topology has a tolerable number of empty cells and large statistics and therefore serves the purpose of the cross section extraction best. 

For both analyzed topologies, a small admixture of the three-pion background is present. This is similar to double-pion production off free protons, where the main background channel is $ep\rightarrow e'p'\pi^{+}\pi^{-}\pi^{0}$.  The free-proton analysis~\cite{Fed_an_note:2017,Fed_paper_2018} that was carried out for the same beam energy $E_{\text beam} = 2.039$~GeV demonstrated that although the admixture of events from this background channel becomes discernible at $W\gtrsim 1.6$~GeV, it remains negligible and well separated from double-pion events via the exclusivity cuts. For double-pion production off protons in deuterium, one more background channel can be distinguished, which is $en(p) \rightarrow e'p'(p')\pi^{+}\pi^{-}\pi^{-}$, but events from this channel follow the same kinematic pattern as events from the aforementioned $ep\rightarrow e'p'\pi^{+}\pi^{-}\pi^{0}$ reaction.

In general, two more topologies can be distinguished, {\it i.e.}~the proton missing topology and the $\pi^{+}$ missing topology. Both require registration of the $\pi^{-}$ in the final state and as a result suffer from the same issues of suppressed statistics and limited acceptance as the fully exclusive topology.  These two topologies are typically ignored in double-pion analyses~\cite{Rip_an_note:2002,Ripani:2002ss,Fed_an_note:2007,Fedotov:2008aa,Isupov:2017lnd}. Nevertheless, as demonstrated in a previous free-proton analysis~\cite{Fed_an_note:2017,Fed_paper_2018}, all four reaction topologies can be used in combination, which allows for an increase in the statistics and a reduction in the number of empty cells.

However, if a $\pi^{+}\pi^{-}$ pair was produced off a proton bound in deuterium, these two additional topologies turn out to be contaminated with events from other reactions. Specifically, in the proton missing topology, missing particle reconstruction fails to determine whether the pion pair was produced off the proton or off the neutron because their masses are almost identical. A similar situation occurs for the $\pi^+$ missing topology, where one can hardly distinguish between the production of a $\pi^{+}\pi^{-}$ pair off the proton and a $\pi^{0}\pi^{-}$ pair off the neutron, if only the proton and the $\pi^{-}$ in the final state were registered. Furthermore, the $\pi^+$ missing topology also has a strong admixture of events from the reaction $en(p)\rightarrow e'p'(p')\pi^{-}$. These circumstances prevented the use of the proton missing and the $\pi^{+}$ missing topologies in this analysis.

\subsubsection{Fermi smearing and FSIs}

Exclusive reactions off bound protons have the following features that are not present in reactions off free protons: (a) Fermi motion of the initial proton and (b) final-state interactions (FSIs) of the reaction final hadrons with the spectator neutron. These features introduce some complications into the exclusive event selection, as discussed below.

Since the momentum of the initial proton was not experimentally measured, this analysis uses the so-called target-at-rest assumption for calculation of some kinematic quantities (such as missing masses, reaction invariant mass $W$, etc.). This leads to the Fermi smearing of the corresponding experimental distributions~\cite{Skorodumina:2015rea,note_mm_distr} (see also Sec.\!~\ref{Sect:smearing_blurring}). To reliably identify the exclusive channel and correctly estimate the detector efficiency, a good match between the distributions of experimental events and reconstructed Monte Carlo events should be observed. This demands the simulated distributions reproduce the Fermi smearing of the experimental distributions, which implies that the effects of the initial proton motion are properly included in the Monte Carlo simulation.

For this reason, the Monte Carlo simulation in this analysis was performed using the TWOPEG-D~\cite{twopeg-d} event generator, which simulates a quasifree process of double-pion electroproduction off a moving proton. This is an extension of TWOPEG, which is an event generator for double-pion electroproduction off a free proton~\cite{twopeg}. For the TWOPEG-D version of the event generator, the Fermi motion of the initial proton is generated according to the Bonn potential~\cite{Machleidt:1987hj} and then is naturally merged into the specific kinematics of double-pion electroproduction.

FSIs of the reaction final hadrons with the spectator nucleon introduce the second intrinsic feature of exclusive reactions off bound nucleons. Such interactions alter the total four-momentum of the reaction final state and, therefore, introduce distortions into distributions of some kinematic quantities (such as missing masses), thus complicating identification of a specific exclusive channel~\cite{note_mm_distr} (see also Sec.\!~\ref{Sect:fsi_probing}).

In contrast to the effects of the initial proton motion, which can be simulated fairly easily, FSI effects can hardly be taken into account in the simulation because of their complex nature. The Monte Carlo simulation is hence not able to reproduce the distortions of some experimental distributions caused by FSIs with the spectator. For this reason, a proper procedure for isolation of quasifree events from the FSI background had to be developed.

The yield of events in FSI-disturbed kinematics turned out to strongly depend on (i) the reaction invariant mass $W$ and (ii) the hadron scattering angles. The latter issue causes FSI effects to manifest themselves differently depending on the reaction topology, since the topologies have nonidentical geometrical acceptance (see Sec.\!~\ref{sect:fsi_top_comp}). For this reason, the channel identification was performed in each topology individually, as described in the following sections.

\subsubsection{Fully exclusive topology}
\label{Sect:excl_cut_fully_excl}

To isolate quasifree double-pion events in the fully exclusive topology, the distributions of the quantities determined by Eq.\!~\eqref{eq:excl_top_quant} were used. The missing momentum $P_{X}$ and the missing mass squared $M^{2}_{X[0]}$ are defined for the reaction $ep(n)\rightarrow e'p'(n')\pi^{+}\pi^{-}X$, where $X$ corresponds to the undetected part.
\begin{equation}
\begin{aligned}
&P_{X}&=&~|\overrightarrow{P}_{e} - \overrightarrow{P}_{e'}- \overrightarrow{P}_{p'} - \overrightarrow{P}_{\pi^{+}} - \overrightarrow{P}_{\pi^{-}}|;\\[-2pt]
&M_{X[0]}^{2}&=&~[P_{e}^{\mu} + P_{p}^{\mu}- P_{e'}^{\mu}- P_{p'}^{\mu}-  P_{\pi^{+}}^{\mu} - P_{\pi^{-}}^{\mu}]^{2}.\\[-2pt]
\end{aligned}\label{eq:excl_top_quant}
\end{equation}

Here $P_{i}^{\mu}$ are the four-momenta and $\overrightarrow{P_{i}}$ the three-momenta of the particle $i$. Both quantities were calculated under the target-at-rest assumption, {\it i.e.}~considering $P^{\mu}_{p} = (0,0,0,m_{p})$ with the proton rest mass~$m_{p}$.

\begin{figure}[htp]
\begin{center}
\includegraphics[width=0.45\textwidth,keepaspectratio]{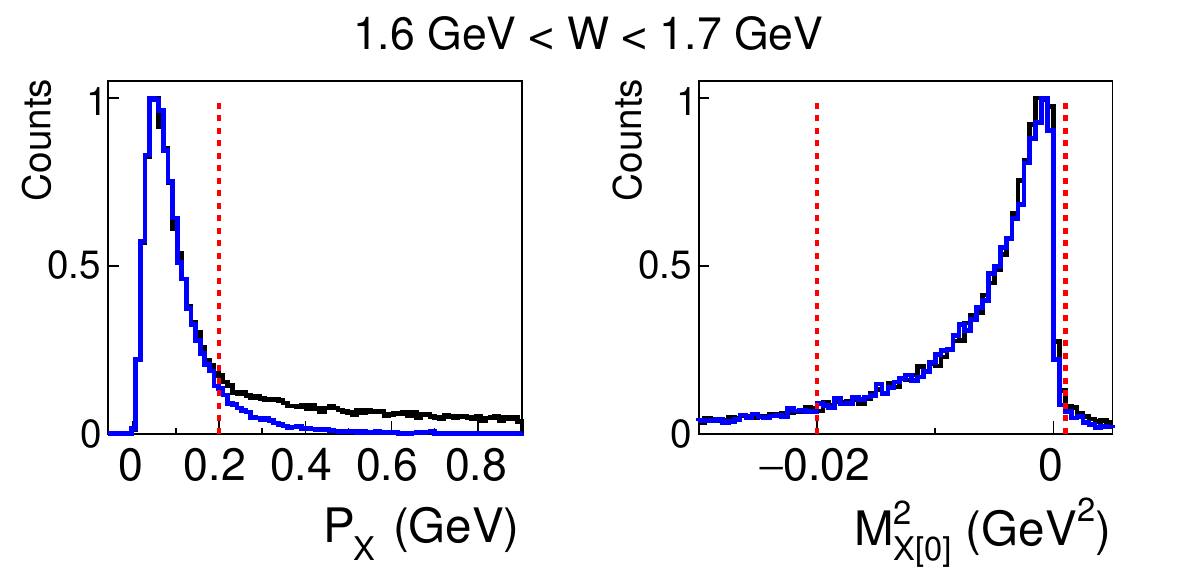}
\caption{Distributions of the quantities $P_{X}$ (left) and $M^{2}_{X[0]}$ (right) defined in Eq.\!~\eqref{eq:excl_top_quant} plotted for the experimental data (black) and the Monte Carlo simulation (blue) in one 100-MeV-wide bin in $W$. Vertical red lines indicate the cuts applied to select exclusive quasifree events. The distributions are normalized to their maxima.}
\label{fig:excl_top_cut}
\end{center}
\end{figure}

The quantities $P_{X}$ and $M^{2}_{X[0]}$ are unique for the fully exclusive topology, as they can be calculated only if all final hadrons were registered. Figure~\ref{fig:excl_top_cut} presents the distributions of $P_{X}$ (left plot) and $M^{2}_{X[0]}$ (right plot) for the experimental data (in black) and the Monte Carlo simulation (in blue) in~a 100-MeV-wide bin~in~$W$.

As seen in Fig.\!~\ref{fig:excl_top_cut}, the simulated $P_{X}$ distribution matches the experimental one for $P_{X} < 0.2$~GeV, while for $P_{X} > 0.2$~GeV the simulation underestimates the data. The mismatch mainly originates from experimental events in which the final hadrons interacted with the spectator neutron. The contribution from such events cannot be reproduced by the Monte Carlo simulation as the latter does not include FSI effects. The background channels, also not included in the simulation, contribute to the mismatch, too.

The cut on the missing momentum $P_{X}$ was applied to select exclusive events in quasifree kinematics, and the cut value was chosen to be $P_{X} = 0.2$~GeV. To further clean up the sample of selected events, the cut on the missing mass squared $M^{2}_{X[0]}$ was applied complementing the cut on the missing momentum. The cuts are shown in Fig.\!~\ref{fig:excl_top_cut} by the vertical red lines.

It is noteworthy that although in the fully exclusive topology the four-momentum of the $\pi^{-}$ was measured, it was not used in the subsequent calculation of kinematic variables for the cross section extraction. Instead, it was replaced by the four-momentum that was calculated as missing (and thus was Fermi smeared) to achieve consistency with the main $\pi^{-}$ missing topology.

\subsubsection{$\pi^{-}$ missing topology}
\label{Sect:excl_cut_pim_miss}

In the $\pi^{-}$ missing topology, the quantities $P_{X}$ and $M^{2}_{X[0]}$ defined in Eq.\!~\eqref{eq:excl_top_quant} are not available due to incomplete knowledge of the reaction final state. The channel identification was therefore performed using the four-momentum $P_{X[\pi^{-}]}^{\mu}$ for the reaction $ep(n)\rightarrow e'p'(n')\pi^{+}X$, which was calculated as
\begin{equation}
\begin{aligned}
P_{X[\pi^{-}]}^{\mu}=P_{e}^{\mu} + P_{p}^{\mu}- P_{e'}^{\mu}- P_{p'}^{\mu}-  P_{\pi^{+}}^{\mu},\\[-2pt]
\end{aligned}\label{eq:pimmiss_top_quant}
\end{equation}
where $P_{i}^{\mu}$ are the four-momenta of the particle $i$ and $X$ corresponds to the undetected part.

To isolate quasifree double-pion events in the $\pi^{-}$ missing topology, a special procedure was developed, in which the following quantity was used to perform the exclusivity cut:
\begin{equation}
\begin{aligned}
 M_{X[\pi^{-}]}=\sqrt{|[P_{X[\pi^{-}]}^{\mu}]^{2}|}.
\end{aligned}\label{eq:main_top_mm_nosq}
\end{equation}

\begin{figure}[htp]
\begin{center}
\includegraphics[width=0.45\textwidth,keepaspectratio]{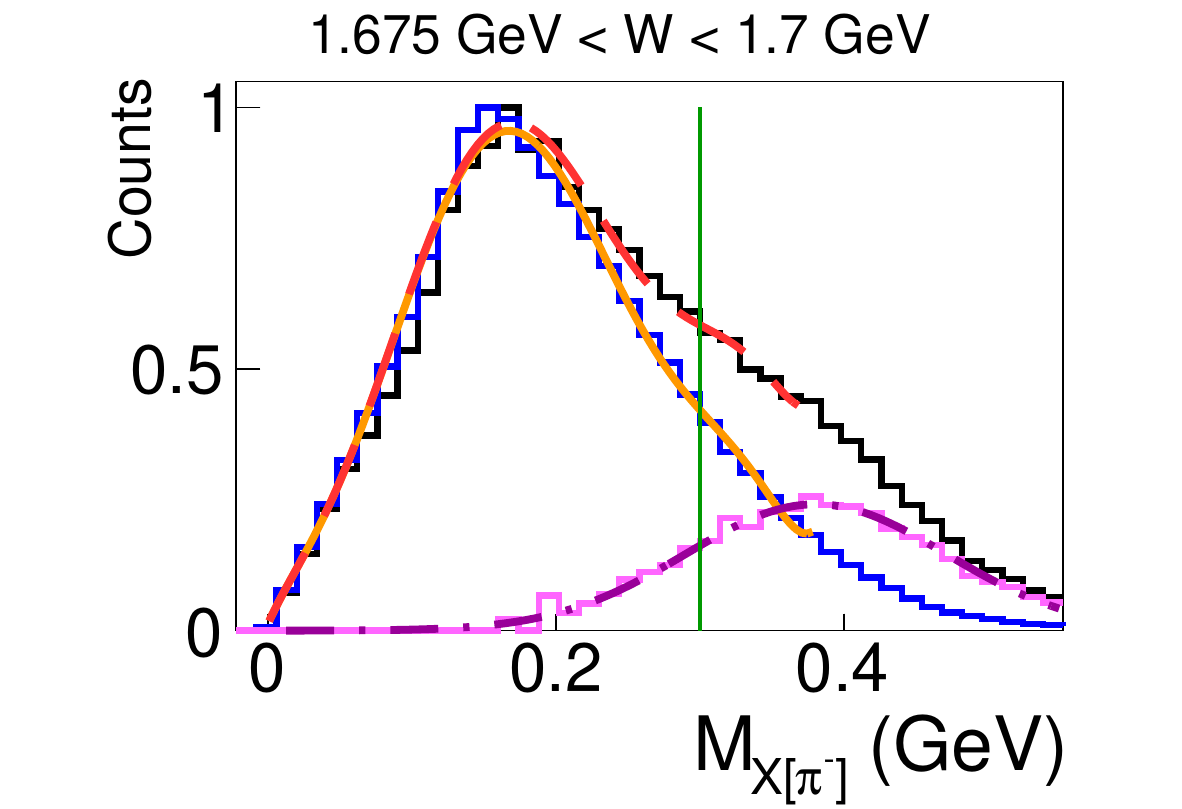}
\caption{Distributions of the quantity $M_{X[\pi^{-}]}$ [defined by Eq.\!~\eqref{eq:main_top_mm_nosq}] in one 25-MeV-wide $W$ bin for the experimental data (black histogram), the Monte Carlo simulation (blue histogram), and their difference (magenta histogram). The black and blue histograms are normalized to their maxima. The explanation of the fit curves is given in the text. The vertical green line shows the applied exclusivity cut.}
\label{fig:pim_miss_top_cut}
\end{center}
\end{figure}

Figure~\ref{fig:pim_miss_top_cut} shows the distribution of the quantity $M_{X[\pi^{-}]}$ plotted for the experimental data (black histogram) and the Monte Carlo simulation (blue histogram) in one 25-MeV-wide $W$ bin. The magenta histogram shows the difference between them and thus represents the distribution of background events, which are mainly events affected by FSIs with the spectator. The green line corresponds to the cut applied to select quasifree events. 

As seen in Fig.\!~\ref{fig:pim_miss_top_cut}, the exclusivity cut does not allow for complete isolation of the quasifree event sample. Tightening the cut would lead to significant reduction in the statistics of selected events, yet without total elimination of the FSI background. Therefore, an ``effective correction" of the FSI-background admixture was performed, which included the following steps.

\begin{itemize}
\item The $M_{X[\pi^{-}]}$ distribution of reconstructed Monte Carlo events (blue histogram) was fit with a polynomial. A typical result of this fit is shown in Fig.\!~\ref{fig:pim_miss_top_cut} by the solid orange curve.
\item The magenta background distribution was fit with a Gaussian. The result of the fit is shown by the dark magenta dash-dotted curve.
\item The orange and dark magenta curves were summed up to produce the red dashed curve that matches the black experimental histogram.
\item The correction factor $F_{\text{\tiny FSI}}$ was determined for the left side of the green cut~line,
\begin{equation}
\begin{aligned}
~~~~F_{\text{\tiny FSI}}(W) = \frac{\text{area~under~the~orange~curve}}{\text{area~under~the~red~curve}} \leq 1.\label{eq:fsi_corr_fact}
\end{aligned}
\end{equation}
\item In each $W$ bin, the experimental event yield was multiplied by the factor $F_{\text{\tiny FSI}}$, which served as an effective correction due to the remaining admixture of FSI-background events.
\end{itemize}

The factor $F_{\text{\tiny FSI}}$ is assumed to be only $W$ dependent as it was not found to exhibit any $Q^{2}$ dependence, and the dependence on the final hadronic variables is neglected due to the statistics limitation. The value of $F_{\text{\tiny FSI}}$ varies from $\approx$0.97 to $\approx$0.93 for the $W$ bins in the range from 1.45 to 1.825~GeV. For $W < 1.45$~GeV, the correction was not needed as no mismatch between the experimental and the simulated distributions was observed in this region (see Sec.\!~\ref{sect:fsi_top_comp}).

Note that the exclusivity cut shown in Fig.\!~\ref{fig:pim_miss_top_cut}, accompanied by the corresponding correction, accounts for all other possible effects that along with FSI effects may contribute to the mismatch between the data and the simulation in this topology (including a minor three-pion background contribution).

\section{Cross section calculation}

\subsection{Fermi smearing of the invariant mass $W$}
\label{Sect:smearing_blurring}

For the process of double-pion electroproduction off protons (as for any other exclusive process), the reaction invariant mass can, in general, be determined in two ways, {\it i.e.}~either from the initial particle  four-momenta ($W_{\text{i}}$) or from the final particle  four-momenta ($W_{\text{f}}$) as Eqs.\!~\eqref{W_fin_1} and~\eqref{W_fin_2} demonstrate. 
\begin{eqnarray}
W_{\text{i}}&= & \sqrt{(P_{p}^{\mu}+P_{\gamma_{v}}^{\mu})^{2}} \label{W_fin_1}. \\
W_{\text{f}}&= & \sqrt{(P_{\pi^{+}}^{\mu}+P_{\pi^{-}}^{\mu}+P_{p'}^{\mu})^{2}}. \label{W_fin_2}
\end{eqnarray}

Here $P_{\pi^{+}}^{\mu}$, $P_{\pi^{-}}^{\mu}$, and $P_{p'}^{\mu}$ are the four-momenta of the final-state hadrons, $P_{p}^{\mu}$ is the four-momentum of the initial proton, and $P_{\gamma_{v}}^{\mu}=P_{e}^{\mu}-P_{e'}^{\mu}$ the four-momentum of the virtual photon with $P_{e}^{\mu}$ and $P_{e'}^{\mu}$ the four-momenta of the incoming and scattered electrons, respectively. 

In general, to determine $W_{\text{f}}$, all final hadrons should be registered, while for calculation of $W_{\text{i}}$, it is sufficient to register just the scattered electron. The latter option allows one to analyze event samples in which information on the reaction final state is incomplete, as, {\it e.g.}~in topologies with one unregistered final hadron (see Sec.\!~\ref{Sect:topologies}). However, in reactions off bound nucleons, this opportunity comes with a complication, for to correctly calculate $W_{\text{i}}$, information on the initial proton momentum ($P_{p}^{\mu}$) is also required. In this analysis, however, this information is not accessible in the $\pi^{-}$ missing topology. This situation brings up a choice to either demand registration of all final hadrons to determine $W_{\text{f}}$ (which reduces the analysis flexibility) or to calculate $W_{\text{i}}$ assuming the initial proton to be at rest.

In this study, the invariant mass $W_{\text{i}}$ calculated under the target-at-rest assumption was used to describe the reaction, based on the statistically dominant $\pi^{-}$ missing topology. For this reason, the extracted cross sections turn out to be Fermi smeared~\cite{Skorodumina:2015rea,twopeg-d}. To retrieve the nonsmeared observable, a correction that unfolds this effect was applied, which is described in Sec.\!~\ref{Sect:fermi_corr}.

\subsection{Lab-to-CMS transformation}
\label{Sect:lab_cms}

Once the double-pion events were selected as described in Sec.~\ref{Sect:select}, the laboratory four-momenta of all final particles are known as they are either registered or reconstructed as missing. These four-momenta were then used to calculate the kinematic variables, which are introduced in Sec.\!~\ref{Sect:kin_var}. 

The cross sections were extracted in the center-of-mass frame of the {\em virtual photon -- initial proton} system (CMS). Therefore, to calculate the kinematic variables, the four-momenta of all particles need to be transformed from the laboratory system (Lab) to the CMS.

The CMS is uniquely defined as the system where the initial proton and the photon move towards each other with the net momentum equal to zero and the $z_{CMS}$ axis pointing along the photon. However, the procedure of the Lab-to-CMS transformation differs depending on the specificity of the reaction initial state (real or virtual photons, at rest or moving initial proton).
 
The correct procedure of the Lab-to-CMS transformation for an electroproduction experiment off a moving proton can be subdivided into two major steps.

\begin{enumerate}
\item First, one needs to perform a transition to the auxiliary system, where the target proton is at rest, while the incoming electron moves along the $z$ axis. This system can be called ``quasiLab" because the initial conditions in this frame imitate those existing in the Lab system for the case of a free proton experiment. The recipe of the Lab-to-quasiLab transformation is given in detail in Ref.\!~\cite{twopeg-d}.

\item Then, the quasiLab-to-CMS transformation should be performed by the standard method used for an electroproduction experiment off a proton at rest~\cite{Fed_an_note:2017,Fed_paper_2018,my_an_note:2020, my_thesis:2021}.
\end{enumerate}

The first step of this procedure (Lab-to-quasiLab transformation) implies that the momentum of the initial proton is known for each analyzed event~\cite{twopeg-d}. In this study, however, information on the initial proton momentum can be accessible only in the fully exclusive topology (it can be deduced via momentum conservation as shown in Sec.\!~\ref{Sect:excl_cut_fully_excl}), while in the $\pi^{-}$ missing topology, this information turns out to be irrevocably lost due to incomplete knowledge about the reaction final state. As a result, for the majority of analyzed events, the correct Lab-to-CMS transformation could not be performed. For this reason, in this analysis, the procedure of the Lab-to-CMS transformation for an electroproduction experiment off a proton at rest was used. The procedure is described in Refs.\!~\cite{Fed_an_note:2017,Fed_paper_2018,my_an_note:2020, my_thesis:2021} and was employed for both the fully exclusive and the $\pi^{-}$ missing topologies for consistency.

This approximation in the Lab-to-CMS transformation introduces a systematic inaccuracy to the extracted cross sections. A correction for this effect is included in the procedure of unfolding the effects of the initial proton motion (see Sec.\!~\ref{Sect:fermi_corr}).

\subsection{Kinematic variables}
\label{Sect:kin_var}

Once the four-momenta of all particles were defined and transformed into the CMS, the kinematic variables that describe the reaction $ep(n) \rightarrow e'p'(n')\pi^{+}\pi^{-}$ were calculated. To define the reaction initial state, only two variables are needed. In this study, they were chosen to be the reaction invariant mass $W$ and the photon virtuality $Q^{2}$.

Meanwhile, the three-body final hadronic state of the reaction is unambiguously determined by five kinematic variables~\cite{Fed_an_note:2017}, and in general there can be different options for their choice. In this analysis, the following generalized set of variables was used~\cite{Fed_an_note:2017,Fed_paper_2018,Byckling:1971vca,Isupov:2017lnd,Fedotov:2008aa,Mokeev:2015lda,my_an_note:2020, my_thesis:2021}:

\begin{itemize}
\item invariant mass of the first pair of hadrons $M_{h_{1}h_{2}}$,
\item invariant mass of the second pair of hadrons $M_{h_{2}h_{3}}$,
\item the first hadron solid angle $\Omega_{h_{1}}\! \!=\! (\theta_{h_{1}}, \varphi_{h_{1}})$, and
\item the angle $\alpha_{h_{1}}$ between the two planes: (a) defined by the three-momenta of the virtual photon (or initial proton) and the first final hadron and (b) defined by the three-momenta of all final hadrons.
\end{itemize}

The cross sections were obtained in three sets of variables depending on various assignments for the first, second, and third final hadrons:
\begin{itemize}
\item[1.] [$p'$, $\pi^{+}$, $\pi^{-}$]
$M_{p'\pi^{+}}$, $M_{\pi^{+}\pi^{-}}$, $\theta_{p'}$~,~$\varphi_{p'}$~,~$\alpha_{p'}$,
\item[2.] [$\pi^{-}$, $\pi^{+}$, $p'$]
$M_{\pi^{-}\pi^{+}}$, $M_{\pi^{+}p'}$, $\theta_{\pi^{-}}$, $\varphi_{\pi^{-}}$, $\alpha_{\pi^{-}}$,~\text{and}
\item[3.]  [$\pi^{+}$, $\pi^{-}$, $p'$]
$M_{\pi^{+}\pi^{-}}$, $M_{\pi^{-}p'}$, $\theta_{\pi^{+}}$, $\varphi_{\pi^{+}}$, $\alpha_{\pi^{+}}$.
\end{itemize}

Details on the calculation of the kinematic variables from the particle four-momenta can be found in Refs.\!~\cite{my_an_note:2020, my_thesis:2021,Fed_an_note:2017,Fedotov:2008aa}.

\begin{table*}[htb]\normalsize
\centering 
  \caption{\small Number of bins for hadronic variables.} \label{tab:summary_bins}
  \begin{tabular}{lm{4cm}cccc}
    \toprule
    & & \multicolumn{4}{c}{$W$ subrange (GeV)} \\
    \multicolumn{2}{c}{\centering Hadronic variable }  & [1.3,~1.35] & [1.35,~1.4] & [1.4,~1.475] & [1.475,~1.825] \\
    \cmidrule(l{5pt}r{15pt}){1-2} \cmidrule(l{5pt}r{5pt}){3-6}
    $M_{h_{1}h_{2}}$   & Invariant mass       &   8  & 10 & 12 & 12  \\
    $M_{h_{2}h_{3}}$   & Invariant mass       &   8  & 10 & 12 & 12  \\
    $\theta_{h_{1}}$   & Polar angle          &   6  & 8  & 10 & 10  \\
    $\varphi_{h_{1}}$  & Azimuthal angle      &   5  & 5  & 5  & 6   \\
    $\alpha_{h_{1}}$   & Angle between planes &   5  & 6  & 8  & 8   \\
    \cmidrule(l{5pt}r{15pt}){1-2} \cmidrule(l{5pt}r{5pt}){3-6}
              & Total~~$\!$number~~$\!$of~~$\!$$\Delta^{5}\tau$ \newline cells in a $\!$$\Delta W \Delta Q^2$ $\!$bin &   9600  & 24000  & 57600  & 69120   \\
    \bottomrule
  \end{tabular}
\end{table*}

\subsection{Binning and kinematic coverage}
\label{Sect:binning}

\begin{figure}[htp]
\begin{center}
\includegraphics[width=0.49\textwidth]{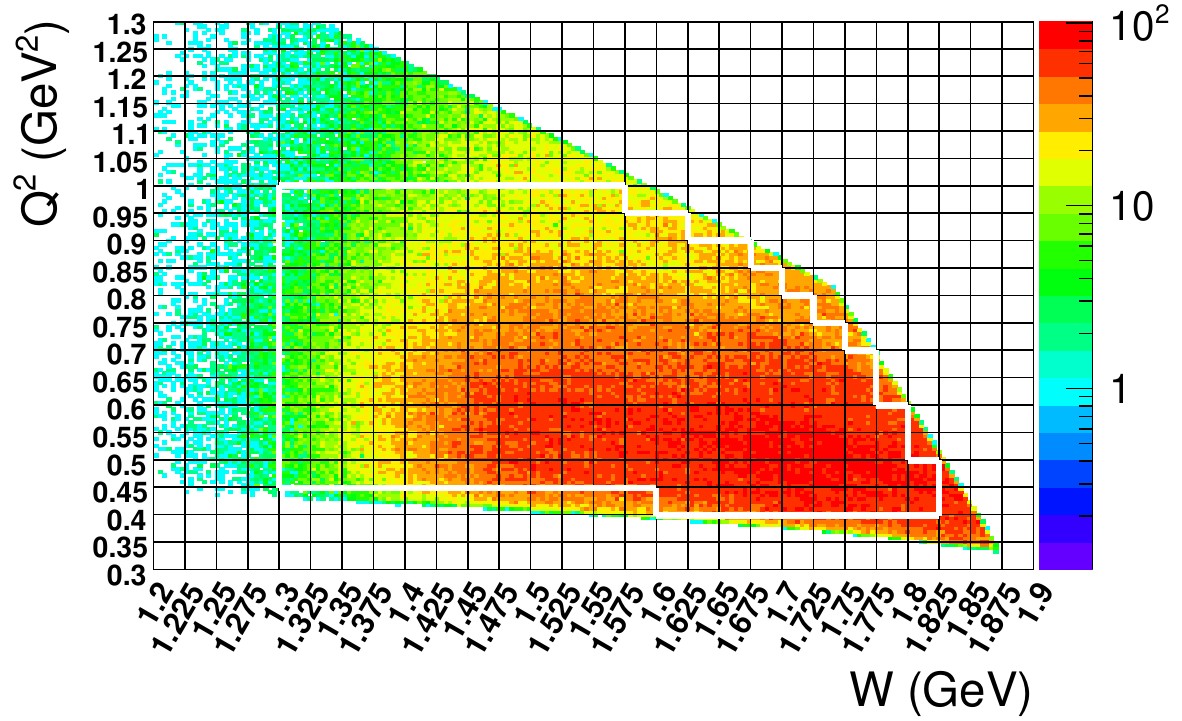}
\caption{\small $Q^2$ versus $W$ distribution populated with selected double-pion events. The black grid demonstrates the chosen binning in the initial-state variables (25~MeV in $W$ and 0.05~GeV$^{2}$ in $Q^{2}$) The cross section was calculated in two-dimensional cells within the white boundaries.} \label{fig:q2_vs_w}
\end{center}
\end{figure}

The available kinematic coverage in the initial-state variables is shown by the $Q^2$ versus $W$ distribution in Fig.\!~\ref{fig:q2_vs_w}. This distribution is filled with the double-pion events that survived after the event selection described in Sec.\!~\ref{Sect:select}. The white boundary shows the analyzed kinematic area, where the double-pion cross sections were extracted. The black grid demonstrates the chosen binning in the initial-state variables (25~MeV in $W$ and 0.05~GeV$^{2}$ in $Q^{2}$).

The kinematic coverage in the final-state variables has the following reaction-related features. The angular variables $\theta_{h_{1}}$, $\varphi_{h_{1}}$, and $\alpha_{h_{1}}$ vary in the fixed ranges of $[0,~\pi]$, $[0,~2\pi]$, and $[0,~2\pi]$, respectively. Meanwhile, the ranges of the invariant masses $M_{h_{1}h_{2}}$ and $M_{h_{2}h_{3}}$ are not fixed --- they are $W$ dependent and broaden as $W$ grows~\cite{my_an_note:2020, my_thesis:2021}.

The binning in the final hadronic variables used in this study is specified in Table~\ref{tab:summary_bins}. In each $W$ and $Q^{2}$ bin, the full range of each final hadronic variable was divided into bins of equal size. The number of bins differs in various $W$ subranges in order to take into account (i) the statistics drop near the reaction threshold and (ii) the aforementioned broadening of the reaction phase space with growing $W$. The chosen number~of~bins in each considered $W$ subrange reflects the intention to maintain reasonable statistical uncertainties of the single-differential cross sections for all $W$ and $Q^2$ bins. 

For the binning in the polar angle, the reader should note the following. The cross section, although being differential in $[-\cos\theta]$, is binned in $\theta$. These $\Delta \theta$ bins are of equal size in each corresponding $W$ subrange. See also Sec.\!~\ref{Sect:cr_sect_formula} on this matter.

The specific organization of the double-pion production phase space in the invariant masses $(M_{h_{1}h_{2}}, M_{h_{2}h_{3}})$ impels the need to pay careful attention to the binning in these variables. Equation~\eqref{eq:inv_mass_boundary} exemplifies the expressions for the lower and upper boundaries of the $M_{h_{1}h_{2}}$ distribution and demonstrates that the upper boundary depends on the value of $W$, while the lower does not.
\begin{equation}
\begin{aligned}
M_{\text{lower}} &= m_{h_1} + m_{h_2}; \\
M_{\text{upper}} (W) &= W - m_{h_3}. \label{eq:inv_mass_boundary}
\end{aligned}  
\end{equation}

Here $m_{h_1}$, $m_{h_2}$, and $m_{h_3}$ are the rest masses of the final hadrons.

Since the cross section is calculated in $W$ bins of a given width, the boundary of $M_{\text{upper}}$ is not distinct. For the purpose of binning in masses, the value of $M_{\text{upper}}$ was calculated using $W_{\text{center}}$, at the center of the $W$ bin, which caused events with $W > W_{\text{center}}$ to be located beyond $M_{\text{upper}}$. For this reason, it was decided to use a specific arrangement of mass bins with the bin width $\Delta M$ determined by
\begin{equation}
\begin{aligned}
\Delta M = \frac{M_{\text{upper}}-M_{\text{lower}}}{N_{\text{bins}}-1}, \\[-10pt]\label{eq:bin_width}
\end{aligned}  
\end{equation} 
where $N_{\text{bins}}$ is the number of bins specified in the first row of Table~\ref{tab:summary_bins}. The left boundary of the first bin was set to $M_{\text{lower}}$.

The chosen arrangement of bins forces the last mass bin to be situated completely out of the boundaries determined according to Eq.\!~\eqref{eq:inv_mass_boundary} using $W_{\text{center}}$. The cross section in this extra bin is not reported. However, the bin was kept when integrating the cross section over the mass distributions so that no events were lost [see  Eqs.\!~\eqref{inegr5diff}].

Note that the cross section in the next-to-last bin in invariant masses needs a correction. See more details in Sec.\!~\ref{Sect:bin_cor}.

\subsection{Cross section formula}
\label{Sect:cr_sect_formula}

\subsubsection{Electron scattering cross section}

The experimental electron scattering cross section $\sigma_{e}$ for the reaction $ep(n) \rightarrow e'p'(n') \pi^{+} \pi^{-}$ is seven-fold differential and determined by
\begin{equation}
\frac{\textrm{d}^{7}\sigma_{e}}{\textrm{d}W\textrm{d}Q^{2}\textrm{d}^{5}\tau} = \frac{1}{ R \! \cdot \! \mathcal{F}}  \cdot 
\frac{\left( \frac{N_{\text{full}}}{Q_{\text{full}}}-\frac{N_{\text{empty}}}{Q_{\text{empty}}} \right)}{
\Delta W \! \cdot \! \Delta Q^{2} \! \cdot \! \Delta^{5} \tau \! \cdot \! \left[ \frac{l \cdot \rho \cdot N_{A}}{q_{e}\cdot \mu_{d}} \right]\! \cdot \!\mathcal{E}},
\label{expcrossect}
\end{equation}
where

\begin{itemize}
\item $\textrm{d}^{5}\tau = \textrm{d}M_{h_{1}h_{2}} \textrm{d}M_{h_{2}h_{3}} \textrm{d}\Omega_{h_1} \textrm{d}\alpha_{h_1}$ is the differential of the five independent variables of the $p\pi^{+}\pi^{-}$ final state, which are described in Sec.\!~\ref{Sect:kin_var};
\item $N_{\text{full}}$ and $N_{\text{empty}}$ are the numbers of selected double-pion events inside a seven-dimensional bin for runs with liquid deuterium and empty target, respectively;
\item the quantity in the square brackets in the denominator corresponds to the luminosity (per charge) of the experiment $\mathcal{L}$ in the units cm$^{-2}\cdot$C$^{-1}$ and its components are
\begin{itemize}
\item[]$\!\!\!\!\!\!\!\!$$l\!=\!2\!$ cm, the target length,\vspace{-0.1em}
\item[]$\!\!\!\!\!\!\!\!$$\rho\!=\!0.169$ g$\cdot$cm$^{-3}$, the liquid-deuterium density,\vspace{-0.1em}
\item[ ]$\!\!\!\!\!\!\!\!$$N_{A}\!=\!6.022$$\cdot 10^{23}$ mol$^{-1}$, Avogadro's number,\vspace{-0.1em}
\item[ ]$\!\!\!\!\!\!\!\!$$q_{e}\!=\!1.602$$\cdot 10^{-19}$ C, the elementary charge,$\:$and\vspace{-0.1em}
\item[ ]$\!\!\!\!\!\!\!\!$$\mu_{d}\!=\!2.014$ g$\cdot$mol$^{-1}$, the deuterium~molar~mass,\vspace{-0.1em}
\end{itemize}
which results in the luminosity value of $\mathcal{L}$ = 0.63$\cdot$10$^{42}$ cm$^{-2}\cdot$C$^{-1}$ = 0.63$\cdot$10$^{12}$ $\mu$b$^{-1}\cdot$C$^{-1}$;

\item $Q_{\text{full}}$ = 3734.69 $\mu$C and $Q_{\text{empty}}$ = 464.797 $\mu$C are the values of the integrated Faraday cup charge for liquid-deuterium and empty-target runs, respectively, which results in the corresponding values of the integrated luminosity $L=\mathcal{L}\cdot Q$ of 2.35$\cdot 10^{9}$ $\mu \text{b}^{-1}$ and 0.29$\cdot 10^{9}$ $\mu \text{b}^{-1}$; 

\item $\mathcal{E} = \mathcal{E}(\Delta W, \Delta Q^{2}, \Delta^{5}\tau)$ is the detector efficiency (which includes the detector acceptance) for each seven-dimensional bin as determined by the Monte Carlo simulation (see Sec.\!~\ref{Sect:eff_eval}); 

\item $R = R(\Delta W, \Delta Q^{2})$ is the radiative correction factor described in Sec.\!~\ref{Sect:rad_corr}; 

\item $\mathcal{F} = \mathcal{F}(\Delta W, \Delta Q^{2}, \Delta^{5}\tau)$ is the correction factor that aims at unfolding the effects of the initial proton motion (see Sec.\!~\ref{Sect:fermi_corr}).

\end{itemize}

\subsubsection{Virtual photoproduction cross section}

The goal of the analysis was to extract the virtual photoproduction cross section $\sigma_{\text{v}}$ of the reaction $\gamma_{v}p(n) \rightarrow p'(n') \pi^{+} \pi^{-}$. This virtual photoproduction cross section $\sigma_{\text{v}}$ is five-fold differential and in the single-photon exchange approximation is connected with the seven-fold differential electron scattering cross section $\sigma_{e}$ via 
\begin{equation}
\begin{aligned}
\frac{\textrm{d}^{5}\sigma_{\text{v}}}{\textrm{d}^{5}\tau} &= \frac{1}{\Gamma_{\text{v}}}\frac{\textrm{d}^{7}\sigma_{e}}{\textrm{d}W\textrm{d}Q^{2}\textrm{d}^{5}\tau}  \textrm{ ,}
\end{aligned} 
\label{fulldiff}
\end{equation}
where $\Gamma_{\text{v}}$ is the virtual photon flux given by
\begin{equation}
\Gamma_{\text{v}} (W, Q^2) =
\frac{\alpha}{4\pi}\frac{1}{E_{\text{beam}}^{2}m_{p}^{2}}\frac{W(W^{2}-m_{p}^{2})}
{(1-\varepsilon_{\text{T}})Q^{2}} \textrm{ .}
\label{flux}
\end{equation}

Here $\alpha$ is the fine structure constant $\left(1/137\right)$, $m_{p}$ the proton mass, $E_{\text{beam}}$ = 2.039 GeV the energy of the incoming electrons in the Lab frame, and $\varepsilon_{\text{T}}$ the virtual photon transverse polarization given by 
\begin{equation}
\varepsilon_{\text{T}} = \left( 1 + 2\left( 1 +
\frac{\nu^{2}}{Q^{2}} \right)
\tan^{2}\left(\frac{\theta_{e'}}{2}\right) \right)^{-1} \textrm{ ,}
\label{polarization}
\end{equation}
where $\nu = E_{\text{beam}} - E_{e'}$ is the virtual photon energy, while $E_{e'}$ and $\theta_{e'}$ are the energy and the polar angle of the scattered electron in the Lab frame, respectively.

The limited statistics of the experiment did not allow for estimates of the five-fold differential cross section $\sigma_{\text{v}}$ with reasonable accuracy. Therefore, the cross section $\sigma_{\text{v}}$ was first calculated on the multi-dimensional grid and then integrated over at least four hadronic variables, which means that only single-differential and fully integrated cross sections were obtained.

For each variable set described in Sec.\!~\ref{Sect:kin_var}, the following cross sections were extracted:

\begin{equation}
\begin{aligned}
\frac{\textrm{d}\sigma_{\text{v}}}{\textrm{d}M_{h_{1}h_{2}}} & =\int\frac{\textrm{d}^{5}\sigma_{\text{v}}}{\textrm{d}^{5}\tau}\textrm{d}M_{h_{2}h_{3}}\textrm{d}\Omega_{h_{1}}\textrm{d}\alpha_{h_{1}}, \\
\frac{\textrm{d}\sigma_{\text{v}}}{\textrm{d}M_{h_{2}h_{3}}} & =\int\frac{\textrm{d}^{5}\sigma_{\text{v}}}{\textrm{d}^{5}\tau}\textrm{d}M_{h_{1}h_{2}}\textrm{d}\Omega_{h_{1}}\textrm{d}\alpha_{h_{1}}, \\
\frac{\textrm{d}\sigma_{\text{v}}}{\textrm{d}[-\cos\theta_{h_{1}}]} & =\int\frac{\textrm{d}^{5}\sigma_{\text{v}}}{\textrm{d}^{5}\tau}\textrm{d}M_{h_{1}h_{2}}\textrm{d}M_{h_{2}h_{3}}\textrm{d}\varphi_{h_{1}}d\alpha_{h_{1}}, \\
\frac{\textrm{d}\sigma_{\text{v}}}{\textrm{d}\alpha_{h_{1}}} & =\int\frac{\textrm{d}^{5}\sigma_{\text{v}}}{\textrm{d}^{5}\tau}\textrm{d}M_{h_{1}h_{2}}\textrm{d}M_{h_{2}h_{3}}\textrm{d}\Omega_{h_{1}},~~\text{and}\\
\sigma_{\text{v}}^{\text{int}} (W, Q^{2}) &= \int \frac{d^{5}\sigma_{\text{v}}}{\textrm{d}^{5}\tau}\textrm{d}M_{h_{1}h_{2}}\textrm{d}M_{h_{2}h_{3}}\textrm{d}\Omega_{h_{1}}\textrm{d}\alpha_{h_{1}}.
\end{aligned}
\label{inegr5diff}
\end{equation}

As a final result for each $W$ and $Q^{2}$ bin, the fully integrated cross section $\sigma_{\text{v}}^{\text{int}}$, averaged over the three variable sets, is reported together with the nine single-differential cross sections given in Eq.\!~\eqref{eq:reported_sec}, where each column is taken from the corresponding variable~set.
\begin{equation}
\begin{aligned}
&~~~~~~\frac{\textrm{\textrm{d}}\sigma_{\text{v}}}{\textrm{d}M_{p'\pi^{+}}}&&~~~~~~\frac{\textrm{d}\sigma_{\text{v}}}{\textrm{d}M_{\pi^{-}\pi^{+}}}&&~~~~~~\frac{\textrm{d}\sigma_{\text{v}}}{\textrm{d}M_{\pi^{-}p'}}\\[8pt] 
&~~\frac{\textrm{d}\sigma_{\text{v}}}{\textrm{d}[-\cos\theta_{p'}]}&&~~\frac{\textrm{d}\sigma_{\text{v}}}{\textrm{d}[-\cos\theta_{\pi^{-}}]}&&~~\frac{\textrm{d}\sigma_{\text{v}}}{\textrm{d}[-\cos\theta_{\pi^{+}}]}\\[8pt] 
&~~~~~~~\frac{\textrm{d}\sigma_{\text{v}}}{\textrm{d}\alpha_{p'}}&&~~~~~~~~\frac{\textrm{d}\sigma_{\text{v}}}{\textrm{d}\alpha_{\pi^{-}}}&&~~~~~~~\frac{\textrm{d}\sigma_{\text{v}}}{\textrm{d}\alpha_{\pi^{+}}}
\end{aligned}
\label{eq:reported_sec}
\end{equation}

Regarding the middle row in Eq.\!~\eqref{eq:reported_sec}, note the following. Although being differential in $[-\cos\theta]$, the cross sections were calculated in $\Delta \theta$ bins, which are of equal size in a corresponding $W$ subrange (see also Sec.\!~\ref{Sect:binning}). This follows the convention used to extract the $\theta$ distributions in studies of double-pion cross sections~\cite{Rip_an_note:2002,Ripani:2002ss,Fed_an_note:2007,Fedotov:2008aa,Isupov:2017lnd,Arjun,Fed_an_note:2017,Fed_paper_2018}.

\subsection{Efficiency evaluation}
\label{Sect:eff_eval}

In this study, the Monte Carlo simulation was performed using the TWOPEG-D event generator~\cite{twopeg-d}, which is capable of simulating a quasifree process of double-pion electroproduction off a moving proton. In this event generator, the Fermi motion of the initial proton is generated according to the Bonn potential~\cite{Machleidt:1987hj} and then is naturally merged into the specific kinematics of double-pion electroproduction. TWOPEG-D accounts for radiative effects according to the approach described in Refs.\!~\cite{Mo:1968cg,twopeg}.

Events generated with TWOPEG-D were passed through a standard multistage procedure of the detector simulation and event reconstruction with the majority of the parameters kept the same as in the studies~\cite{Fed_an_note:2017,Fed_paper_2018,Markov:2014,Ye_Tian:2017,CLAS:2022kta} that were also devoted to the ``e1e" run period. More information on the procedure and the related parameters can be found in Ref.\!~\cite{my_an_note:2020}.

In studies of double-pion cross sections, it is important to generate enough Monte Carlo statistics in order to saturate each multi-dimensional bin of the reaction phase space with events (see Table~\ref{tab:summary_bins}). Insufficient Monte Carlo statistics will lead to an improper efficiency evaluation and an unnecessary rise in the number of empty cells (see Sec.\!~\ref{Sect:empt_cells}), thus systematically affecting the accuracy of the extracted cross sections. For this study, the total of about 4$\cdot$10$^{10}$ double-pion events were generated in the investigated kinematic region, which was found to be sufficient.

TWOPEG-D performs a weighted event generation~\cite{twopeg}. This means that all kinematic variables are randomly generated according to the double-pion production phase space, while each event generated at a particular kinematic point acquires an individual weight that reflects the cross section value at that point.  The efficiency factor $\mathcal{E}$ from Eq.\!~\eqref{expcrossect} was then calculated in each $\Delta W\Delta Q^2\Delta^{5}\tau$ bin by
\begin{equation}
\begin{aligned}
\mathcal{E}(\Delta W, \Delta Q^2, \Delta^{5}\tau) = \frac{\mathbb{N}_{\text{rec}}}{\mathbb{N}_{\text{gen}}} =  \frac{\sum\limits_{i=1}^{N_{\text{rec}}} w_{i}}{\sum\limits_{j=1}^{N_{\text{gen}}} w_{j}} ,
\end{aligned}
\label{eq:eff}
\end{equation}
where $N_{\text{gen}}$ is the number of generated double-pion events (without any cuts) inside a multi-dimensional bin and $N_{\text{rec}}$ is the number of reconstructed double-pion events that survived in the bin after the event selection, while $\mathbb{N}_{\text{gen}}$ and  $\mathbb{N}_{\text{rec}}$ are the weighted numbers of the corresponding events and $w$ is the weight of an individual event.

\begin{figure*}[htp]
\begin{center}
\includegraphics[width=0.75\textwidth]{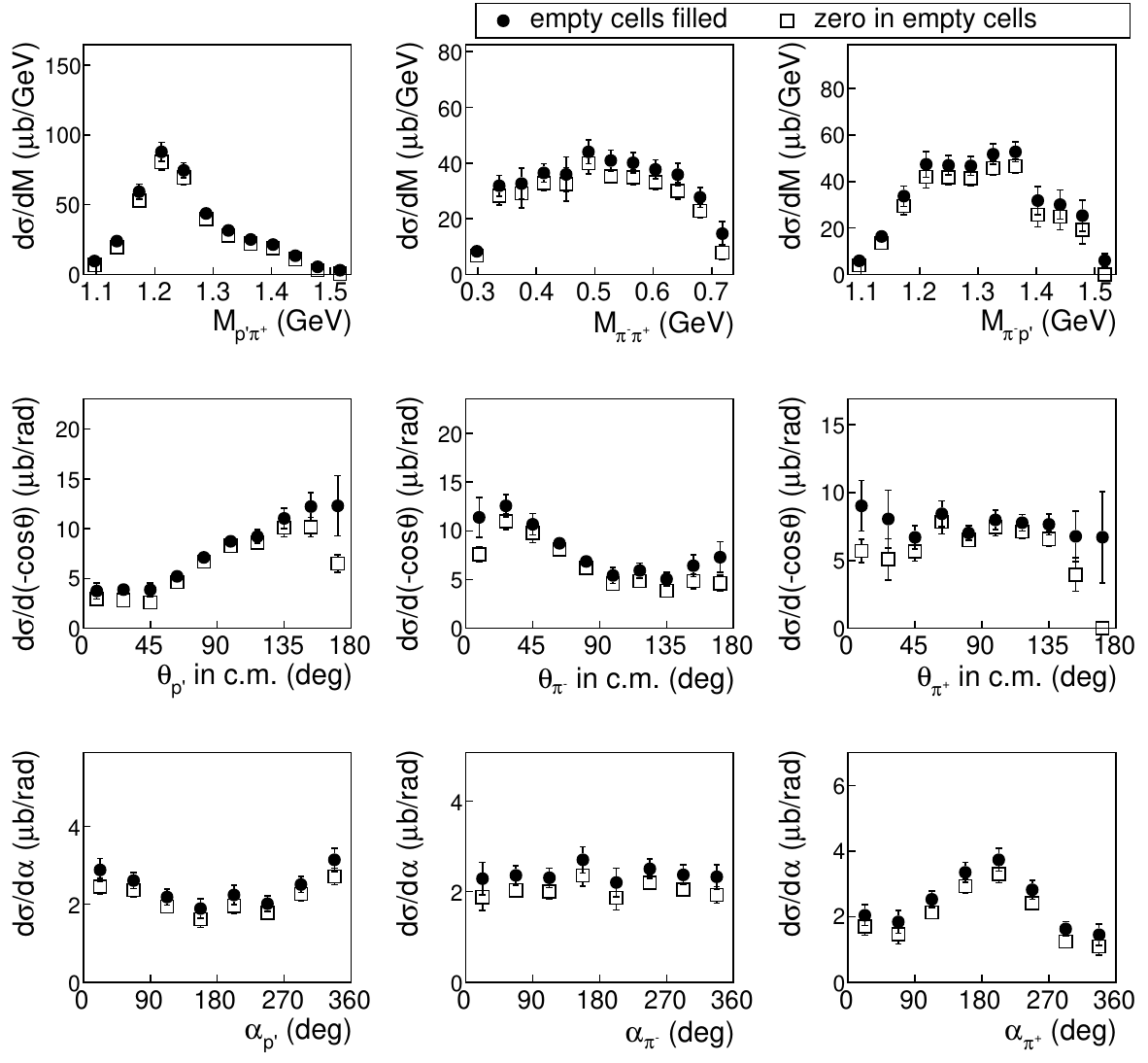}
\caption{\small Extracted single-differential cross sections for the cases when the contribution from the empty cells was ignored (open squares) and when it was taken into account (black circles). The former are reported with the uncertainty $\delta_{\text{stat}}^{\text{tot}}$ given by Eq.\!~\eqref{errortot} and the latter -- with the uncertainty $\delta_{\text{stat,mod}}^{\text{tot}}$ given by Eq.\!~\eqref{eq:error_stat_mod}. All distributions are given for one particular bin in $W$ and $Q^2$ ({\it i.e.}~$W = $1.6375 GeV, $Q^2 = $0.625 GeV$^2$).} \label{fig:empt_corr}
\end{center}
\end{figure*}

\everypar{\looseness=-1}
In some kinematic bins, the efficiency $\mathcal{E}$ could not be reliably determined due to boundary effects, bin-to-bin event migration, and/or limited Monte Carlo statistics. In such bins, the relative efficiency uncertainty $\delta \mathcal{E}/\mathcal{E}$ is typically large. In this study, a cut on the relative efficiency uncertainty $\delta \mathcal{E}/\mathcal{E}$ was performed that excluded from consideration all multi-dimensional cells with uncertainties greater than 30\%. The excluded cells were ranked as ``empty cells" and, along with the other empty cells, were subject to the filling procedure (see Sec.\!~\ref{Sect:empt_cells}).

It is noteworthy that the impact of the $\delta \mathcal{E}/\mathcal{E}$ cut on the cross section uncertainties is not straightforward. On the one hand, the cut eliminates $\Delta^{5} \tau$ bins with large $\delta \mathcal{E}/\mathcal{E}$ values, hence reducing the uncertainty due to the Monte Carlo statistics and,~consequently, the total statistical uncertainty of the extracted cross sections (see Sec.\!~\ref{Sect:stat_uncert}). On the other hand, this cut increases the number of empty cells, thus increasing the cross section model dependence and the uncertainty associated with it (see Sec.\!~\ref{Sect:mod_dep}). The cut value was chosen as a compromise between these two effects. 

The idea of this cut was taken from the free-proton study~\cite{Fed_an_note:2017,Fed_paper_2018}, which also sets the cut value at 30\%. More details can be found in Refs.\!~\cite{my_an_note:2020, my_thesis:2021,Fed_an_note:2017,Fed_paper_2018}.

\section{Corrections to the cross sections}

\subsection{Filling kinematic cells with zero acceptance}
\label{Sect:empt_cells}

Due to blind areas in the geometrical coverage of the CLAS detector, some kinematic bins of the double-pion production phase space turned out to have zero acceptance. In such bins, which are usually called empty cells, the cross section cannot be experimentally defined. These cells contribute to the integrals in Eqs.\!~\eqref{inegr5diff} along with all other kinematic bins. If ignored, the contribution from the empty cells causes a systematic underestimation of the cross section, and therefore some assumptions for their contents are needed. This situation causes some model dependence of the extracted cross sections.

The map of the empty cells was determined using the Monte Carlo simulation. A multi-dimensional cell was treated as empty if it contained generated events ($N_{\text{gen}} >$~0) but did not contain any reconstructed events ($N_{\text{rec}}$ = 0). The cells with unreliable efficiencies, ruled out based on the 30\% cut on the efficiency uncertainty, were also treated as empty (see Sec.\!~\ref{Sect:eff_eval}).

For studies of double-pion cross sections with CLAS, it has become conventional to fill the empty cells by means of a Monte Carlo event generator in order to account for their contribution. See more details in Refs.\!~\cite{Rip_an_note:2002,Ripani:2002ss,Fed_an_note:2007,Fedotov:2008aa,Isupov:2017lnd,Arjun,Fed_an_note:2017,Fed_paper_2018}.

In the present work, empty multi-dimensional cells were filled with generated Monte Carlo events produced by TWOPEG-D~\cite{twopeg-d}. These events were subject to integral scaling, in order to adjust their yield to the yield of experimental events in regular (nonempty) cells. The scaling was performed individually in each $\Delta W\Delta Q^2$ bin according to the ratio of the integrated yields of experimental events and reconstructed Monte Carlo events in all nonempty cells of that bin (see Refs.\!~\cite{my_an_note:2020, my_thesis:2021} for details).

Figure~\ref{fig:empt_corr} introduces the single-differential cross sections given by Eqs.\!~\eqref{inegr5diff} and \eqref{eq:reported_sec}. The open squares correspond to the case when the contribution from the empty cells was ignored, and the black circles are for the case when it was taken into account as described above. The figure indicates a satisfactory small contribution from the empty cells for the majority of data points (and therefore a small model dependence of the results). Only the edge points in the $\theta$ distributions (middle row) reveal pronounced contributions due to the negligible or zero CLAS acceptance in the corresponding directions.

For most of the $(W,Q^2)$ points, the contribution from the empty cells to the fully integrated cross section is kept to a low level of $\approx$15\%, slightly rising at the low-$Q^{2}$ and high-$W$ boundaries. Besides this, the empty cells contribution grows towards the reaction threshold ({\it i.e.}~for $W\lesssim 1.4$~GeV). This behavior was also observed in Refs.\!~\cite{Fed_an_note:2017,Fed_paper_2018,Fed_an_note:2007,Fedotov:2008aa} devoted to double-pion electroproduction off free protons.

To account for the model dependence, the approach established in previous studies of double-pion cross sections was followed~\cite{Isupov:2017lnd,Fed_an_note:2017,Golovach}, {\it i.e.}~the part of the single-differential cross section that came from the empty cells was assigned a 50\% relative uncertainty (more details are in Sec.\!~\ref{Sect:mod_dep}).

\subsection{Radiative corrections}
\label{Sect:rad_corr}

The incoming and scattered electrons are subject to radiative effects, which means that they can emit photons, thereby reducing their energy. Information on such emissions is typically experimentally inaccessible, and therefore, these changes in electron energy cannot be directly taken into account in the cross section calculation. As a result, measured cross sections acquire distortions.

The common way of handling this problem is to apply radiative corrections to the extracted cross sections. In this study, radiative corrections were performed using TWOPEG-D~\cite{twopeg-d}, which is an event generator for double-pion electroproduction off a moving proton. TWOPEG-D accounts for radiative effects by means of the well-known Mo and Tsai approach~\cite{Mo:1968cg}, which has traditionally been used for radiative corrections in studies of double-pion cross sections~\cite{Rip_an_note:2002,Ripani:2002ss,Fed_an_note:2007,Fedotov:2008aa,Fed_an_note:2017,Fed_paper_2018,Isupov:2017lnd,Arjun}. In Ref.\!~\cite{Mo:1968cg}, the approach was applied to the inclusive case, while in TWOPEG-D, the fully integrated double-pion cross sections are used instead~\cite{twopeg,twopeg-d}. 

In the employed approach~\cite{Mo:1968cg,twopeg,twopeg-d}, the radiative photons are considered to be emitted collinearly either in the direction of the incoming or scattered electron (termed the ``peaking approximation"). The calculation of the radiative cross section is split into two parts. The ``soft" part assumes the energy of the emitted radiative photon to be less than a certain minimal value (10 MeV), while the ``hard" part is for the photons with an energy greater than that value. The ``soft" part is evaluated explicitly, while for the calculation of the ``hard" part, an inclusive hadronic tensor is assumed. Based on previous experience with radiative corrections for double-pion cross sections, the latter assumption is considered to be adequate~\cite{Rip_an_note:2002,Ripani:2002ss,Fed_an_note:2007,Fedotov:2008aa,Fed_an_note:2017,Fed_paper_2018,Isupov:2017lnd,Arjun}. Also note that approaches that are capable of describing radiative processes in exclusive double-pion electroproduction are not yet available.

\begin{figure}[htp]
\begin{center}
\includegraphics[width=0.5\textwidth]{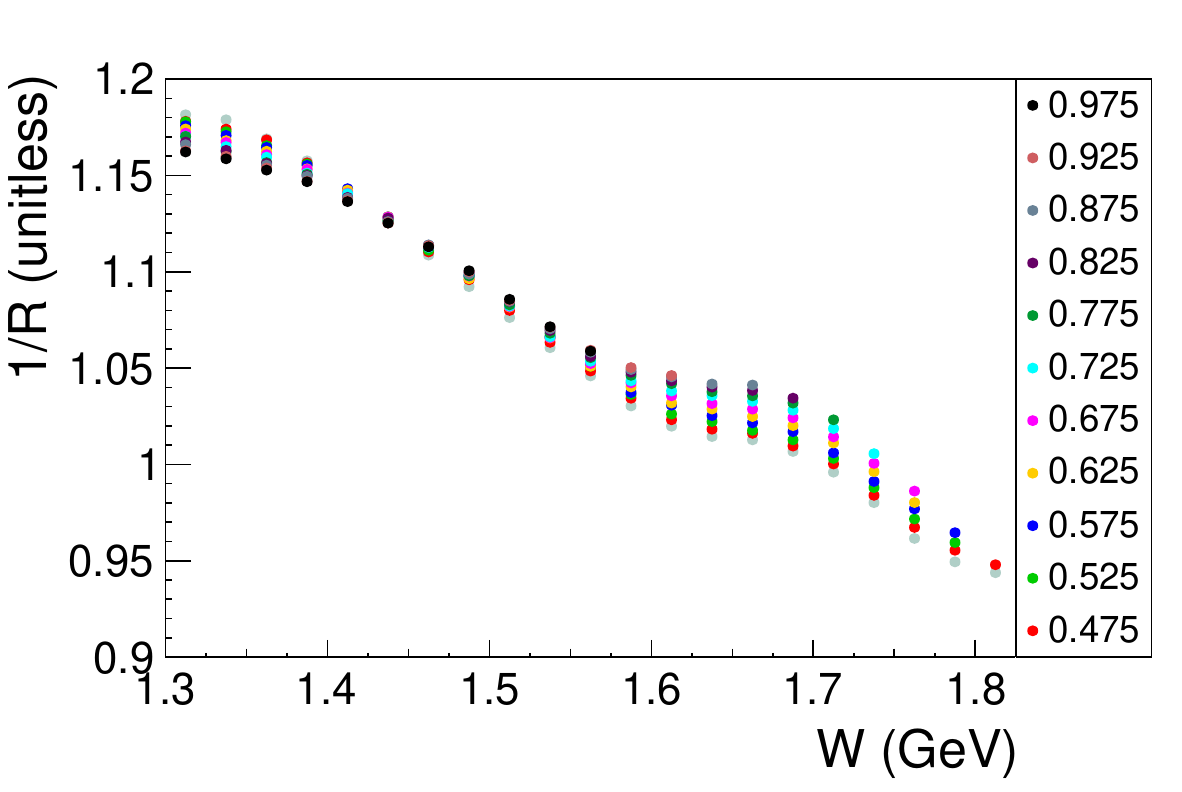}
\caption{\small Reciprocal of the radiative correction factor ({\it i.e.}~$1/R$) as a function of $W$ for different $Q^{2}$ bins [see Eq.\!~\eqref{eq:rad_corr}].} \label{fig:radcorr}
\end{center}
\end{figure}

The radiative correction factor $R$ in Eq.\!~\eqref{expcrossect} was determined in the following way. Double-pion events either with or without radiative effects were generated with TWOPEG-D. Both radiated and nonradiated events were subjected to the Fermi smearing. Then the ratio given by Eq.\!~\eqref{eq:rad_corr} was taken in each $\Delta W \Delta Q^{2}$ bin.
\begin{equation}
R(\Delta W, \Delta Q^{2}) = \frac{\mathbb{N}_{\text{rad}}}{\mathbb{N}_{\text{norad}}}
\label{eq:rad_corr}
\end{equation}

Here $\mathbb{N}_{\text{rad}}$ and $\mathbb{N}_{\text{norad}}$ are the weighted numbers of generated events in each $\Delta W \Delta Q^{2}$ bin with and without radiative effects, respectively. Neither $\mathbb{N}_{\text{rad}}$ nor $\mathbb{N}_{\text{norad}}$ were subject to any cuts.

This approach gives the correction factor $R$ only as a function of $W$ and $Q^{2}$, disregarding its dependence on the hadronic variables. However, the need to integrate the cross section at least over four hadronic variables [see Eq.\!~\eqref{inegr5diff}] considerably reduces the influence of the final-state hadron kinematics on the radiative correction factor, thus justifying the applicability of the procedure~\cite{Mo:1968cg,twopeg,twopeg-d}.

The quantity $1/R$ is plotted in Fig.\!~\ref{fig:radcorr} as a function of $W$ for different $Q^{2}$ bins. The uncertainties associated with the statistics of generated events are very small and therefore not visible in the plot.

\subsection{\!\!\!Unfolding the effects of\! the\! initial proton motion}
\label{Sect:fermi_corr}

In this study, information on the initial proton momentum is inaccessible for the majority of analyzed experimental events, as discussed above. For this reason, the invariant mass $W_{\textrm{i}}$ calculated under the target-at-rest assumption was used for the cross section binning (see Sec.\!~\ref{Sect:smearing_blurring}), which leads to the Fermi smearing of the fully integrated and single-differential cross sections. The same reason necessitates the use of an approximate procedure of the Lab-to-CMS transformation (see Sec.\!~\ref{Sect:lab_cms}). This approximation introduces some inaccuracy to the measured angular ($\theta$, $\varphi$, and $\alpha$) distributions but does not affect the invariant mass distributions and $W$ and $Q^{2}$ cross section dependences due to the Lorentz invariance of these variables.

Being folded with the aforementioned effects of the initial proton motion, the extracted cross sections require a corresponding unfolding correction. This correction was performed by means of two Monte Carlo event generators TWOPEG~\cite{twopeg} and TWOPEG-D~\cite{twopeg-d}. TWOPEG is an event generator for double-pion electroproduction off a free proton that currently provides the best cross section estimation in the investigated kinematic region. TWOPEG-D is an event generator for the same exclusive reaction but off a proton that moves in a deuterium nucleus. This event generator was specifically developed for the studies where experimental information on the initial proton momentum is inaccessible and the target-at-rest assumption has to be employed. TWOPEG-D convolutes double-pion cross sections with the effects of the initial proton motion, thus imitating the conditions of experimental cross section extraction.

To calculate the correction factor, two samples of double-pion events, produced off protons at rest and off moving protons, were generated with TWOPEG and TWOPEG-D, respectively. For the latter, the smeared value of $W$ was used for the binning and the approximate Lab-to-CMS transformation was applied, so that the sample incorporated the same inaccuracies as the experimentally extracted cross sections.

The unfolding correction was performed in each multi-dimensional bin of the double-pion production phase space, {\it i.e.}~in each $\Delta W \Delta Q^{2}\Delta^{5}\tau$ bin the cross section was divided by the correction factor $\mathcal{F}$ that was calculated as
\begin{equation}
\mathcal{F}(\Delta W, \Delta Q^{2},\Delta^{5}\tau) = \frac{\mathbb{N}_{\text{Fermi}}}{\mathbb{N}_{\text{noFermi}}},
\label{eq:ferm_corr}
\end{equation}
where $\mathbb{N}_{\text{noFermi}}$ and $\mathbb{N}_{\text{Fermi}}$ are the weighted numbers of generated double-pion events in a $\Delta W \Delta Q^{2}\Delta^{5}\tau$ bin produced off protons at rest and off moving protons, respectively.

For most points of the fully integrated cross sections at $W>$~1.4~GeV, the correction does not exceed $\pm 5$\%. However, for $W$ closer to the reaction threshold, the correction grows rapidly, reaching up to about 40\% for the first $W$ bin (see Refs.\!~\cite{my_an_note:2020, my_thesis:2021} for details).

The value of the correction factor in Eq.\!~\eqref{eq:ferm_corr} depends on both the free proton cross sections and the deuteron wave function implemented into the event generators. The former relies strongly on the JM model fit of the available data on double-pion cross sections, whereas for the latter, the Bonn model was used (see Refs.\!~\cite{twopeg,twopeg-d}). Therefore, the uncertainty of the extracted cross sections that comes from the unfolding correction was attributed to the model-dependent uncertainty as discussed in Sec.\!~\ref{Sect:mod_dep}.

\subsection{Corrections for binning effects}
\label{Sect:bin_cor}

Being extracted in a finite bin, the cross section naturally undergoes averaging within this bin. In the present work, this averaged value is then assigned to the bin central point. Any nonlinear behavior of the cross section within the bin will likely result in an offset of the obtained cross section value. To cure this effect, binning corrections were applied to the fully integrated and single-differential cross sections that included a cubic spline approximation for the cross section shape~\cite{my_an_note:2020, my_thesis:2021}. Due to the relatively fine binning in all kinematic variables used in this study, the influence of binning effects on the cross sections is marginal: the typical value of the correction is $\approx$1\% rising up to 5\% for some data points at low $W$.

In addition to that, the cross section in the next-to-last point of the invariant mass distributions was subject to a separate correction, which is described in detail in Refs.\!~\cite{Fed_an_note:2017,Fed_paper_2018,my_an_note:2020, my_thesis:2021}. The need for this correction follows from the broadening of the reaction phase space with $W$, which causes the upper boundary of the invariant mass distributions to be $W$ dependent.

\section{Cross section uncertainties}
\label{Sect:uncert}

In this study (like in other studies of double-pion cross sections~\cite{Rip_an_note:2002,Ripani:2002ss,Fed_an_note:2007,Fedotov:2008aa,Isupov:2017lnd,Arjun,Fed_an_note:2017,Fed_paper_2018}) three separate types of cross section uncertainties were considered, {\it i.e.}~statistical uncertainties, uncertainties due to the model dependence, and systematic uncertainties. 

\subsection{Statistical uncertainties}
\label{Sect:stat_uncert}

The limited statistics of both the experimental data and the Monte Carlo simulation are the two sources of statistical fluctuations of the extracted 
cross sections.

The absolute statistical  uncertainty due to the limited statistics of the experimental data was calculated in a nonempty multi-dimensional bin as
\begin{equation}
\delta_{\text{stat}}^{\text{expt}}(\Delta^{5} \tau) = \frac{1}{\mathcal{E} \! \cdot \! R \! \cdot \! \mathcal{F} \! \cdot \! \Gamma_{\text{v}} }  \cdot  \frac{\sqrt{\left( \frac{N_{\text{full}}}{Q_{\text{full}}^{2}}+\frac{N_{\text{empty}}}{Q_{\text{empty}}^{2}} \right) } }{
\Delta W \! \cdot \!  \Delta Q^{2} \! \cdot \!  \Delta^{5} \tau \! \cdot \! \left [\mathcal{L} \right ]},
\label{staterrors}
\end{equation}
where $\Gamma_{\text{v}}$ is the virtual photon flux given by Eq.\!~\eqref{flux}, while the other variables are explained in the context of Eq.\!~\eqref{expcrossect}.

The absolute uncertainty due to the limited Monte Carlo statistics was estimated in nonempty bins as 
\begin{equation}
\delta_{\text{stat}}^{\text{MC}}(\Delta^{5} \tau) = \frac{\textrm{d}^{5}\sigma_{\text{v}}}{\textrm{d}^{5}\tau} \left( \frac{\delta \mathcal{E}}{\mathcal{E}} \right),
\label{montecarloerror}
\end{equation}
where $\frac{\textrm{d}^{5}\sigma_{\text{v}}}{\textrm{d}^{5}\tau}$ is the virtual photoproduction cross section given by Eq.\!~\eqref{fulldiff}, $\mathcal{E}$ is the efficiency inside a multi-dimensional bin defined by Eq.\!~\eqref{eq:eff}, and $\delta \mathcal{E}$ is the absolute efficiency uncertainty.

The calculation of the efficiency uncertainty $\delta \mathcal{E}$ is not straightforward because (i) $N_{\text{gen}}$ and $N_{\text{rec}}$ in Eq.\!~\eqref{eq:eff} are not independent and (ii) Monte Carlo events in this equation are subject to weighting. Therefore, a special approach described in Ref.\!~\cite{Laforge:1996ts} was used to calculate $\delta \mathcal{E}$. Neglecting event migration between bins, this approach gives the following expression for the absolute statistical uncertainty of the efficiency in a bin for the case of a weighted Monte Carlo simulation:
\begin{equation}
\begin{aligned}
\delta \mathcal{E}(\Delta^{5} \tau) = \sqrt{\frac{\mathbb{N}_{\text{gen}} - 2\mathbb{N}_{\text{rec}}}{\mathbb{N}_{\text{gen}}^{3}}\sum\limits_{i=1}^{N_{\text{rec}}} w_{i}^{2} + \frac{\mathbb{N}_{\text{rec}}^{2}}{\mathbb{N}_{\text{gen}}^{4}}\sum\limits_{j=1}^{N_{\text{gen}}} w_{j}^{2}},
\end{aligned}
\label{eq:eff_err_weighted}
\end{equation}
where $N_{\text{gen}}$ and $N_{\text{rec}}$ are the numbers of generated and reconstructed Monte Carlo events inside a multi-dimensional bin, respectively, $\mathbb{N}_{\text{gen}}$ and  $\mathbb{N}_{\text{rec}}$ are the corresponding weighted event numbers, and $w$ is the weight of an individual event.

The two parts of the statistical uncertainty given by Eqs.\!~\eqref{staterrors} and \eqref{montecarloerror} were combined quadratically into the total absolute statistical uncertainty in each nonempty $\Delta^{5} \tau$ bin,
\begin{equation}
\delta_{\text{stat}}^{\text{tot}}(\Delta^{5} \tau) =
\sqrt{\left (\delta_{\text{stat}}^{\text{expt}} \right )^{2} + \left (\delta_{\text{stat}}^{\text{MC}}\right )^{2}}.
\label{errortot}
\end{equation}

The cross section assigned to the empty $\Delta^{5} \tau$ cells acquires zero statistical uncertainty.

For the extracted  single-differential cross sections, the statistical uncertainty $\delta_{\text{stat}}^{\text{tot}}(\Delta X)$ was obtained from the uncertainty  $\delta_{\text{stat}}^{\text{tot}}(\Delta^{5} \tau)$ of the five-fold differential cross section according to the standard error propagation rules (where $X$ denotes one of the final-state variables, {\it e.g.}~$M_{h_{1}h_{2}}$, $\theta_{h_1}$, $\alpha_{h_1}$).

\subsection{Model-dependent uncertainties}
\label{Sect:mod_dep}

In studies of double-pion production off free protons, the cross section model dependence originates from the filling of the empty cells, and the corresponding cross section uncertainty is commonly treated as a separate uncertainty type~\cite{Rip_an_note:2002,Ripani:2002ss,Fed_an_note:2007,Fedotov:2008aa,Isupov:2017lnd,Arjun,Fed_an_note:2017,Fed_paper_2018}. In this analysis, one further source of the model dependence appears, which is the correction that unfolds the effects of the initial proton motion (see Sec.\!~\ref{Sect:fermi_corr}). The two sources were found to give comparable uncertainties for the two lowest $W$ bins, while for the other bins the dominant part of the model-dependent uncertainty comes from the empty cells filling.

Both the contribution from the empty cells and the value of the unfolding correction vary greatly (from insignificant to considerable) for different bins in the final hadronic variables. Therefore, the model-dependent uncertainties were estimated in each $\Delta X$ bin of the single-differential cross sections (where $X$ is one of the final-state variables introduced in Sec.\!~\ref{Sect:kin_var}).

The absolute uncertainty $\delta^{\text{cells}}_{\text{model}}(\Delta X)$ due to the filling of the empty cells was calculated by
\begin{equation}
\delta^{\text{cells}}_{\text{model}} (\Delta X) = \frac{1}{2}\left ( \left [ \frac{\textrm{d}\sigma}{\textrm{d}X} \right ]_{\text{filled}}\!\!\!\! - \left [\frac{\textrm{d}\sigma}{\textrm{d}X} \right ]_{\text{not~\!filled}}\!~\!\right ),
\label{eq:error_mod_abs}
\end{equation}
where the parentheses contain the difference between the cross section values calculated with the empty cells contributions (``filled") and without them (``not filled"). See also Fig.\!~\ref{fig:empt_corr}.

For each $\Delta X$ bin of the single-differential distributions, the relative uncertainty due to the unfolding correction was estimated by
\begin{equation}
\varepsilon^{\text{unfold}}_{\text{model}} (\Delta X) = \left |\dfrac{\left [ \frac{\textrm{d}\sigma}{\textrm{d}X} \right ]_{\text{folded}} - \left [ \frac{\textrm{d}\sigma}{\textrm{d}X} \right ]_{\text{unfolded}}}{\left [ \frac{\textrm{d}\sigma}{\textrm{d}X} \right ]_{\text{folded}} + \left [ \frac{\textrm{d}\sigma}{\textrm{d}X} \right ]_{\text{unfolded}}} \right |.
\label{eq:rel_mod_err_fermi}
\end{equation}

The corresponding absolute uncertainty is then given by
\begin{equation}
\delta^{\text{unfold}}_{\text{model}} (\Delta X) = \left [ \frac{\textrm{d}\sigma_{\text{v}}}{\textrm{d}X} \right ]_{\text{final}}\!\! \cdot \varepsilon^{\text{unfold}}_{\text{model}}.
\label{eq:error_stat_mod_fermi}
\end{equation}

\subsection{Systematic uncertainties}
\label{Sect:sys_uncert}

The systematic uncertainty of the extracted cross sections was estimated in each bin in $W$ and $Q^{2}$. The following sources are considered to contribute to the total systematic uncertainty.

The presence of quasielastic events in the dataset facilitates the verification of both the overall cross section normalization and the quality of electron selection. The former may lack accuracy due to potential miscalibrations of the Faraday cup, fluctuations in the target density, and imprecise knowledge of other parameters involved in the luminosity calculation [see Eq.\!~\eqref{expcrossect}]. The electron selection quality in turn may suffer from potential miscalibrations of different detector parts, inaccuracies in electron tracking and identification, and uncertainties of the cuts and corrections involved in the electron selection.

To verify the cross section normalization and the quality of the electron selection, the quasielastic cross section was estimated and compared with the Bosted  parameterization of the quasielastic cross section off the deuteron~\cite{Bosted_fit,Bosted:2007xd}. This comparison indicated a better than 5\% agreement between the experimental and the parameterized cross sections, and therefore a 5\% global uncertainty was assigned to the extracted double-pion cross sections to account for potential inaccuracies in the normalization and electron selection~\cite{my_an_note:2020, my_thesis:2021}.

In this study, the cross sections were extracted in three sets of kinematic variables, as described in Sec.\!~\ref{Sect:kin_var}. The fully integrated cross sections $\sigma_{\text{v}}^{\text{int}}$ were found to slightly differ among the sets due to different data and efficiency propagation to various kinematic grids. For this reason, the arithmetic mean of the cross sections $\sigma_{\text{v}}^{\text{int}}$ obtained in the three variable sets is reported as a final result, and the standard error of the mean is interpreted as a systematic uncertainty~\cite{my_an_note:2020, my_thesis:2021}. Since different variable sets correspond to different registered final hadrons (and, therefore, to different combinations of the hadron cuts), this uncertainty includes potential inaccuracies in the hadron identification. The average value of this uncertainty among all $W$ and $Q^{2}$ bins is 1.6\%.

The cut on the relative efficiency uncertainty performed in this study excluded from consideration all multi-dimensional cells with uncertainties greater than 30\% (see Sec.\!~\ref{Sect:eff_eval}). To estimate the systematic effect of this cut, the fully integrated cross sections were also calculated for the cut values 25\% and 35\%. As a final result, the arithmetic mean of the cross sections $\sigma_{\text{v}}^{\text{int}}$ calculated for these three cut values is reported, and the standard error of the mean is interpreted as a systematic uncertainty~\cite{my_an_note:2020, my_thesis:2021}. The systematic effect of the relative efficiency uncertainty cut was estimated for each bin in $W$ and $Q^{2}$, and the average uncertainty value among all bins was found to be 0.8\%.

One more part of the systematic inaccuracies comes from the effective correction due to the FSI-background admixture. This correction was performed for experimental events in the $\pi^{-}$ missing topology as described in Sec.\!~\ref{Sect:excl_cut_pim_miss}. The fit shown in Fig.\!~\ref{fig:pim_miss_top_cut} and the corresponding correction factor given by Eq.\!~\eqref{eq:fsi_corr_fact} were found to be slightly dependent on the histogram binning. To account for this uncertainty, the correction factor was calculated for five different histogram bin sizes, and the arithmetic mean of these five individual values was used for the correction. The absolute uncertainty of the resulting correction factor was calculated as the standard error of the mean. The corresponding cross section uncertainty was estimated according to the standard error propagation rules~\cite{my_an_note:2020, my_thesis:2021}. The systematic uncertainty due to the FSI-background correction was estimated for each bin in $W$ and $Q^{2}$ where the correction was applied, and its average value was found to be 0.4\%.

As a common practice in studies of double-pion cross sections with CLAS~\cite{Rip_an_note:2002,Ripani:2002ss,Fed_an_note:2007,Fedotov:2008aa,Isupov:2017lnd,Arjun,Fed_an_note:2017,Fed_paper_2018}, a 5\% global uncertainty was assigned to the cross sections due to the inclusive radiative correction procedure (see Sec.\!~\ref{Sect:rad_corr}).

The uncertainties due to these sources were summed up in quadrature in each $W$ and $Q^{2}$ bin to obtain the total systematic uncertainty for the fully integrated cross sections. The average value of the total relative systematic uncertainty $\varepsilon_{\text{sys}}^{\text{tot}}$ is 7.4\%.

\subsection{Summary for the cross section uncertainties}
\label{Sect:uncert_resume}

Finally, the model-dependent uncertainties $\delta^{\text{cells}}_{\text{model}}(\Delta X)$ and $\delta^{\text{unfold}}_{\text{model}}(\Delta X)$ defined by Eqs.\!~\eqref{eq:error_mod_abs} and\!~\eqref{eq:error_stat_mod_fermi}, respectively, were combined with the total statistical uncertainty $\delta_{\text{stat}}^{\text{tot}}(\Delta X)$ defined in Sec.\!~\ref{Sect:stat_uncert} as follows:
\begin{equation}
\delta_{\text{stat,mod}}^{\text{tot}} (\Delta X) =
\sqrt{\left (\delta_{\text{stat}}^{\text{tot}} \right )^{2} + \left (\delta^{\text{cells}}_{\text{model}}\right )^{2} + \left (\delta^{\text{unfold}}_{\text{model}}\right )^{2}}.
\label{eq:error_stat_mod}
\end{equation}

The extracted cross sections are reported with the uncertainty $\delta_{\text{stat,mod}}^{\text{tot}}$, which for the single-differential distributions is given by Eq.\!~\eqref{eq:error_stat_mod}. For the fully integrated cross sections, $\delta_{\text{stat,mod}}^{\text{tot}}$ is obtained from the uncertainty of the single-differential distributions according to the standard error propagation rules. For the majority of integral $(W,Q^{2})$ points, the uncertainty $\delta_{\text{stat,mod}}^{\text{tot}}$ stays on a level of 4-6\%.

For the fully integrated cross sections, in addition~to the uncertainty $\delta_{\text{stat,mod}}^{\text{tot}}$, the total systematic uncertainty is also reported as a separate quantity. If necessary, the relative systematic uncertainty $\varepsilon_{\text{sys}}^{\text{tot}}$ can be propagated as a global factor to the corresponding single-differential distributions in each $W$ and $Q^{2}$ bin.

For most points of the fully integrated cross sections, the uncertainty $\delta_{\text{stat,mod}}^{\text{tot}}$ is less than the total systematic uncertainty, exceeding the latter only for $W \lesssim 1.4$~GeV. This is because $\delta_{\text{stat,mod}}^{\text{tot}}$ rises near the reaction threshold due to small experimental statistics, large contributions from the empty cells (see Sec.\!~\ref{Sect:empt_cells}), and pronounced impact of the unfolding correction (see Sec.\!~\ref{Sect:fermi_corr}).

The extracted cross sections with their estimated uncertainties are presented in Sec.\!~\ref{Sect:cr_sect_qf}.

\section{Final-state interactions}
\label{Sect:discuss_fsi}

\subsection{FSIs for $\gamma_{v}p(n) \rightarrow p' (n')\pi^{+}\pi^{-}$}
\label{Sect:intro_fsi}

Hadrons produced in exclusive reactions are subject to final-state interactions (FSIs). The nature of this phenomenon is complicated due to numerous mechanisms being involved, most of which are driven by the strong interaction~\cite{Darwish:2002qu,PhysRevC.84.035203}. 

For reactions occurring off nucleons contained in nuclei, one can separate FSIs into two general types:

\begin{itemize}
\item interactions among the final hadrons\footnote[3]{Here the term ``final hadrons" denotes $p'$, $\pi^{+}$, and $\pi^{-}$, which define the reaction final state.} and
\item interaction of the final hadrons with the spectator nucleon\footnote[4]{Here the term ``spectator" denotes the neutron, which is the spectator of the original exclusive reaction.}.
\end{itemize}

Both FSI types can involve simple momentum exchanges between the hadrons, as well as far more complicated processes, such as nucleon resonance excitations or charge exchange.

Clearly, FSIs in conventional reactions off free protons are limited to the first type.

The reaction final hadrons are produced in one vertex, and after the production, they fly apart in radial directions.  The spectator neutron, which was not involved in the reaction, is located aside from the production vertex, so that the final hadrons can scatter off the neutron. Therefore, for double-pion production off protons bound in deuterium, FSIs with the spectator correspond to a combination of proton-neutron~\cite{Shirokov_Yudin:1980,PhysRev.75.705} and pion-neutron scattering~\cite{PhysRevD.20.2804,Gasparyan:2003fp,Vrana:1999nt} and thus represent a superposition of a broad spectrum of mechanisms inherent for these two scattering types.

Due to the relatively low energy of the final hadrons in this experiment, the majority of FSIs in the investigated reaction are thought to happen elastically, which implies that the quantum numbers of the participating hadrons do not change, and no new particles are produced in such interactions~\cite{Shirokov_Yudin:1980}.

\subsection{Distortions due to FSIs}
\label{Sect:fsi_probing}

The two FSI types introduced above have some kinematic distinctions from each other and their impacts on the extracted cross sections also differ.%

An important feature of FSIs among the final hadrons is that they preserve the net four-momentum of all three final hadrons (as long as no new particles are produced in such interactions). For this reason, this FSI type does not affect missing mass distributions~\cite{my_thesis:2021}. However, interactions among the final hadrons still alter the momenta of the individual particles and thus introduce distortions into the measured cross sections (no matter whether off a free or bound nucleon). These cross section distortions can hardly be avoided at the level of experimental data analysis due to insensitivity of missing mass distributions to this FSI type. This issue needs to be accounted for at the level of theoretical and/or phenomenological cross section interpretation.

Meanwhile, interactions with the spectator nucleon have one distinctive difference from FSIs among the final hadrons. Specifically, as the spectator nucleon is extrinsic to the original exclusive reaction, any interaction with it changes the total four-momentum of the reaction final state. As a consequence, events in which the final hadrons interacted with the spectator introduce distortions into missing mass distributions~\cite{note_mm_distr}. The distortions reveal agglomerations of FSI-affected events, which allows for their separation from the quasifree event sample, so that the quasifree cross sections can be extracted (see more details in Sec.\!~\ref{Sect:excl_cut}).

In this analysis, distortions due to FSIs with the spectator neutron are clearly visible in experimental distributions of the missing quantities $P_{X}$ and  $M^{2}_{X[\pi^{-}]}$, where the former is defined by Eq.\eqref{eq:excl_top_quant} and the latter is $[P_{X[\pi^{-}]}^{\mu}]^2$ with the $P_{X[\pi^{-}]}^{\mu}$ defined by Eq.\eqref{eq:pimmiss_top_quant}. These distortions were found to differ in different regions of the reaction phase space and also to be topology dependent. Illustrations and further details are given in the following sections.

\begin{figure}[htp]
\begin{center}
\includegraphics[width=0.45\textwidth]{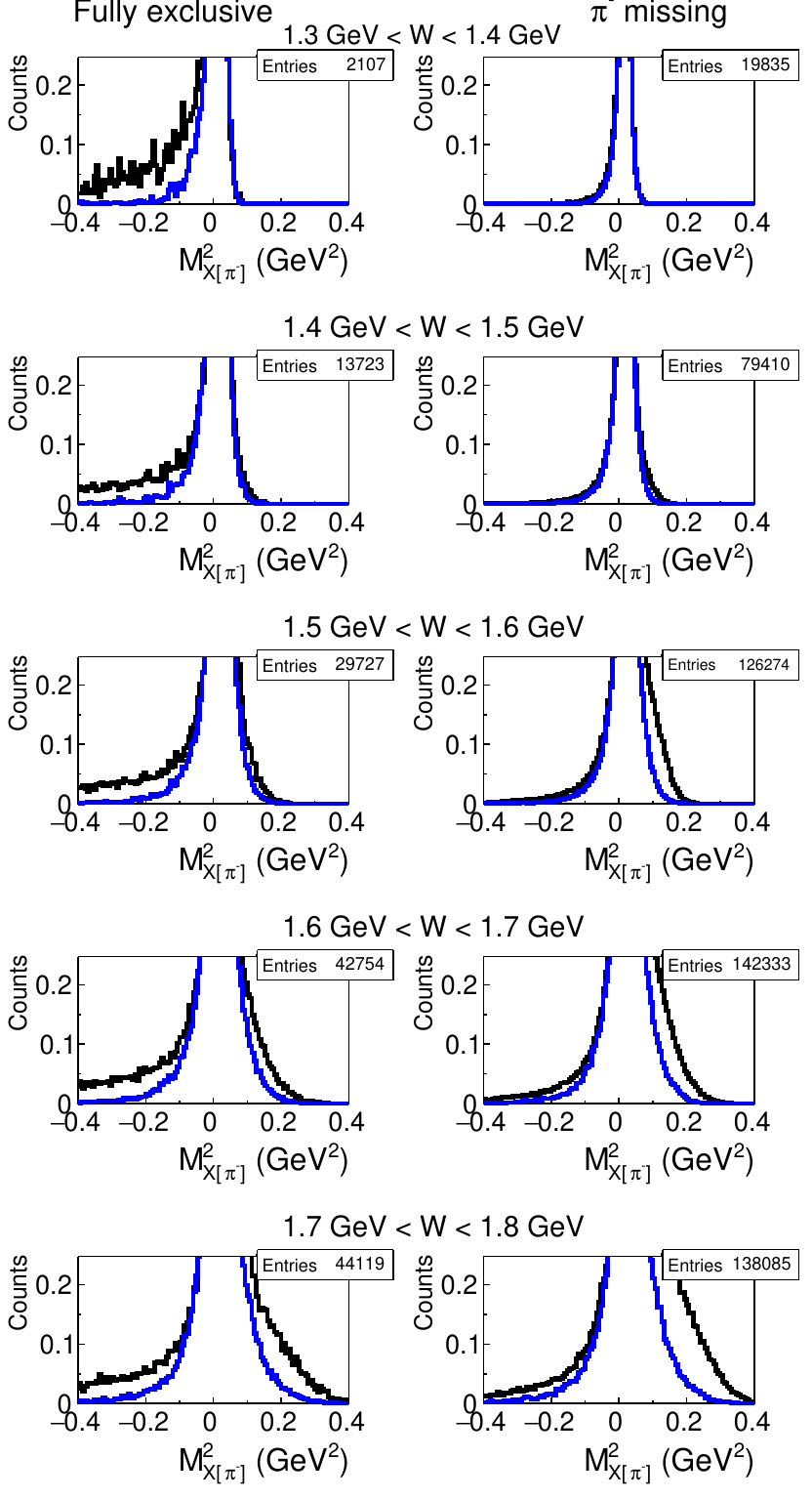}
\caption{\small $M^{2}_{X[\pi^{-}]}$ distributions for the fully exclusive (left) and the $\pi^{-}$ missing (right) topologies in five 100-MeV-wide bins in $W$. The mismatch between the experimental (black) and the simulated (blue) histograms reveals agglomerations of events in which the final hadrons interacted with the spectator neutron. The distributions are normalized in a way that the peak maxima are equal to one and then zoomed in on the range [0,~0.25] on the $y$ axis, to better visualize the mismatch. The presented statistics correspond to the unzoomed experimental distributions.
} \label{fig:top_comp}
\end{center}
\end{figure}

\subsection{Comparison between the two topologies}
\label{sect:fsi_top_comp}

Figure~\ref{fig:top_comp} presents the $M^{2}_{X[\pi^{-}]}$ distributions for the fully exclusive (left) and the $\pi^{-}$ missing (right) topologies in five 100-MeV-wide bins in $W$. Experimental distributions are shown in black, while the blue histograms correspond to simulated distributions of pure quasifree events. The mismatch between them therefore reveals agglomerations of events in which the final hadrons interacted with the spectator neutron. The Monte Carlo simulation was performed on the basis of the TWOPEG-D event generator~\cite{twopeg-d}, which successfully reproduces the Fermi smearing of the missing quantities but does not include FSI effects. The distributions are normalized in a way that the peak maxima are equal to one and then zoomed in on the range [0,~0.25] on the $y$ axis, to better visualize the mismatch.

Final-state hadrons attributed to various topologies have different kinematics and therefore different probabilities to interact with the spectator neutron. For this reason, events affected by FSIs with the spectator are distributed differently in the two reaction topologies as illustrated by Fig.\!~\ref{fig:top_comp}. More details can be found in Ref.\!~\cite{my_thesis:2021}.

\subsection{FSIs in the fully exclusive topology}
\label{Sect:fsi_fully_excl}

To better understand the redistribution of events with FSIs in the fully exclusive topology, experimental $P_{X}$ and  $M^{2}_{X[\pi^{-}]}$ distributions were examined in different slices of the final hadron momentum magnitudes and polar angles (in the Lab system).

\begin{figure}[htp]
\begin{center}
\includegraphics[width=0.45\textwidth]{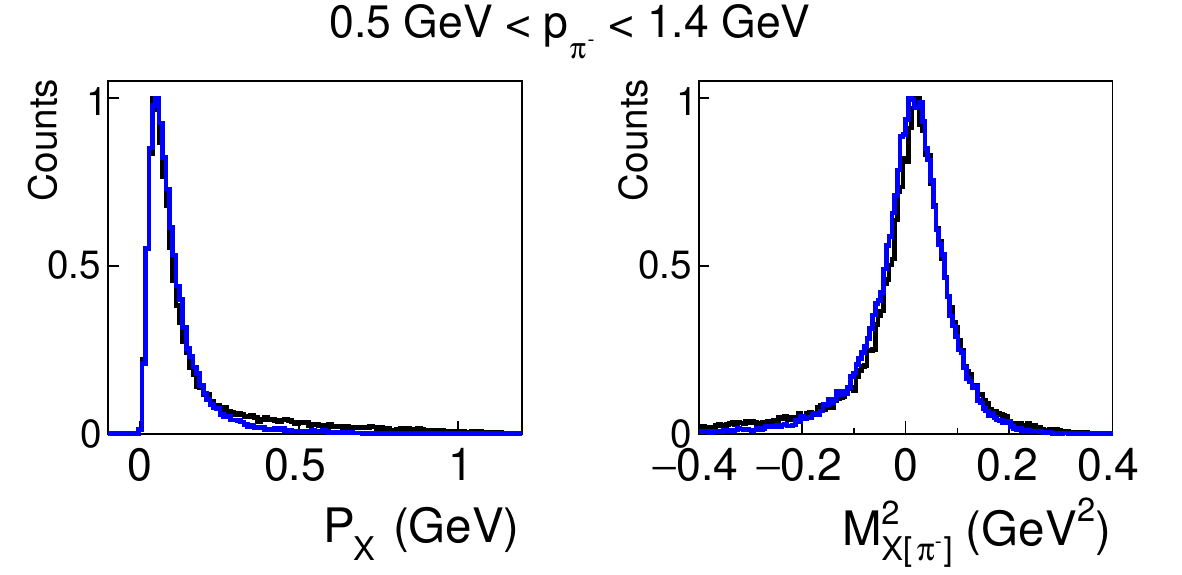}
\caption{\small Experimental (black) and simulated (blue) distributions of the quantities $P_{X}$ (left) and $M^{2}_{X[\pi^{-}]}$ (right) for the $\pi^{-}$ momentum magnitude ranging from 0.5 to 1.4~GeV. The distributions are plotted for the fully exclusive topology and normalized in a way that the maxima of the main peaks are equal to one.} \label{fig:fsi_pim_mom}
\end{center}
\end{figure}

In general, the proportion of events affected by FSIs with the spectator was found to vary greatly (from negligible to considerable) in different ranges of the final hadron momenta and/or angles.  

Remarkably, some regions of the reaction phase space were found to be completely dominated by quasifree events.  For example, Fig.\!~\ref{fig:fsi_pim_mom} shows a good match between the experimental (black) and the simulated (blue) distributions of the quantities $P_{X}$ (left) and $M^{2}_{X[\pi^{-}]}$ (right) for the $\pi^{-}$ momentum magnitude from 0.5 to 1.4~GeV, which indicates the dominance of quasifree events in this momentum range.

\begin{figure}[htp]
\begin{center}
\includegraphics[width=0.45\textwidth]{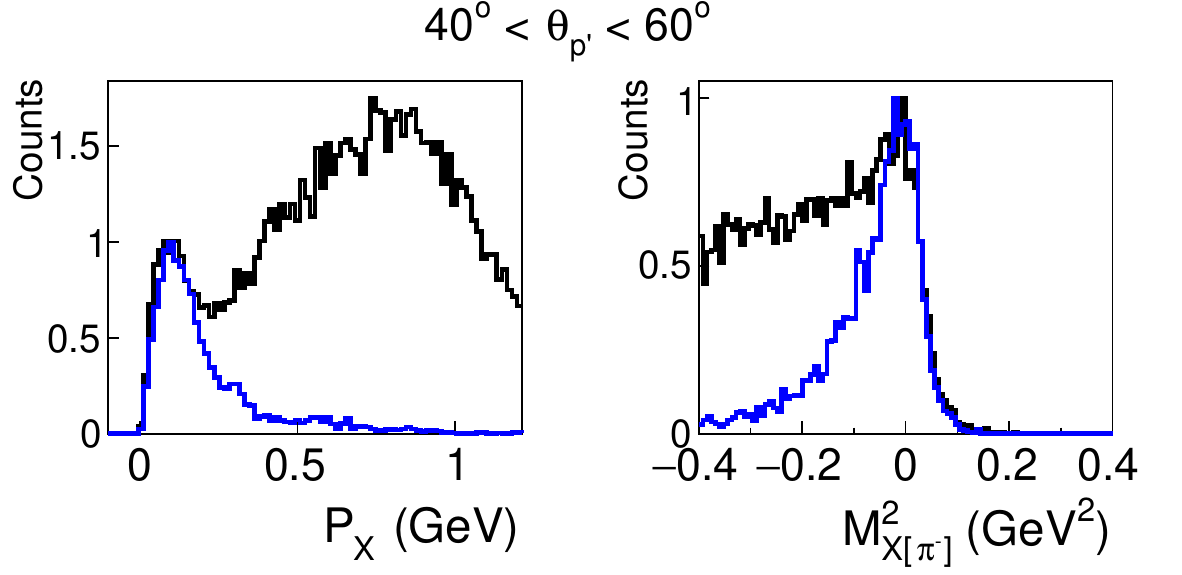}
\caption{\small Experimental (black) and simulated (blue) distributions of the quantities $P_{X}$ (left) and $M^{2}_{X[\pi^{-}]}$ (right) for the proton polar angle ranging from 40$^{\circ}$ to 60$^{\circ}$. The distributions are plotted for the fully exclusive topology and normalized in a way that the maxima of the main peaks are equal to one.} \label{fig:fsi_pr_ang}
\end{center}
\end{figure}

Conversely, other regions of the reaction phase space were revealed to be mostly populated with events in FSI-disturbed kinematics. Figure~\ref{fig:fsi_pr_ang} illustrates this effect for the proton polar angle range from 40$^{\circ}$ to 60$^{\circ}$, where a large mismatch between the experimental (black) and the simulated (blue) distributions is observed, which reveals a considerable fraction of events affected by FSIs with the spectator in this kinematic region. More illustrations can be found in Ref.\!~\cite{my_thesis:2021}.

\subsection{FSIs in topologies with a missing hadron}
\label{sect:fsi_top_miss}

In topologies with an unregistered hadron $i$, the quantity $M^{2}_{X[i]}$ is typically used for the channel identification (it is defined analogously to $M^{2}_{X[\pi^{-}]}$ introduced above). For reactions occurring off bound nucleons, interactions between the final hadrons and the spectator nucleon have different impacts on $M^{2}_{X[i]}$ depending on which final hadron experienced the FSI.

\begin{figure*}[htp]
\begin{center}
\includegraphics[width=16.25cm]{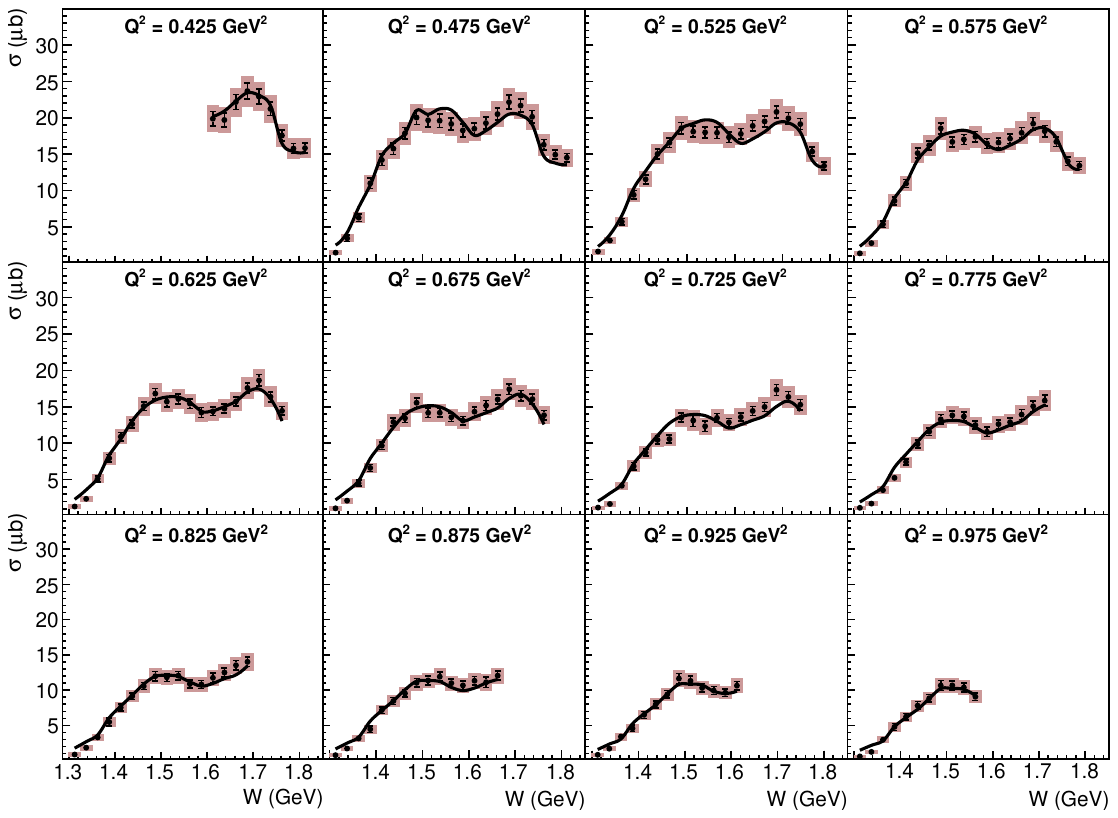}
\caption{\small $W$ dependences of the fully integrated cross sections in various bins in $Q^{2}$. The pink shadowed areas show the total cross section uncertainty, which is the uncertainty $\delta_{\text{stat,mod}}^{\text{tot}}$ (see Sec.\!~\ref{Sect:uncert_resume}) summed up in quadrature with the total systematic uncertainty (see Sec.\!~\ref{Sect:sys_uncert}). The error bars correspond to the $\delta_{\text{stat,mod}}^{\text{tot}}$ uncertainty only. The cross section estimation shown by the solid curves is based on the free proton event generator TWOPEG~\cite{twopeg} (see text for more details).} \label{fig:w_dep_model}
\end{center}
\end{figure*}

In general, the following three possibilities can be distinguished for events from the topologies with an unregistered hadron for reactions off a proton bound in deuterium (assuming that at most one final hadron in an event interacted with the neutron).

\begin{itemize}
\item[1.] All final hadrons in an event avoided interactions with the neutron. Then this event is a true quasifree event and the four-momentum of the unregistered hadron can be successfully reconstructed as missing.
\item[2.] The unregistered hadron avoided FSIs, while one of the registered hadrons interacted with the neutron, changing its four-momentum and hence losing kinematic affiliation to the original exclusive reaction. This does not allow for proper reconstruction of the missing hadron~four-momentum, causing the event to contribute to the FSI background in the $M^{2}_{X[i]}$~distributions~\cite{note_mm_distr}.
\item[3.] The unregistered hadron interacted with the neutron and the registered hadrons did not. In this case, the missing four-momentum of the unregistered hadron corresponds to its four-momentum before the FSI. Such an event then kinematically mimics a quasifree event.
\end{itemize}

This disposition reveals that for reactions off bound nucleons, topologies with a missing hadron suffer from the presence of falsely defined quasifree events, which are events of the third type. Such events are kinematically indistinguishable from true quasifree events and for this reason this effect can hardly be corrected for.

\section{Extracted quasifree cross sections}
\label{Sect:cr_sect_qf}

\begin{figure*}[htp]
\begin{center}
\includegraphics[width=0.75\textwidth]{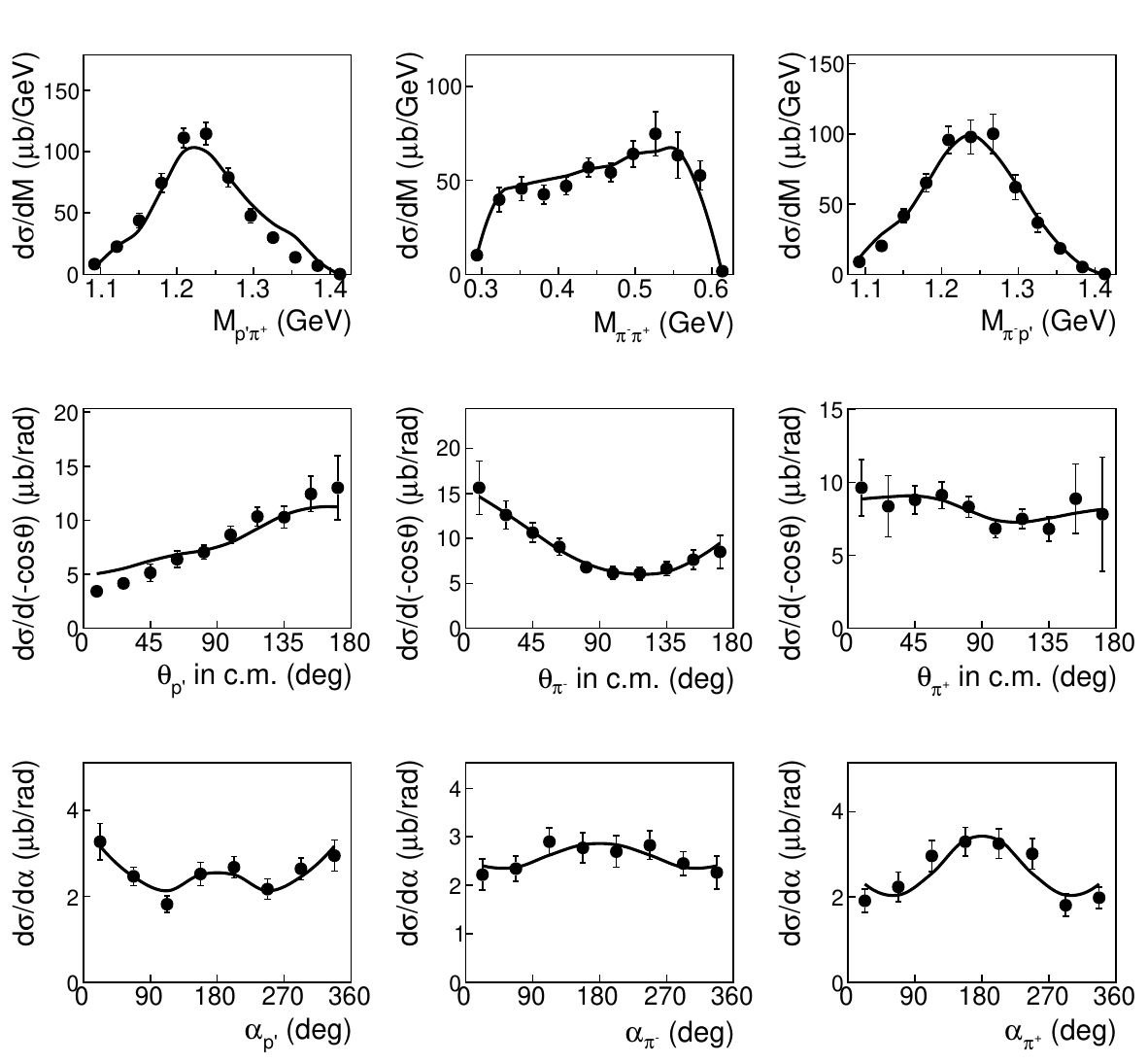}
\caption{\small Single-differential cross sections for $W= 1.5375$~GeV and $Q^{2}= 0.625$~GeV$^{2}$. The error bars correspond to the uncertainty $\delta_{\text{stat,mod}}^{\text{tot}}$ defined in Sec.\!~\ref{Sect:uncert_resume}. The cross section estimation shown by the solid curves is based on the free proton event generator TWOPEG~\cite{twopeg} (see text for more details).} \label{fig:diff_model}
\end{center}
\end{figure*}

In Fig.\!~\ref{fig:w_dep_model}, the $W$ dependences of the extracted fully integrated cross sections of the reaction $\gamma_{v} p (n) \rightarrow p' (n') \pi^{+} \pi^{-}$ are shown by the solid black circles for 12 analyzed $Q^{2}$ bins. For each point, the pink shadowed area is the total cross section uncertainty, which is the uncertainty $\delta_{\text{stat,mod}}^{\text{tot}}$ (see Sec.\!~\ref{Sect:uncert_resume}) summed up in quadrature with the total systematic uncertainty (see Sec.\!~\ref{Sect:sys_uncert}). The error bars correspond to the $\delta_{\text{stat,mod}}^{\text{tot}}$ uncertainty~only.

For each integral cross section point, a set of nine single-differential cross sections was obtained (as described in Sec.\!~\ref{Sect:cr_sect_formula}). As a typical example, Figure~\ref{fig:diff_model} presents the single-differential cross sections for $W= 1.5375$~GeV and $Q^{2}= 0.625$~GeV$^{2}$. The cross sections are reported with the uncertainty $\delta_{\text{stat,mod}}^{\text{tot}}$ shown by the error bars. The full set of extracted single-differential cross sections is available in the CLAS physics database~\cite{CLAS_DB} and also on GitHub~\cite{Github:data}.

The extracted cross sections are quasifree, meaning that the contribution from events in which the final hadrons interacted with the spectator neutron was reduced to the kinematically achievable minimum. The cross sections, however, are still convoluted with effects of FSIs among the final hadrons, which is like conventional free proton cross sections (see Sec.\!~\ref{Sect:fsi_probing}).

In general, the admixture of the FSI background left after the exclusivity cut in the $\pi^{-}$ missing topology may potentially affect the shape of the extracted single-differential distributions (because this effect was corrected only in an integral sense as described in Sec.\!~\ref{Sect:excl_cut_pim_miss}). However, as this admixture is present only in the $\pi^{-}$ missing topology for $W>$~1.45~GeV and stays at a level of 3-7\%, its impact is not thought to be discernible against the total cross section uncertainty.

\begin{figure*}[htp]
\begin{center}
\includegraphics[width=16.25cm]{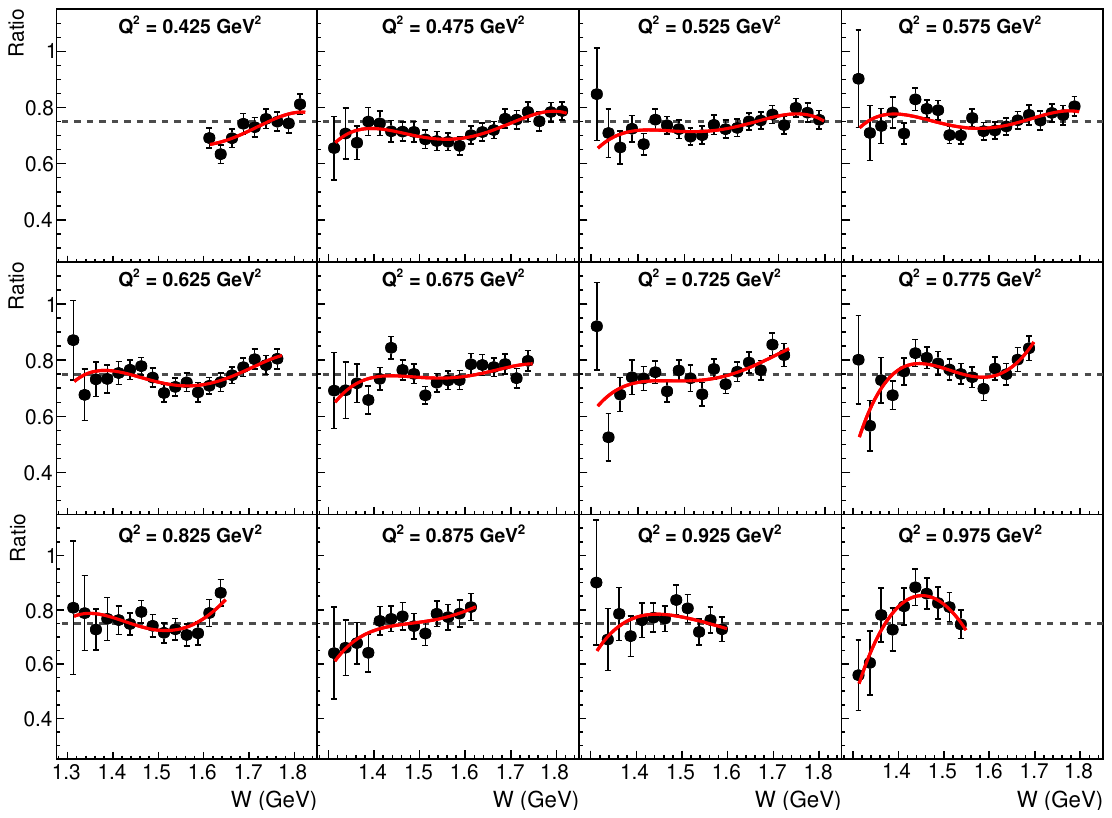}
\caption{\small Ratio of the fully integrated quasifree cross sections obtained in this study to the free proton cross sections from Refs.\!~\cite{Fed_an_note:2017,Fed_paper_2018}. The red curves correspond to polynomial fits. The dashed line marks the value of 0.75. } \label{fig:fed_w}
\end{center}
\end{figure*}

One more potential uncertainty source for the extracted quasifree cross sections is the presence of falsely identified quasifree events in the $\pi^{-}$ missing topology. Unfortunately, as true quasifree events are kinematically identical to those that are falsely identified, no corresponding correction to the cross sections can be developed (see Sec.\!~\ref{sect:fsi_top_miss} for details). Also note that in this study, the initial proton is assumed to be on shell.

The solid curves in Figs.~\ref{fig:w_dep_model} and~\ref{fig:diff_model} correspond to the cross section estimation performed by means of TWOPEG~\cite{twopeg}, which is an event generator for double-pion electroproduction off free protons. This event generator currently provides the best estimation of free proton cross sections in the investigated kinematic region.

The cross section approximation implemented into the TWOPEG event generator is based on the meson-baryon reaction model JM~\cite{Mokeev:2008iw,Mokeev:2012vsa,Mokeev:2015lda}. The generator employs the five-differential structure functions from the JM model fit to the existing CLAS results on double-pion photo- and electroproduction off free protons~\cite{Ripani:2002ss,Mokeev:2012vsa,Fedotov:2008aa,Golovach}. In the kinematic areas already covered by the CLAS data, TWOPEG performs interpolation of the model structure functions and successfully reproduces the available fully integrated and single-differential cross sections. In the areas not yet covered by the CLAS data, special extrapolation procedures were applied that included additional world data on the fully integrated photoproduction cross sections~\cite{Wu:2005wf,ABBHHM:1968aa}.

For the comparison shown in Figs.~\ref{fig:w_dep_model} and~\ref{fig:diff_model}, cross section distributions obtained by TWOPEG were normalized to integrally match the quasifree cross sections extracted in this study.

Figures~\ref{fig:w_dep_model} and~\ref{fig:diff_model} indicate that apart from the overall integral scaling, free proton cross sections may serve as an adequate zeroth-order approximation for quasifree cross sections off protons in deuterium. More elaborate insight into the interpretation of the extracted cross sections is given in the next section, which presents their comparison with the free proton measurements from Refs.\!~\cite{Fed_an_note:2017,Fed_paper_2018}.

\section{Comparison with the free proton measurements}
\label{Sect:with_fed_comp}

\begin{figure*}[htp]
\begin{center}
\includegraphics[width=16.25cm]{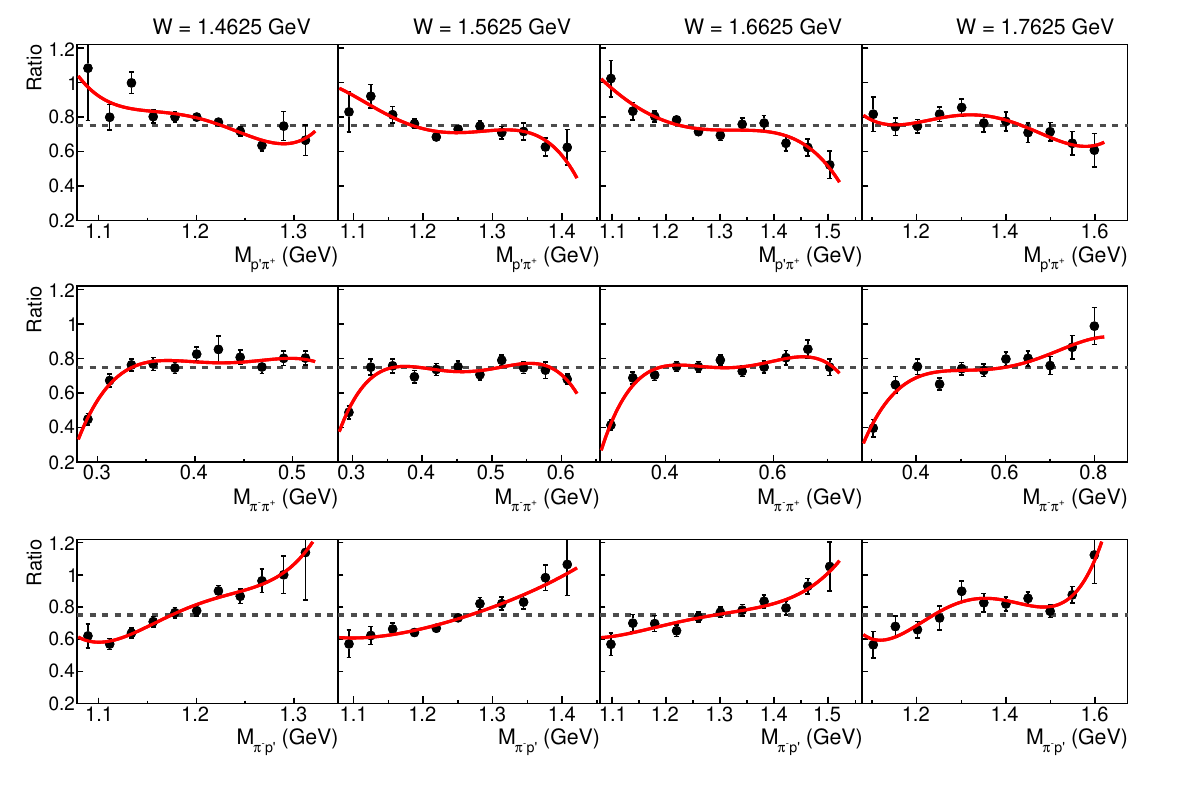}
\caption{\small Ratio of the invariant mass distributions obtained in this study to their free proton counterparts from Refs.\!~\cite{Fed_an_note:2017,Fed_paper_2018}. Rows from top to bottom correspond to $M_{p'\pi^{+}}$, $M_{\pi^{-}\pi^{+}}$, and $M_{\pi^{-}p'}$, respectively. The ratios were averaged over $Q^{2}$ to decrease the resulting uncertainties. The red curves correspond to polynomial fits. The dashed line marks the value of 0.75.} \label{fig:fed_mass}
\end{center}
\end{figure*}

\begin{figure*}[htp]
\begin{center}
\includegraphics[width=12cm]{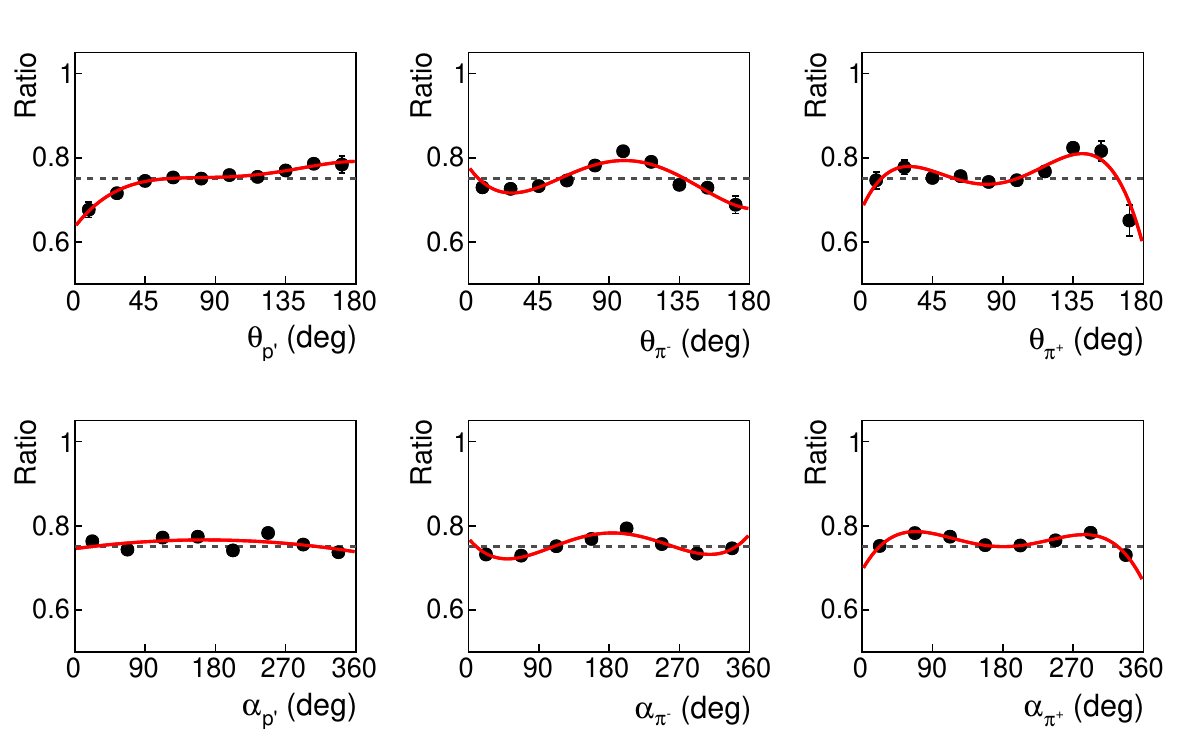}
\caption{\small Ratio of the angular distributions from this study to their free proton counterparts from Refs.\!~\cite{Fed_an_note:2017,Fed_paper_2018}. The first row shows the $\theta_{p'}$, $\theta_{\pi^{-}}$, and $\theta_{\pi^{+}}$ distributions, while the second row shows the $\alpha_{p'}$, $\alpha_{\pi^{-}}$, and $\alpha_{\pi^{+}}$ distributions, respectively. The ratios were averaged over $W$ and $Q^{2}$ to minimize the resulting uncertainties. The red curves correspond to polynomial fits. The dashed line marks the value of 0.75.} \label{fig:fed_ang}
\end{center}
\end{figure*}

This study benefits from the fact that the cross sections of the same exclusive reaction off free protons have been recently extracted from CLAS data~\cite{Fed_an_note:2017,Fed_paper_2018}. These free proton measurements were performed under the same experimental conditions as in this study, including the beam energy value and the target setup. For most points of the fully integrated cross sections, the uncertainty $\delta_{\text{stat,mod}}^{\text{tot}}$ is at a level of 1-3\% for~the~free proton measurements and at a level of 4-6\% for the cross sections obtained in this study (see Sec.\!~\ref{Sect:uncert_resume}). The two measurements have identical binning in all kinematic variables and similar inherent systematic inaccuracies (with a typical value of the total systematic uncertainty in a $W$ and $Q^2$ bin of about 7-8\%). Therefore, a direct comparison of these two cross section sets would provide the experimentally best possible opportunity to explore distinctions between the $\pi^{+}\pi^{-}$ electroproduction off protons in deuterium and the corresponding reaction off free protons.

This section compares the cross sections extracted in this study with their free proton counterparts from Refs.\!~\cite{Fed_an_note:2017,Fed_paper_2018} and examines the difference between them over the entire reaction phase space. In this~comparison, only statistical and model-dependent uncertainties of the two measurements were considered ($\delta_{\text{stat,mod}}^{\text{tot}}$), while systematic effects were assumed to cancel out.

Figure~\ref{fig:fed_w} presents the ratio of the fully integrated quasifree cross sections obtained in this study to the free proton cross sections of Refs.\!~\cite{Fed_an_note:2017,Fed_paper_2018}. The ratio was fit with a fourth-order polynomial. The dashed line marks the value of 0.75.  

As seen in Fig.\!~\ref{fig:fed_w}, the ratio of the two cross section sets demonstrates a modest $W$ dependence with an average level of 70-75\%, appearing to drop slightly near the reaction threshold, as well as in the dip region between the two integral resonance peaks.

Figure~\ref{fig:fed_mass} shows the ratio of the invariant mass distributions from this study to those from the free proton study~\cite{Fed_an_note:2017,Fed_paper_2018}. Rows from top to bottom correspond to $M_{p'\pi^{+}}$, $M_{\pi^{-}\pi^{+}}$, and $M_{\pi^{-}p'}$, respectively. The ratios were obtained individually for each $(W,Q^{2})$ point and then averaged over $Q^{2}$ to decrease the resulting uncertainties. Kinematic broadening of the invariant mass distributions with $W$ (described in Sec.\!~\ref{Sect:binning}) and the consequent nonidentical distribution of data points in different $W$ bins did not allow for further averaging over $W$. The red curves correspond to the fit with a fourth-order polynomial and the dashed line marks the value of 0.75.  

As seen in Fig.\!~\ref{fig:fed_mass}, the cross section ratio demonstrates different consistent patterns for the three invariant mass distributions. For $M_{p'\pi^{+}}$ (top row), it gives a rise near the left distribution edge, then gradually decreases towards the right edge, featuring a small plateau in the middle. For $M_{\pi^{-}\pi^{+}}$ (middle row), the situation is different as the cross section ratio shows a pronounced drop down to $\approx$40\% at the left edge, then rises abruptly up to $\approx$75\% and stays on this constant level further on. For the third invariant mass, $M_{\pi^{-}p'}$ (bottom row), the ratio continuously and almost linearly grows from $\approx$60\% at the left edge to $\approx$100\% at the right.

Figure~\ref{fig:fed_ang} presents the ratio of the angular distributions from this study to those from the free proton study~\cite{Fed_an_note:2017,Fed_paper_2018}. The first row shows the $\theta_{p'}$, $\theta_{\pi^{-}}$, and $\theta_{\pi^{+}}$ distributions, while the second row shows the $\alpha_{p'}$, $\alpha_{\pi^{-}}$, and $\alpha_{\pi^{+}}$ distributions, respectively. The ratios were obtained individually for each $(W,Q^{2})$ point and then averaged over $W$ and $Q^{2}$ to minimize the resulting uncertainties. The red curves correspond to polynomial fits and the dashed line marks the value of 0.75.

As seen in Fig.\!~\ref{fig:fed_ang}, the behavior of the cross section ratio differs for various angular distributions. Its dependence on the hadron polar angles appears to be of the most interest. For $\theta_{p'}$ (top left plot), the ratio grows from $\approx$70\% at small angles up to $\approx$75\% at 50$^{\circ}$ and further stays on a distinct plateau up to 120$^{\circ}$, showing then a mild rise up to $\approx$80\% at backward angles. For $\theta_{\pi^{-}}$ (top middle plot), the ratio stays at $\approx$75\% at small angles, then grows up to $\approx$80\% giving a broad peak at around 100$^{\circ}$, and drops down to $\approx$70\% at backward angles. For the third polar angle, $\theta_{\pi^{+}}$ (top right plot), the ratio maintains the level of $\approx$75\% up to around 120$^{\circ}$, then peaks up to more than 80\% at 150$^{\circ}$, and finally shows a steep drop.

Note that conventional free proton cross sections represent all reaction events, while the cross sections extracted in this study are quasifree (up to the accuracy with which quasifree events could be kinematically isolated) and hence do not include contributions from events in which the final hadrons interacted with the spectator neutron. The latter events thus are mainly responsible for the difference between the two cross section sets. 

Therefore, the performed comparison allows us to estimate the proportion of events affected by FSIs with the spectator neutron for the reaction off protons in deuterium. From Figs.~\ref{fig:fed_w},~\ref{fig:fed_mass}, and~\ref{fig:fed_ang}, one can conclude that the contribution from such events to the total number of reaction events varies from $\approx$60\% to a few percent in different regions of the reaction phase space. However, for the most part of the phase space, one can estimate the contribution from events affected by FSIs with the spectator to be on a level of $\approx$25\%.

Meanwhile, a small part of the difference between the two cross section sets may come from other sources, as for example from possible modifications of nucleons and their excited states inside the nuclear medium~\cite{Mokeev:1995fy,Bianchi:1994ax,Ahrens:1986hn,Krusche:2004xz,Noble:1980my}. To make any conclusions on this matter and also to gain insight into different involved FSI mechanisms, a further more comprehensive investigation is needed, which should employ a theoretical interpretation of the obtained cross section ratios.

\section{CONCLUSIONS}

\everypar{\looseness=-1}

This paper reports the results of the experimental data analysis for the process of charged double-pion electroproduction off protons bound in deuterium.

The fully integrated and single-differential cross sections~of the reaction $\gamma_{v}p(n) \rightarrow p' (n')\pi^{+}\pi^{-}$ have been obtained for the first time. The measurements were performed in the kinematic region of the invariant mass $W$ from 1.3 to 1.825~GeV and the photon virtuality $Q^{2}$ from 0.4~GeV to 1.0~GeV$^2$. The results benefit from fine binning in all kinematic variables, small statistical uncertainties, and modest model dependence. The extracted cross sections are quasifree, meaning that the admixture of events in which the final hadrons interacted with the spectator neutron is reduced to the kinematically achievable minimum. The whole set of the obtained cross sections is available in the CLAS physics database~\cite{CLAS_DB} and also on GitHub~\cite{Github:data}.

Due to the Fermi motion that initial protons undergo in deuterium nuclei, this study encountered a set of peculiarities that were not relevant for free proton studies. Effects of the initial proton motion turned out to be intertwined with many analysis aspects: they led to the smearing of some kinematic quantities, altered the common procedure~of the Lab-to-CMS transformation, caused the need to perform an unfolding correction to the extracted cross sections, and more~\cite{my_an_note:2020, my_thesis:2021, twopeg-d, note_mm_distr}. To deal with these issues, special methods and techniques were developed, which go beyond the conventional analysis framework elaborated in previous free proton studies~\cite{Rip_an_note:2002,Ripani:2002ss,Fed_an_note:2007,Fedotov:2008aa,Isupov:2017lnd,Golovach,Arjun,Fed_an_note:2017,Fed_paper_2018}.

Interactions of the reaction final hadrons with the spectator nucleon represent another peculiar aspect of this analysis. These interactions introduce distortions into distributions of some kinematic quantities (such as missing masses), thus complicating the exclusive event selection. FSI effects were found to differ depending on the reaction topology due to nonidentical geometrical acceptance of the topologies in CLAS. For this study, isolation of quasifree events from the FSI background was performed in each topology individually according to the specially developed procedures.

The paper also presents the comparison of the obtained cross sections with the corresponding free proton cross sections recently extracted from CLAS data~\cite{Fed_an_note:2017,Fed_paper_2018}. Assuming that the difference between the two cross section sets mostly originates from events in which FSIs between the final hadrons and the spectator neutron took place, this comparison allowed us to make an estimate of the contribution from such events to the total number of reaction events. For the most part of the reaction phase space, the contribution from events affected by FSIs with the spectator was found to be around 25\%. Further theoretical interpretation of the obtained cross section ratios will allow one to investigate various involved FSI mechanisms and to explore other potential reasons that may contribute to the difference between the cross section sets, which includes possible in-medium modifications of properties of nucleons and their excited states~\cite{Mokeev:1995fy,Bianchi:1994ax,Ahrens:1986hn,Krusche:2004xz,Noble:1980my}.

\vspace{2em}

\begin{acknowledgments}
The authors thank the technical staff at Jefferson Lab and at all the participating institutions for their invaluable contributions to the success of the experiment.
This work was supported in part by the National Science Foundation (NSF) under Grant No.~PHY 10011349, Jefferson Science Associates (JSA), the U.S. Department of Energy (DOE) under Contract No.~DE-AC05-06OR23177, University of South Carolina (USC), the Skobeltsyn Institute of Nuclear Physics (SINP), the Physics Department at Lomonosov Moscow State University (MSU), Ohio University (OU), the Chilean Agencia Nacional de Investigacion y Desarrollo (ANID), the Italian Istituto Nazionale di Fisica Nucleare (INFN), the French Centre National de la Recherche Scientifique (CNRS), the French Commissariat \'a l'Energie Atomique (CEA), the Scottish Universities Physics Alliance (SUPA), the National Research Foundation of Korea (NRF), and the UK Science and Technology Facilities Council (STFC). 
\end{acknowledgments}

\bibliography{paper}{}

\begin{thebibliography}{63}%
\makeatletter
\providecommand \@ifxundefined [1]{%
 \@ifx{#1\undefined}
}%
\providecommand \@ifnum [1]{%
 \ifnum #1\expandafter \@firstoftwo
 \else \expandafter \@secondoftwo
 \fi
}%
\providecommand \@ifx [1]{%
 \ifx #1\expandafter \@firstoftwo
 \else \expandafter \@secondoftwo
 \fi
}%
\providecommand \natexlab [1]{#1}%
\providecommand \enquote  [1]{``#1''}%
\providecommand \bibnamefont  [1]{#1}%
\providecommand \bibfnamefont [1]{#1}%
\providecommand \citenamefont [1]{#1}%
\providecommand \href@noop [0]{\@secondoftwo}%
\providecommand \href [0]{\begingroup \@sanitize@url \@href}%
\providecommand \@href[1]{\@@startlink{#1}\@@href}%
\providecommand \@@href[1]{\endgroup#1\@@endlink}%
\providecommand \@sanitize@url [0]{\catcode `\\12\catcode `\$12\catcode
  `\&12\catcode `\#12\catcode `\^12\catcode `\_12\catcode `\%12\relax}%
\providecommand \@@startlink[1]{}%
\providecommand \@@endlink[0]{}%
\providecommand \url  [0]{\begingroup\@sanitize@url \@url }%
\providecommand \@url [1]{\endgroup\@href {#1}{\urlprefix }}%
\providecommand \urlprefix  [0]{URL }%
\providecommand \Eprint [0]{\href }%
\providecommand \doibase [0]{http://dx.doi.org/}%
\providecommand \selectlanguage [0]{\@gobble}%
\providecommand \bibinfo  [0]{\@secondoftwo}%
\providecommand \bibfield  [0]{\@secondoftwo}%
\providecommand \translation [1]{[#1]}%
\providecommand \BibitemOpen [0]{}%
\providecommand \bibitemStop [0]{}%
\providecommand \bibitemNoStop [0]{.\EOS\space}%
\providecommand \EOS [0]{\spacefactor3000\relax}%
\providecommand \BibitemShut  [1]{\csname bibitem#1\endcsname}%
\let\auto@bib@innerbib\@empty
\bibitem [{\citenamefont {Krusche}\ and\ \citenamefont
  {Schadmand}(2003)}]{Krusche:2003ik}%
  \BibitemOpen
  \bibfield  {author} {\bibinfo {author} {\bibfnamefont {B.}~\bibnamefont
  {Krusche}}\ and\ \bibinfo {author} {\bibfnamefont {S.}~\bibnamefont
  {Schadmand}},\ }\href {\doibase 10.1016/S0146-6410(03)90005-6} {\bibfield
  {journal} {\bibinfo  {journal} {Prog. Part. Nucl. Phys.}\ }\textbf {\bibinfo
  {volume} {51}},\ \bibinfo {pages} {399} (\bibinfo {year} {2003})}\BibitemShut
  {NoStop}%
\bibitem [{\citenamefont {Aznauryan}\ and\ \citenamefont
  {Burkert}(2012)}]{Aznauryan:2011qj}%
  \BibitemOpen
  \bibfield  {author} {\bibinfo {author} {\bibfnamefont {{\relax
  I.G}.}~\bibnamefont {Aznauryan}}\ and\ \bibinfo {author} {\bibfnamefont
  {{\relax V.D}.}~\bibnamefont {Burkert}},\ }\href {\doibase
  10.1016/j.ppnp.2011.08.001} {\bibfield  {journal} {\bibinfo  {journal} {Prog.
  Part. Nucl. Phys.}\ }\textbf {\bibinfo {volume} {67}},\ \bibinfo {pages} {1}
  (\bibinfo {year} {2012})}\BibitemShut {NoStop}%
\bibitem [{\citenamefont {Skorodumina}\ \emph
  {et~al.}(2015{\natexlab{a}})\citenamefont {Skorodumina} \emph
  {et~al.}}]{Skorodumina:2016pnb}%
  \BibitemOpen
  \bibfield  {author} {\bibinfo {author} {\bibfnamefont {{\relax
  Iu.A}.}~\bibnamefont {Skorodumina}} \emph {et~al.},\ }\href {\doibase
  10.3103/S002713491506017X} {\bibfield  {journal} {\bibinfo  {journal} {Moscow
  Univ. Phys. Bull.}\ }\textbf {\bibinfo {volume} {70}},\ \bibinfo {pages}
  {429} (\bibinfo {year} {2015}{\natexlab{a}})},\ \bibinfo {note} {[Vestn.
  Mosk. Univ.,no.6,3(2015)]}\BibitemShut {NoStop}%
\bibitem [{\citenamefont {Ireland}\ \emph {et~al.}(2020)\citenamefont
  {Ireland}, \citenamefont {Pasyuk},\ and\ \citenamefont
  {Strakovsky}}]{Ireland:2019uwn}%
  \BibitemOpen
  \bibfield  {author} {\bibinfo {author} {\bibfnamefont {{\relax
  D.G}.}~\bibnamefont {Ireland}}, \bibinfo {author} {\bibfnamefont
  {E.}~\bibnamefont {Pasyuk}}, \ and\ \bibinfo {author} {\bibfnamefont
  {I.}~\bibnamefont {Strakovsky}},\ }\href {\doibase
  10.1016/j.ppnp.2019.103752} {\bibfield  {journal} {\bibinfo  {journal} {Prog.
  Part. Nucl. Phys.}\ }\textbf {\bibinfo {volume} {111}},\ \bibinfo {pages}
  {103752} (\bibinfo {year} {2020})}\BibitemShut {NoStop}%
\bibitem [{\citenamefont {Thiel}\ \emph {et~al.}(2022)\citenamefont {Thiel},
  \citenamefont {Afzal},\ and\ \citenamefont {Wunderlich}}]{Thiel:2022xtb}%
  \BibitemOpen
  \bibfield  {author} {\bibinfo {author} {\bibfnamefont {A.}~\bibnamefont
  {Thiel}}, \bibinfo {author} {\bibfnamefont {F.}~\bibnamefont {Afzal}}, \ and\
  \bibinfo {author} {\bibfnamefont {Y.}~\bibnamefont {Wunderlich}},\ }\href
  {\doibase 10.1016/j.ppnp.2022.103949} {\bibfield  {journal} {\bibinfo
  {journal} {Prog. Part. Nucl. Phys.}\ }\textbf {\bibinfo {volume} {125}},\
  \bibinfo {pages} {103949} (\bibinfo {year} {2022})}\BibitemShut {NoStop}%
\bibitem [{\citenamefont {Mecking}\ \emph {et~al.}(2003)\citenamefont {Mecking}
  \emph {et~al.}}]{Mecking:2003zu}%
  \BibitemOpen
  \bibfield  {author} {\bibinfo {author} {\bibfnamefont {{\relax
  B.A}.}~\bibnamefont {Mecking}} \emph {et~al.},\ }\href {\doibase
  10.1016/S0168-9002(03)01001-5} {\bibfield  {journal} {\bibinfo  {journal}
  {Nucl. Instrum. Meth.}\ }\textbf {\bibinfo {volume} {A503}},\ \bibinfo
  {pages} {513} (\bibinfo {year} {2003})}\BibitemShut {NoStop}%
\bibitem [{\citenamefont {{CLAS Physics Database}}()}]{CLAS_DB}%
  \BibitemOpen
  \bibfield  {author} {\bibinfo {author} {\bibnamefont {{CLAS Physics
  Database}}},\ }\href@noop {} {}\bibinfo {howpublished}
  {\url{http://clas.sinp.msu.ru/cgi-bin/jlab/db.cgi}}\BibitemShut {NoStop}%
\bibitem [{\citenamefont {Mokeev}\ \emph {et~al.}(1995)\citenamefont {Mokeev},
  \citenamefont {Santopinto}, \citenamefont {Giannini},\ and\ \citenamefont
  {Ricco}}]{Mokeev:1995fy}%
  \BibitemOpen
  \bibfield  {author} {\bibinfo {author} {\bibfnamefont {{\relax
  V.I}.}~\bibnamefont {Mokeev}}, \bibinfo {author} {\bibfnamefont
  {E.}~\bibnamefont {Santopinto}}, \bibinfo {author} {\bibfnamefont {{\relax
  M.M}.}~\bibnamefont {Giannini}}, \ and\ \bibinfo {author} {\bibfnamefont
  {G.}~\bibnamefont {Ricco}},\ }\href {\doibase 10.1142/S0218301395000237}
  {\bibfield  {journal} {\bibinfo  {journal} {Int. J. Mod. Phys.}\ }\textbf
  {\bibinfo {volume} {E4}},\ \bibinfo {pages} {607} (\bibinfo {year} {1995})},\
  \Eprint {http://arxiv.org/abs/(see also Refs.[1-5] therein)} {(see also
  Refs.[1-5] therein)} \BibitemShut {NoStop}%
\bibitem [{\citenamefont {Bianchi}\ \emph {et~al.}(1994)\citenamefont {Bianchi}
  \emph {et~al.}}]{Bianchi:1994ax}%
  \BibitemOpen
  \bibfield  {author} {\bibinfo {author} {\bibfnamefont {N.}~\bibnamefont
  {Bianchi}} \emph {et~al.},\ }\href {\doibase 10.1016/0370-2693(94)90021-3}
  {\bibfield  {journal} {\bibinfo  {journal} {Phys. Lett.}\ }\textbf {\bibinfo
  {volume} {B325}},\ \bibinfo {pages} {333} (\bibinfo {year}
  {1994})}\BibitemShut {NoStop}%
\bibitem [{\citenamefont {Ahrens}(1985)}]{Ahrens:1986hn}%
  \BibitemOpen
  \bibfield  {author} {\bibinfo {author} {\bibfnamefont {J.}~\bibnamefont
  {Ahrens}},\ }\href {\doibase 10.1016/0375-9474(85)90591-3} {\bibfield
  {journal} {\bibinfo  {journal} {Nucl. Phys.}\ }\textbf {\bibinfo {volume}
  {A446}},\ \bibinfo {pages} {229C} (\bibinfo {year} {1985})}\BibitemShut
  {NoStop}%
\bibitem [{\citenamefont {Osipenko}\ \emph {et~al.}(2005)\citenamefont
  {Osipenko} \emph {et~al.}}]{Osipenko_2005_note}%
  \BibitemOpen
  \bibfield  {author} {\bibinfo {author} {\bibfnamefont {M.}~\bibnamefont
  {Osipenko}} \emph {et~al.},\ }\href@noop {} {\bibfield  {journal} {\bibinfo
  {journal} {{\normalfont CLAS-NOTE-2005-013,
  \url{https://arxiv.org/abs/hep-ex/0507098} }}\ } (\bibinfo {year}
  {2005})}\BibitemShut {NoStop}%
\bibitem [{\citenamefont {Osipenko}\ \emph {et~al.}(2006)\citenamefont
  {Osipenko} \emph {et~al.}}]{Osipenko:2005gt}%
  \BibitemOpen
  \bibfield  {author} {\bibinfo {author} {\bibfnamefont {M.}~\bibnamefont
  {Osipenko}} \emph {et~al.} (\bibinfo {collaboration} {CLAS Collaboration}),\
  }\href {\doibase 10.1103/PhysRevC.73.045205} {\bibfield  {journal} {\bibinfo
  {journal} {Phys. Rev.}\ }\textbf {\bibinfo {volume} {C73}},\ \bibinfo {pages}
  {045205} (\bibinfo {year} {2006})}\BibitemShut {NoStop}%
\bibitem [{\citenamefont {Osipenko}\ \emph {et~al.}(2010)\citenamefont
  {Osipenko} \emph {et~al.}}]{Osipenko:2010sb}%
  \BibitemOpen
  \bibfield  {author} {\bibinfo {author} {\bibfnamefont {M.}~\bibnamefont
  {Osipenko}} \emph {et~al.} (\bibinfo {collaboration} {CLAS Collaboration}),\
  }\href {\doibase 10.1016/j.nuclphysa.2010.05.059} {\bibfield  {journal}
  {\bibinfo  {journal} {Nucl. Phys.}\ }\textbf {\bibinfo {volume} {A845}},\
  \bibinfo {pages} {1} (\bibinfo {year} {2010})}\BibitemShut {NoStop}%
\bibitem [{\citenamefont {Mokeev}\ \emph {et~al.}(2009)\citenamefont {Mokeev}
  \emph {et~al.}}]{Mokeev:2008iw}%
  \BibitemOpen
  \bibfield  {author} {\bibinfo {author} {\bibfnamefont {{\relax
  V.I}.}~\bibnamefont {Mokeev}} \emph {et~al.},\ }\href {\doibase
  10.1103/PhysRevC.80.045212} {\bibfield  {journal} {\bibinfo  {journal} {Phys.
  Rev.}\ }\textbf {\bibinfo {volume} {C80}},\ \bibinfo {pages} {045212}
  (\bibinfo {year} {2009})}\BibitemShut {NoStop}%
\bibitem [{\citenamefont {Mokeev}\ \emph {et~al.}(2012)\citenamefont {Mokeev}
  \emph {et~al.}}]{Mokeev:2012vsa}%
  \BibitemOpen
  \bibfield  {author} {\bibinfo {author} {\bibfnamefont {{\relax
  V.I}.}~\bibnamefont {Mokeev}} \emph {et~al.} (\bibinfo {collaboration} {CLAS
  Collaboration}),\ }\href {\doibase 10.1103/PhysRevC.86.035203} {\bibfield
  {journal} {\bibinfo  {journal} {Phys. Rev.}\ }\textbf {\bibinfo {volume}
  {C86}},\ \bibinfo {pages} {035203} (\bibinfo {year} {2012})}\BibitemShut
  {NoStop}%
\bibitem [{\citenamefont {Mokeev}\ \emph {et~al.}(2016)\citenamefont {Mokeev}
  \emph {et~al.}}]{Mokeev:2015lda}%
  \BibitemOpen
  \bibfield  {author} {\bibinfo {author} {\bibfnamefont {{\relax
  V.I}.}~\bibnamefont {Mokeev}} \emph {et~al.},\ }\href {\doibase
  10.1103/PhysRevC.93.025206} {\bibfield  {journal} {\bibinfo  {journal} {Phys.
  Rev.}\ }\textbf {\bibinfo {volume} {C93}},\ \bibinfo {pages} {025206}
  (\bibinfo {year} {2016})}\BibitemShut {NoStop}%
\bibitem [{\citenamefont {Fedotov}\ \emph {et~al.}(2017)\citenamefont {Fedotov}
  \emph {et~al.}}]{Fed_an_note:2017}%
  \BibitemOpen
  \bibfield  {author} {\bibinfo {author} {\bibfnamefont {{\relax
  G.V}.}~\bibnamefont {Fedotov}} \emph {et~al.},\ }\href@noop {} {\bibfield
  {journal} {\bibinfo  {journal} {CLAS-Analysis-2017-101 (CLAS-NOTE-2018-001),
  \url{https://misportal.jlab.org/ul/Physics/Hall-B/clas/viewFile.cfm/2018-001.pdf?documentId=776}}\
  } (\bibinfo {year} {2017})}\BibitemShut {NoStop}%
\bibitem [{\citenamefont {Fedotov}\ \emph {et~al.}(2018)\citenamefont {Fedotov}
  \emph {et~al.}}]{Fed_paper_2018}%
  \BibitemOpen
  \bibfield  {author} {\bibinfo {author} {\bibfnamefont {{\relax
  G.V}.}~\bibnamefont {Fedotov}} \emph {et~al.} (\bibinfo {collaboration} {CLAS
  Collaboration}),\ }\href {\doibase 10.1103/PhysRevC.98.025203} {\bibfield
  {journal} {\bibinfo  {journal} {Phys. Rev.}\ }\textbf {\bibinfo {volume}
  {C98}},\ \bibinfo {pages} {025203} (\bibinfo {year} {2018})}\BibitemShut
  {NoStop}%
\bibitem [{\citenamefont {Krusche}(2005)}]{Krusche:2004xz}%
  \BibitemOpen
  \bibfield  {author} {\bibinfo {author} {\bibfnamefont {B.}~\bibnamefont
  {Krusche}},\ }\href {\doibase 10.1016/j.ppnp.2004.12.002} {\bibfield
  {journal} {\bibinfo  {journal} {Prog. Part. Nucl. Phys.}\ }\textbf {\bibinfo
  {volume} {55}},\ \bibinfo {pages} {46} (\bibinfo {year} {2005})}\BibitemShut
  {NoStop}%
\bibitem [{\citenamefont {Noble}(1981)}]{Noble:1980my}%
  \BibitemOpen
  \bibfield  {author} {\bibinfo {author} {\bibfnamefont {{\relax
  J.V}.}~\bibnamefont {Noble}},\ }\href {\doibase 10.1103/PhysRevLett.46.412}
  {\bibfield  {journal} {\bibinfo  {journal} {Phys. Rev. Lett.}\ }\textbf
  {\bibinfo {volume} {46}},\ \bibinfo {pages} {412} (\bibinfo {year}
  {1981})}\BibitemShut {NoStop}%
\bibitem [{\citenamefont {Amarian}\ \emph {et~al.}(2001)\citenamefont {Amarian}
  \emph {et~al.}}]{Amarian:2001zs}%
  \BibitemOpen
  \bibfield  {author} {\bibinfo {author} {\bibfnamefont {M.}~\bibnamefont
  {Amarian}} \emph {et~al.},\ }\href {\doibase 10.1016/S0168-9002(00)00996-7}
  {\bibfield  {journal} {\bibinfo  {journal} {Nucl. Instrum. Meth.}\ }\textbf
  {\bibinfo {volume} {A460}},\ \bibinfo {pages} {239} (\bibinfo {year}
  {2001})}\BibitemShut {NoStop}%
\bibitem [{\citenamefont {Mestayer}\ \emph {et~al.}(2000)\citenamefont
  {Mestayer} \emph {et~al.}}]{Mestayer:2000we}%
  \BibitemOpen
  \bibfield  {author} {\bibinfo {author} {\bibfnamefont {{\relax
  M.D}.}~\bibnamefont {Mestayer}} \emph {et~al.},\ }\href@noop {} {\bibfield
  {journal} {\bibinfo  {journal} {Nucl. Instrum. Meth. A}\ }\textbf {\bibinfo
  {volume} {449}},\ \bibinfo {pages} {81} (\bibinfo {year} {2000})}\BibitemShut
  {NoStop}%
\bibitem [{\citenamefont {Adams}\ \emph {et~al.}(2001)\citenamefont {Adams}
  \emph {et~al.}}]{Adams:2001kk}%
  \BibitemOpen
  \bibfield  {author} {\bibinfo {author} {\bibfnamefont {G.}~\bibnamefont
  {Adams}} \emph {et~al.},\ }\href {\doibase 10.1016/S0168-9002(00)01313-9}
  {\bibfield  {journal} {\bibinfo  {journal} {Nucl. Instrum. Meth.}\ }\textbf
  {\bibinfo {volume} {A465}},\ \bibinfo {pages} {414} (\bibinfo {year}
  {2001})}\BibitemShut {NoStop}%
\bibitem [{\citenamefont {Smith}\ \emph {et~al.}(1999)\citenamefont {Smith}
  \emph {et~al.}}]{Smith:1999ii}%
  \BibitemOpen
  \bibfield  {author} {\bibinfo {author} {\bibfnamefont {{\relax
  E.S}.}~\bibnamefont {Smith}} \emph {et~al.},\ }\href {\doibase
  10.1016/S0168-9002(99)00484-2} {\bibfield  {journal} {\bibinfo  {journal}
  {Nucl. Instrum. Meth.}\ }\textbf {\bibinfo {volume} {A432}},\ \bibinfo
  {pages} {265} (\bibinfo {year} {1999})}\BibitemShut {NoStop}%
\bibitem [{\citenamefont {Mutchler}\ \emph {et~al.}(1998)\citenamefont
  {Mutchler}, \citenamefont {Taylor},\ and\ \citenamefont
  {Smith}}]{clas_tof_paddles}%
  \BibitemOpen
  \bibfield  {author} {\bibinfo {author} {\bibfnamefont {G.}~\bibnamefont
  {Mutchler}}, \bibinfo {author} {\bibfnamefont {S.}~\bibnamefont {Taylor}}, \
  and\ \bibinfo {author} {\bibfnamefont {E.}~\bibnamefont {Smith}},\
  }\href@noop {} {\bibfield  {journal} {\bibinfo  {journal} {{\normalfont
  CLAS-NOTE-1998-008,
  \url{https://www.jlab.org/Hall-B/notes/clas_notes98/note98-008.pdf}}}\ }
  (\bibinfo {year} {1998})}\BibitemShut {NoStop}%
\bibitem [{\citenamefont {{e1e target assembly}}()}]{target}%
  \BibitemOpen
  \bibfield  {author} {\bibinfo {author} {\bibnamefont {{e1e target
  assembly}}},\ }\href@noop {} {}\bibinfo {howpublished}
  {\url{https://userweb.jlab.org/~skorodum/e1e_target/tar_e1e_web.pdf}}\BibitemShut
  {NoStop}%
\bibitem [{\citenamefont {Skorodumina}\ \emph
  {et~al.}(2017{\natexlab{a}})\citenamefont {Skorodumina}, \citenamefont
  {Fedotov},\ and\ \citenamefont {Gothe}}]{twopeg-d}%
  \BibitemOpen
  \bibfield  {author} {\bibinfo {author} {\bibfnamefont {{\relax
  Iu}.}~\bibnamefont {Skorodumina}}, \bibinfo {author} {\bibfnamefont {{\relax
  G.V}.}~\bibnamefont {Fedotov}}, \ and\ \bibinfo {author} {\bibfnamefont
  {{\relax R.W}.}~\bibnamefont {Gothe}},\ }\href@noop {} {\bibfield  {journal}
  {\bibinfo  {journal} {CLAS12-NOTE-2017-014}\ } (\bibinfo {year}
  {2017}{\natexlab{a}})},\ \Eprint {http://arxiv.org/abs/1712.07712}
  {arXiv:1712.07712 [physics.data-an]} \BibitemShut {NoStop}%
\bibitem [{\citenamefont {Egiyan}\ \emph {et~al.}(1999)\citenamefont {Egiyan}
  \emph {et~al.}}]{Egian:007}%
  \BibitemOpen
  \bibfield  {author} {\bibinfo {author} {\bibfnamefont {K.}~\bibnamefont
  {Egiyan}} \emph {et~al.},\ }\href@noop {} {\bibfield  {journal} {\bibinfo
  {journal} {CLAS-NOTE-99-007,
  \url{https://www.jlab.org/Hall-B/notes/clas_notes99.html}}\ } (\bibinfo
  {year} {1999})}\BibitemShut {NoStop}%
\bibitem [{\citenamefont {Skorodumina}\ \emph {et~al.}(2020)\citenamefont
  {Skorodumina} \emph {et~al.}}]{my_an_note:2020}%
  \BibitemOpen
  \bibfield  {author} {\bibinfo {author} {\bibfnamefont {{\relax
  Iu}.}~\bibnamefont {Skorodumina}} \emph {et~al.},\ }\href@noop {} {\bibfield
  {journal} {\bibinfo  {journal} {CLAS-Analysis-2021-102 (CLAS-Note-2021-002),
  \url{https://misportal.jlab.org/ul/Physics/Hall-B/clas/index.cfm?note_year=2021}}\
  } (\bibinfo {year} {2020})}\BibitemShut {NoStop}%
\bibitem [{\citenamefont {Skorodumina}()}]{my_thesis:2021}%
  \BibitemOpen
  \bibfield  {author} {\bibinfo {author} {\bibfnamefont {{\relax
  Iu}.}~\bibnamefont {Skorodumina}},\ }\href@noop {} {\emph {\bibinfo {title}
  {\href{}{\normalfont Ph.D. thesis, University of South Carolina, 2021, }}}}\
  (\bibinfo  {publisher}
  {\url{https://www.jlab.org/Hall-B/general/clas_thesis.html}})\BibitemShut
  {NoStop}%
\bibitem [{\citenamefont {Osipenko}\ \emph {et~al.}(2004)\citenamefont
  {Osipenko}, \citenamefont {Vlassov},\ and\ \citenamefont
  {Taiuti}}]{Osipenko:2004}%
  \BibitemOpen
  \bibfield  {author} {\bibinfo {author} {\bibfnamefont {M.}~\bibnamefont
  {Osipenko}}, \bibinfo {author} {\bibfnamefont {A.}~\bibnamefont {Vlassov}}, \
  and\ \bibinfo {author} {\bibfnamefont {M.}~\bibnamefont {Taiuti}},\
  }\href@noop {} {\bibfield  {journal} {\bibinfo  {journal}
  {CLAS-NOTE-2004-020,
  \url{https://www.jlab.org/Hall-B/notes/clas_notes04/2004-020.pdf}}\ }
  (\bibinfo {year} {2004})}\BibitemShut {NoStop}%
\bibitem [{\citenamefont {Khetarpal}()}]{Khetarpal:2010}%
  \BibitemOpen
  \bibfield  {author} {\bibinfo {author} {\bibfnamefont {{\relax
  P.K}.}~\bibnamefont {Khetarpal}},\ }\href@noop {} {\emph {\bibinfo {title}
  {\href{}{\normalfont Ph.D. thesis, Rensselaer Polytechnic Institute, Troy,
  NY, 2010,}}}}\ (\bibinfo  {publisher}
  {\url{https://www.jlab.org/Hall-B/general/clas_thesis.html}})\BibitemShut
  {NoStop}%
\bibitem [{\citenamefont {Ungaro}\ and\ \citenamefont {Joo}()}]{Ungaro:2010}%
  \BibitemOpen
  \bibfield  {author} {\bibinfo {author} {\bibfnamefont {M.}~\bibnamefont
  {Ungaro}}\ and\ \bibinfo {author} {\bibfnamefont {K.}~\bibnamefont {Joo}},\
  }\href@noop {} {\emph {\bibinfo {title}
  {\href{https://userweb.jlab.org/~ungaro/maureepage/proj/pi0/e\_pid/e\_pid.html}{e1-6
  Electron Identification}}}}\ (\bibinfo  {publisher}
  {\url{https://userweb.jlab.org/~ungaro/maureepage/proj/pi0/e\_pid/e\_pid.html}},\
  \bibinfo {address} {CLAS web page})\BibitemShut {NoStop}%
\bibitem [{\citenamefont {Park}\ \emph {et~al.}(2003)\citenamefont {Park} \emph
  {et~al.}}]{KPark:momcorr}%
  \BibitemOpen
  \bibfield  {author} {\bibinfo {author} {\bibfnamefont {K.}~\bibnamefont
  {Park}} \emph {et~al.},\ }\href@noop {} {\bibfield  {journal} {\bibinfo
  {journal} {CLAS-Note-2003-012,
  \url{https://www.jlab.org/Hall-B/notes/clas_notes03/03-012.pdf}}\ } (\bibinfo
  {year} {2003})}\BibitemShut {NoStop}%
\bibitem [{\citenamefont {Ripani}\ \emph {et~al.}(2002)\citenamefont {Ripani}
  \emph {et~al.}}]{Rip_an_note:2002}%
  \BibitemOpen
  \bibfield  {author} {\bibinfo {author} {\bibfnamefont {M.}~\bibnamefont
  {Ripani}} \emph {et~al.},\ }\href@noop {} {\bibfield  {journal} {\bibinfo
  {journal} {CLAS-Analysis-2002-109,
  \url{https://www.jlab.org/Hall-B/secure/analysis/clas_analysis02/2002-109.pdf}}\
  } (\bibinfo {year} {2002})}\BibitemShut {NoStop}%
\bibitem [{\citenamefont {Ripani}\ \emph {et~al.}(2003)\citenamefont {Ripani}
  \emph {et~al.}}]{Ripani:2002ss}%
  \BibitemOpen
  \bibfield  {author} {\bibinfo {author} {\bibfnamefont {M.}~\bibnamefont
  {Ripani}} \emph {et~al.} (\bibinfo {collaboration} {CLAS Collaboration}),\
  }\href {\doibase 10.1103/PhysRevLett.91.022002} {\bibfield  {journal}
  {\bibinfo  {journal} {Phys. Rev. Lett.}\ }\textbf {\bibinfo {volume} {91}},\
  \bibinfo {pages} {022002} (\bibinfo {year} {2003})}\BibitemShut {NoStop}%
\bibitem [{\citenamefont {Fedotov}\ \emph {et~al.}(2007)\citenamefont {Fedotov}
  \emph {et~al.}}]{Fed_an_note:2007}%
  \BibitemOpen
  \bibfield  {author} {\bibinfo {author} {\bibfnamefont {{\relax
  G.V}.}~\bibnamefont {Fedotov}} \emph {et~al.},\ }\href@noop {} {\bibfield
  {journal} {\bibinfo  {journal} {CLAS-Analysis-2007-117,
  \url{https://misportal.jlab.org/ul/Physics/Hall-B/clas/analysisIndex.cfm?note_year=2007}}\
  } (\bibinfo {year} {2007})}\BibitemShut {NoStop}%
\bibitem [{\citenamefont {Fedotov}\ \emph {et~al.}(2009)\citenamefont {Fedotov}
  \emph {et~al.}}]{Fedotov:2008aa}%
  \BibitemOpen
  \bibfield  {author} {\bibinfo {author} {\bibfnamefont {{\relax
  G.V}.}~\bibnamefont {Fedotov}} \emph {et~al.} (\bibinfo {collaboration} {CLAS
  Collaboration}),\ }\href {\doibase 10.1103/PhysRevC.79.015204} {\bibfield
  {journal} {\bibinfo  {journal} {Phys. Rev.}\ }\textbf {\bibinfo {volume}
  {C79}},\ \bibinfo {pages} {015204} (\bibinfo {year} {2009})}\BibitemShut
  {NoStop}%
\bibitem [{\citenamefont {Isupov}\ \emph {et~al.}(2017)\citenamefont {Isupov}
  \emph {et~al.}}]{Isupov:2017lnd}%
  \BibitemOpen
  \bibfield  {author} {\bibinfo {author} {\bibfnamefont {{\relax
  E.L}.}~\bibnamefont {Isupov}} \emph {et~al.} (\bibinfo {collaboration} {CLAS
  Collaboration}),\ }\href {\doibase 10.1103/PhysRevC.96.025209} {\bibfield
  {journal} {\bibinfo  {journal} {Phys. Rev.}\ }\textbf {\bibinfo {volume}
  {C96}},\ \bibinfo {pages} {025209} (\bibinfo {year} {2017})}\BibitemShut
  {NoStop}%
\bibitem [{\citenamefont {Skorodumina}\ \emph
  {et~al.}(2015{\natexlab{b}})\citenamefont {Skorodumina} \emph
  {et~al.}}]{Skorodumina:2015rea}%
  \BibitemOpen
  \bibfield  {author} {\bibinfo {author} {\bibfnamefont {{\relax
  Yu}.}~\bibnamefont {Skorodumina}} \emph {et~al.},\ }\href {\doibase
  10.3103/S1062873815040292} {\bibfield  {journal} {\bibinfo  {journal} {Bull.
  Russ. Acad. Sci. Phys.}\ }\textbf {\bibinfo {volume} {79}},\ \bibinfo {pages}
  {532} (\bibinfo {year} {2015}{\natexlab{b}})},\ \bibinfo {note} {[Izv. Ross.
  Akad. Nauk Ser. Fiz.79,no.4,575(2015)]}\BibitemShut {NoStop}%
\bibitem [{\citenamefont {Skorodumina}\ \emph {et~al.}(2021)\citenamefont
  {Skorodumina}, \citenamefont {Fedotov},\ and\ \citenamefont
  {Gothe}}]{note_mm_distr}%
  \BibitemOpen
  \bibfield  {author} {\bibinfo {author} {\bibfnamefont {{\relax
  Iu}.}~\bibnamefont {Skorodumina}}, \bibinfo {author} {\bibfnamefont {{\relax
  G.V}.}~\bibnamefont {Fedotov}}, \ and\ \bibinfo {author} {\bibfnamefont
  {{\relax R.W}.}~\bibnamefont {Gothe}},\ }\href@noop {} {\bibfield  {journal}
  {\bibinfo  {journal} {CLAS12-NOTE-2021-002}\ } (\bibinfo {year} {2021})},\
  \Eprint {http://arxiv.org/abs/2105.13532} {arXiv:2105.13532 [nucl-ex]}
  \BibitemShut {NoStop}%
\bibitem [{\citenamefont {Skorodumina}\ \emph
  {et~al.}(2017{\natexlab{b}})\citenamefont {Skorodumina}, \citenamefont
  {Fedotov} \emph {et~al.}}]{twopeg}%
  \BibitemOpen
  \bibfield  {author} {\bibinfo {author} {\bibfnamefont {{\relax
  Iu}.}~\bibnamefont {Skorodumina}}, \bibinfo {author} {\bibfnamefont {{\relax
  G.V}.}~\bibnamefont {Fedotov}},  \emph {et~al.},\ }\href@noop {} {\bibfield
  {journal} {\bibinfo  {journal} {CLAS12-NOTE-2017-001}\ } (\bibinfo {year}
  {2017}{\natexlab{b}})},\ \Eprint {http://arxiv.org/abs/1703.08081}
  {arXiv:1703.08081 [physics.data-an]} \BibitemShut {NoStop}%
\bibitem [{\citenamefont {Machleidt}\ \emph {et~al.}(1987)\citenamefont
  {Machleidt}, \citenamefont {Holinde},\ and\ \citenamefont
  {Elster}}]{Machleidt:1987hj}%
  \BibitemOpen
  \bibfield  {author} {\bibinfo {author} {\bibfnamefont {R.}~\bibnamefont
  {Machleidt}}, \bibinfo {author} {\bibfnamefont {K.}~\bibnamefont {Holinde}},
  \ and\ \bibinfo {author} {\bibfnamefont {C.}~\bibnamefont {Elster}},\ }\href
  {\doibase 10.1016/S0370-1573(87)80002-9} {\bibfield  {journal} {\bibinfo
  {journal} {Phys. Rept.}\ }\textbf {\bibinfo {volume} {149}},\ \bibinfo
  {pages} {1} (\bibinfo {year} {1987})}\BibitemShut {NoStop}%
\bibitem [{\citenamefont {Byckling}\ and\ \citenamefont
  {Kajantie}()}]{Byckling:1971vca}%
  \BibitemOpen
  \bibfield  {author} {\bibinfo {author} {\bibfnamefont {E.}~\bibnamefont
  {Byckling}}\ and\ \bibinfo {author} {\bibfnamefont {K.}~\bibnamefont
  {Kajantie}},\ }\href@noop {} {\emph {\bibinfo {title} {{Particle
  Kinematics}\normalfont, University of Jyvaskyla, Jyvaskyla, Finland,
  1971}}}\BibitemShut {NoStop}%
\bibitem [{\citenamefont {Trivedi}\ and\ \citenamefont {Gothe}(2018)}]{Arjun}%
  \BibitemOpen
  \bibfield  {author} {\bibinfo {author} {\bibfnamefont {A.}~\bibnamefont
  {Trivedi}}\ and\ \bibinfo {author} {\bibfnamefont {{\relax
  R.W}.}~\bibnamefont {Gothe}},\ }\href@noop {} {\bibfield  {journal} {\bibinfo
   {journal} {CLAS-Analysis-2019-102,
  \url{https://misportal.jlab.org/ul/Physics/Hall-B/clas/viewFile.cfm/2019-102.pdf?documentId=781}}\
  } (\bibinfo {year} {2018})}\BibitemShut {NoStop}%
\bibitem [{\citenamefont {Mo}\ and\ \citenamefont {Tsai}(1969)}]{Mo:1968cg}%
  \BibitemOpen
  \bibfield  {author} {\bibinfo {author} {\bibfnamefont {{\relax
  L.W}.}~\bibnamefont {Mo}}\ and\ \bibinfo {author} {\bibfnamefont {{\relax
  Y.-S}.}~\bibnamefont {Tsai}},\ }\href {\doibase 10.1103/RevModPhys.41.205}
  {\bibfield  {journal} {\bibinfo  {journal} {Rev. Mod. Phys.}\ }\textbf
  {\bibinfo {volume} {41}},\ \bibinfo {pages} {205} (\bibinfo {year}
  {1969})}\BibitemShut {NoStop}%
\bibitem [{\citenamefont {Markov}\ \emph {et~al.}(2014)\citenamefont {Markov}
  \emph {et~al.}}]{Markov:2014}%
  \BibitemOpen
  \bibfield  {author} {\bibinfo {author} {\bibfnamefont {N.}~\bibnamefont
  {Markov}} \emph {et~al.},\ }\href@noop {} {\bibfield  {journal} {\bibinfo
  {journal} {CLAS-Analysis-2014-106,
  \url{https://misportal.jlab.org/ul/physics/hall-b/clas/viewFile.cfm/2014-106.pdf?documentId=732}}\
  } (\bibinfo {year} {2014})}\BibitemShut {NoStop}%
\bibitem [{\citenamefont {{\relax Ye Tian}}\ and\ \citenamefont
  {Gothe}(2020)}]{Ye_Tian:2017}%
  \BibitemOpen
  \bibfield  {author} {\bibinfo {author} {\bibnamefont {{\relax Ye Tian}}}\
  and\ \bibinfo {author} {\bibfnamefont {{\relax R.W}.}~\bibnamefont {Gothe}},\
  }\href@noop {} {\bibfield  {journal} {\bibinfo  {journal}
  {CLAS-Analysis-2021-101,
  \url{https://misportal.jlab.org/ul/Physics/Hall-B/clas/analysisIndex.cfm?note_year=2021}}\
  } (\bibinfo {year} {2020})}\BibitemShut {NoStop}%
\bibitem [{\citenamefont {{\relax Ye Tian}}\ \emph {et~al.}(2023)\citenamefont
  {{\relax Ye Tian}} \emph {et~al.}}]{CLAS:2022kta}%
  \BibitemOpen
  \bibfield  {author} {\bibinfo {author} {\bibnamefont {{\relax Ye Tian}}}
  \emph {et~al.} (\bibinfo {collaboration} {CLAS Collaboration}),\ }\href
  {\doibase 10.1103/PhysRevC.107.015201} {\bibfield  {journal} {\bibinfo
  {journal} {Phys. Rev. C}\ }\textbf {\bibinfo {volume} {107}},\ \bibinfo
  {pages} {015201} (\bibinfo {year} {2023})}\BibitemShut {NoStop}%
\bibitem [{\citenamefont {Golovatch}\ \emph {et~al.}(2019)\citenamefont
  {Golovatch} \emph {et~al.}}]{Golovach}%
  \BibitemOpen
  \bibfield  {author} {\bibinfo {author} {\bibfnamefont {E.}~\bibnamefont
  {Golovatch}} \emph {et~al.} (\bibinfo {collaboration} {CLAS Collaboration}),\
  }\href {\doibase 10.1016/j.physletb.2018.10.013} {\bibfield  {journal}
  {\bibinfo  {journal} {Phys. Lett.}\ }\textbf {\bibinfo {volume} {B788}},\
  \bibinfo {pages} {371} (\bibinfo {year} {2019})}\BibitemShut {NoStop}%
\bibitem [{\citenamefont {Laforge}\ and\ \citenamefont
  {Schoeffel}(1997)}]{Laforge:1996ts}%
  \BibitemOpen
  \bibfield  {author} {\bibinfo {author} {\bibfnamefont {B.}~\bibnamefont
  {Laforge}}\ and\ \bibinfo {author} {\bibfnamefont {L.}~\bibnamefont
  {Schoeffel}},\ }\href {\doibase 10.1016/S0168-9002(97)00649-9} {\bibfield
  {journal} {\bibinfo  {journal} {Nucl. Instrum. Meth.}\ }\textbf {\bibinfo
  {volume} {A394}},\ \bibinfo {pages} {115} (\bibinfo {year}
  {1997})}\BibitemShut {NoStop}%
\bibitem [{\citenamefont {Bosted}\ \emph {et~al.}()\citenamefont {Bosted} \emph
  {et~al.}}]{Bosted_fit}%
  \BibitemOpen
  \bibfield  {author} {\bibinfo {author} {\bibfnamefont {P.}~\bibnamefont
  {Bosted}} \emph {et~al.},\ }\href@noop {} {}\bibinfo {howpublished}
  {\url{https://userweb.jlab.org/~bosted/fits.html}}\BibitemShut {NoStop}%
\bibitem [{\citenamefont {Bosted}\ and\ \citenamefont
  {Christy}(2008)}]{Bosted:2007xd}%
  \BibitemOpen
  \bibfield  {author} {\bibinfo {author} {\bibfnamefont {{\relax
  P.E}.}~\bibnamefont {Bosted}}\ and\ \bibinfo {author} {\bibfnamefont {{\relax
  M.E}.}~\bibnamefont {Christy}},\ }\href {\doibase 10.1103/PhysRevC.77.065206}
  {\bibfield  {journal} {\bibinfo  {journal} {Phys. Rev.}\ }\textbf {\bibinfo
  {volume} {C77}},\ \bibinfo {pages} {065206} (\bibinfo {year}
  {2008})}\BibitemShut {NoStop}%
\bibitem [{\citenamefont {Darwish}\ \emph {et~al.}(2003)\citenamefont
  {Darwish}, \citenamefont {Arenhovel},\ and\ \citenamefont
  {Schwamb}}]{Darwish:2002qu}%
  \BibitemOpen
  \bibfield  {author} {\bibinfo {author} {\bibfnamefont {{\relax
  E.M}.}~\bibnamefont {Darwish}}, \bibinfo {author} {\bibfnamefont
  {H.}~\bibnamefont {Arenhovel}}, \ and\ \bibinfo {author} {\bibfnamefont
  {M.}~\bibnamefont {Schwamb}},\ }\href {\doibase 10.1140/epja/i2002-10071-3}
  {\bibfield  {journal} {\bibinfo  {journal} {Eur. Phys. J. A}\ }\textbf
  {\bibinfo {volume} {16}},\ \bibinfo {pages} {111} (\bibinfo {year}
  {2003})}\BibitemShut {NoStop}%
\bibitem [{\citenamefont {Tarasov}\ \emph {et~al.}(2011)\citenamefont
  {Tarasov}, \citenamefont {Briscoe}, \citenamefont {Gao}, \citenamefont
  {Kudryavtsev},\ and\ \citenamefont {Strakovsky}}]{PhysRevC.84.035203}%
  \BibitemOpen
  \bibfield  {author} {\bibinfo {author} {\bibfnamefont {{\relax
  V.E}.}~\bibnamefont {Tarasov}}, \bibinfo {author} {\bibfnamefont {{\relax
  W.J}.}~\bibnamefont {Briscoe}}, \bibinfo {author} {\bibfnamefont
  {H.}~\bibnamefont {Gao}}, \bibinfo {author} {\bibfnamefont {{\relax
  A.E}.}~\bibnamefont {Kudryavtsev}}, \ and\ \bibinfo {author} {\bibfnamefont
  {{\relax I.I}.}~\bibnamefont {Strakovsky}},\ }\href {\doibase
  10.1103/PhysRevC.84.035203} {\bibfield  {journal} {\bibinfo  {journal} {Phys.
  Rev. C}\ }\textbf {\bibinfo {volume} {84}},\ \bibinfo {pages} {035203}
  (\bibinfo {year} {2011})}\BibitemShut {NoStop}%
\bibitem [{\citenamefont {Shirokov}\ and\ \citenamefont
  {Yudin}()}]{Shirokov_Yudin:1980}%
  \BibitemOpen
  \bibfield  {author} {\bibinfo {author} {\bibfnamefont {{\relax
  Yu.M}.}~\bibnamefont {Shirokov}}\ and\ \bibinfo {author} {\bibfnamefont
  {{\relax N.P}.}~\bibnamefont {Yudin}},\ }\href@noop {} {\emph {\bibinfo
  {title} {{Yadernaya fizika (Nuclear Physics)\normalfont, Nauka, Moscow,
  1980}}}}\BibitemShut {NoStop}%
\bibitem [{\citenamefont {Rohrlich}\ and\ \citenamefont
  {Eisenstein}(1949)}]{PhysRev.75.705}%
  \BibitemOpen
  \bibfield  {author} {\bibinfo {author} {\bibfnamefont {F.}~\bibnamefont
  {Rohrlich}}\ and\ \bibinfo {author} {\bibfnamefont {J.}~\bibnamefont
  {Eisenstein}},\ }\href {\doibase 10.1103/PhysRev.75.705} {\bibfield
  {journal} {\bibinfo  {journal} {Phys. Rev.}\ }\textbf {\bibinfo {volume}
  {75}},\ \bibinfo {pages} {705} (\bibinfo {year} {1949})}\BibitemShut
  {NoStop}%
\bibitem [{\citenamefont {Cutkosky}\ \emph {et~al.}(1979)\citenamefont
  {Cutkosky}, \citenamefont {Hendrick}, \citenamefont {Alcock}, \citenamefont
  {Chao}, \citenamefont {Lipes}, \citenamefont {Sandusky},\ and\ \citenamefont
  {Kelly}}]{PhysRevD.20.2804}%
  \BibitemOpen
  \bibfield  {author} {\bibinfo {author} {\bibfnamefont {{\relax
  R.E}.}~\bibnamefont {Cutkosky}}, \bibinfo {author} {\bibfnamefont {{\relax
  R.E}.}~\bibnamefont {Hendrick}}, \bibinfo {author} {\bibfnamefont {{\relax
  J.W}.}~\bibnamefont {Alcock}}, \bibinfo {author} {\bibfnamefont {{\relax
  Y.A}.}~\bibnamefont {Chao}}, \bibinfo {author} {\bibfnamefont {{\relax
  R.G}.}~\bibnamefont {Lipes}}, \bibinfo {author} {\bibfnamefont {{\relax
  J.C}.}~\bibnamefont {Sandusky}}, \ and\ \bibinfo {author} {\bibfnamefont
  {{\relax R.L}.}~\bibnamefont {Kelly}},\ }\href {\doibase
  10.1103/PhysRevD.20.2804} {\bibfield  {journal} {\bibinfo  {journal} {Phys.
  Rev. D}\ }\textbf {\bibinfo {volume} {20}},\ \bibinfo {pages} {2804}
  (\bibinfo {year} {1979})}\BibitemShut {NoStop}%
\bibitem [{\citenamefont {Gasparyan}\ \emph {et~al.}(2003)\citenamefont
  {Gasparyan}, \citenamefont {Haidenbauer}, \citenamefont {Hanhart},\ and\
  \citenamefont {Speth}}]{Gasparyan:2003fp}%
  \BibitemOpen
  \bibfield  {author} {\bibinfo {author} {\bibfnamefont {{\relax
  A.M}.}~\bibnamefont {Gasparyan}}, \bibinfo {author} {\bibfnamefont
  {J.}~\bibnamefont {Haidenbauer}}, \bibinfo {author} {\bibfnamefont
  {C.}~\bibnamefont {Hanhart}}, \ and\ \bibinfo {author} {\bibfnamefont
  {J.}~\bibnamefont {Speth}},\ }\href {\doibase 10.1103/PhysRevC.68.045207}
  {\bibfield  {journal} {\bibinfo  {journal} {Phys. Rev. C}\ }\textbf {\bibinfo
  {volume} {68}},\ \bibinfo {pages} {045207} (\bibinfo {year}
  {2003})}\BibitemShut {NoStop}%
\bibitem [{\citenamefont {Vrana}\ \emph {et~al.}(2000)\citenamefont {Vrana},
  \citenamefont {Dytman},\ and\ \citenamefont {Lee}}]{Vrana:1999nt}%
  \BibitemOpen
  \bibfield  {author} {\bibinfo {author} {\bibfnamefont {{\relax
  T.P}.}~\bibnamefont {Vrana}}, \bibinfo {author} {\bibfnamefont {{\relax
  S.A}.}~\bibnamefont {Dytman}}, \ and\ \bibinfo {author} {\bibfnamefont
  {{\relax T.S.H}.}~\bibnamefont {Lee}},\ }\href {\doibase
  10.1016/S0370-1573(99)00108-8} {\bibfield  {journal} {\bibinfo  {journal}
  {Phys. Rept.}\ }\textbf {\bibinfo {volume} {328}},\ \bibinfo {pages} {181}
  (\bibinfo {year} {2000})}\BibitemShut {NoStop}%
\bibitem [{Git()}]{Github:data}%
  \BibitemOpen
  \href@noop {} {}\bibinfo {howpublished}
  {\url{https://github.com/gleb811/two\_pi\_exp\_data/tree/master/Skorodumina\_data\_Q2\_0425\_0975}}\BibitemShut
  {NoStop}%
\bibitem [{\citenamefont {Wu}\ \emph {et~al.}(2005)\citenamefont {Wu} \emph
  {et~al.}}]{Wu:2005wf}%
  \BibitemOpen
  \bibfield  {author} {\bibinfo {author} {\bibfnamefont {C.}~\bibnamefont {Wu}}
  \emph {et~al.},\ }\href {\doibase 10.1140/epja/i2004-10093-9} {\bibfield
  {journal} {\bibinfo  {journal} {Eur. Phys. J.}\ }\textbf {\bibinfo {volume}
  {A23}},\ \bibinfo {pages} {317} (\bibinfo {year} {2005})}\BibitemShut
  {NoStop}%
\bibitem [{\citenamefont {{\relax ABBHHM
  Collaboration}}(1968)}]{ABBHHM:1968aa}%
  \BibitemOpen
  \bibfield  {author} {\bibinfo {author} {\bibnamefont {{\relax ABBHHM
  Collaboration}}} (\bibinfo {collaboration}
  {Aachen-Berlin-Bonn-Hamburg-Heidelberg-Munich}),\ }\href {\doibase
  10.1103/PhysRev.175.1669} {\bibfield  {journal} {\bibinfo  {journal} {Phys.
  Rev.}\ }\textbf {\bibinfo {volume} {175}},\ \bibinfo {pages} {1669} (\bibinfo
  {year} {1968})}\BibitemShut {NoStop}%
\end{thebibliography}%
\bibliographystyle{apsrev4-1}
\end{document}